\documentclass[amsmath,amssymb, aps, prb, twocolumn, notitlepage]{revtex4-2}
\usepackage{graphicx}
\usepackage{dcolumn}
\usepackage{bm}
\usepackage[usenames]{color}
\usepackage{slashed}
\usepackage[utf8]{inputenc}
\usepackage{amsfonts}
\usepackage{amsmath}
\usepackage{amssymb}
\usepackage{hyperref}
\usepackage{latexsym}
\usepackage{color}
\usepackage{mhchem}
\usepackage{braket}
\usepackage{diagbox}
\usepackage{booktabs}
\usepackage{MnSymbol}
\usepackage{float}
\makeatletter
\let\newfloat\newfloat@ltx

\newcommand{\ti}{\text{i}}

\usepackage{algorithm}
\usepackage{algorithmicx}
\usepackage{algpseudocode}

\newcommand{\printfnsymbol}[1]{%
  \textsuperscript{\@fnsymbol{#1}}%
}

\begin{document}
\title{Efficient and quantum-adaptive machine learning with fermion neural networks}
\author{Pei-Lin Zheng}
\affiliation{International Center for Quantum Materials, School of Physics, Peking University, Beijing, 100871, China}
\author{Jia-Bao Wang}
\affiliation{International Center for Quantum Materials, School of Physics, Peking University, Beijing, 100871, China}
\author{Yi Zhang}
\email{frankzhangyi@gmail.com}
\affiliation{International Center for Quantum Materials, School of Physics, Peking University, Beijing, 100871, China}

\date{\today}

\begin{abstract}

Classical artificial neural networks have witnessed widespread successes in machine-learning applications. Here, we propose fermion neural networks (FNNs) whose physical properties, such as local density of states or conditional conductance, serve as outputs, once the inputs are incorporated as an initial layer. Comparable to back-propagation, we establish an efficient optimization, which entitles FNNs to competitive performance on challenging machine-learning benchmarks. FNNs also directly apply to quantum systems, including hard ones with interactions, and offer in-situ analysis without preprocessing or presumption. Following machine learning, FNNs precisely determine topological phases and emergent charge orders. Their quantum nature also brings various advantages: quantum correlation entitles more general network connectivity and insight into the vanishing gradient problem, quantum entanglement opens up novel avenues for interpretable machine learning, etc.

\end{abstract}

\maketitle

\section{Introduction}

Artificial neural networks (ANNs) lay the foundation of cutting-edge machine learning research and artificial intelligence applications \cite{Lecun2015Deep, Jordan2015Machine, MLbook, Silver2016, Silver2017, Jumper2021, Davies2021, agostinelli2019solving}. Their successes rely on their versatile expression and efficient optimization, as back-propagation determines descending gradients collectively in deep architectures \cite{Lecun2015Deep, Jordan2015Machine, MLbook}. Their complexity, however, comes at the price of increasing obscurity, obstructing interpretable machine learning especially desired for scientific utilities. Recently, various studies have applied classical ANNs to quantum data and systems \cite{Carleo2019, Carleo2016, Deng2016, Torlai2018, Melnikov2018, Luo2019, Zhang2020TQC, Fournier2020, Huang2022ML, qlt2016, Melko20161, Ohtsuki2016, Ohtsuki2017, vanNieuwenburg2017, Zhaihui2018, FrankMLZ2, Lian2019, mlstm2019, Bohrdt2019, Rem2019, Peano2021, Arnold2022, Zheng2022, Koolstra2022, Tibaldi2202, Chang2023, Link2023}, which generally require suitable bridging or preprocessing for compatibility \cite{qlt2016, Melko20161, Ohtsuki2016, Ohtsuki2017, vanNieuwenburg2017, Zhaihui2018, FrankMLZ2, Lian2019, mlstm2019, Bohrdt2019, Rem2019, Peano2021, Arnold2022, Zheng2022, Koolstra2022, Tibaldi2202, Chang2023, Link2023}.

\begin{figure}
    \includegraphics[width=1.0\linewidth]{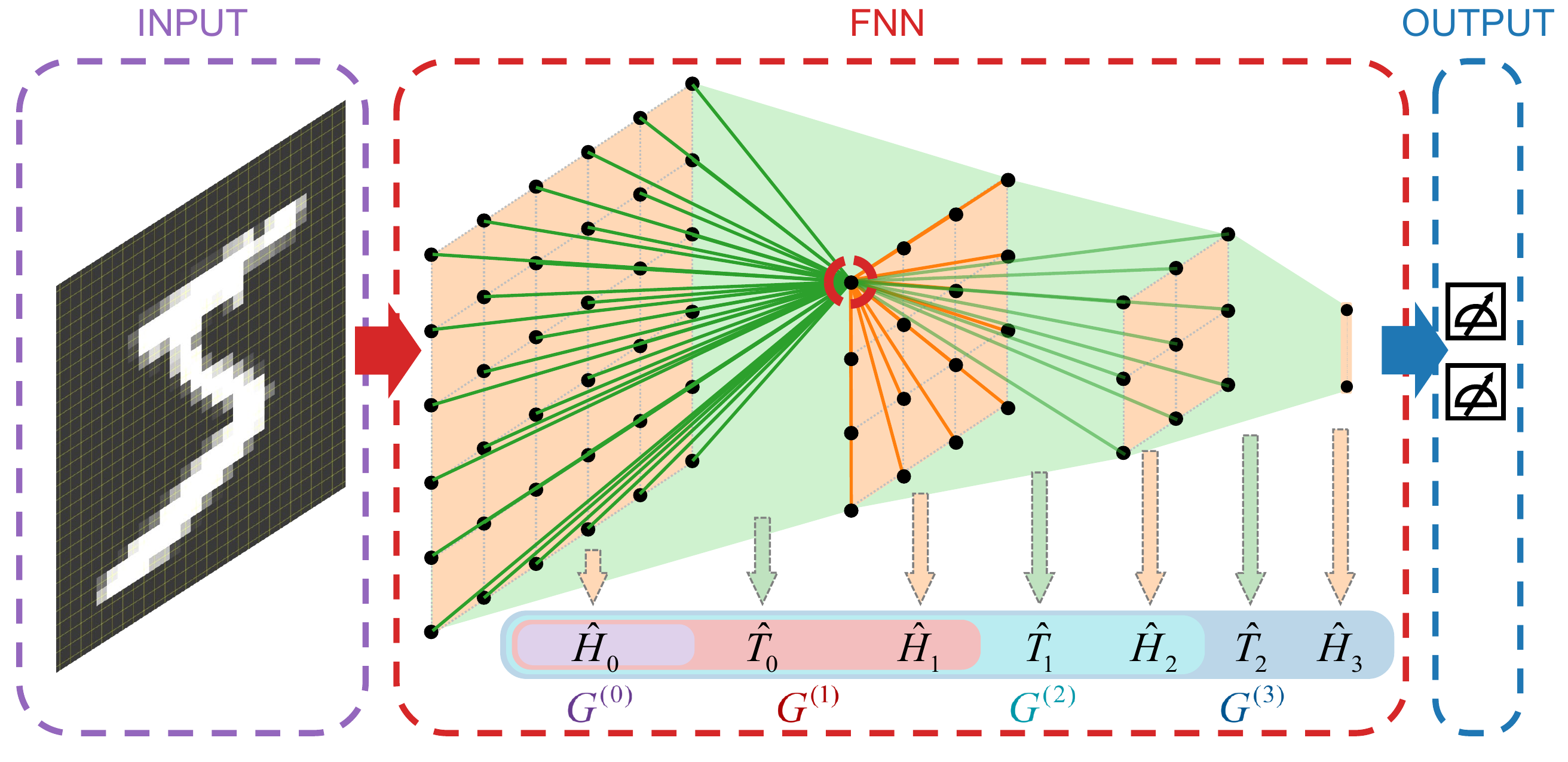}
    \caption{FNN is a fermion model on a network of sites (black dots), whose hopping amplitudes $t_{rr'}$ and onsite potentials $\mu_r$ are parameters for supervised machine learning. It incorporates inputs as the $l=0$ layer. Resulting physical properties, such as LDOS, serve as outputs, available via Green's function $\hat {\bf G}^{(l)}$ recursively, adding one layer ($\hat T_{l-1}$ and $\hat H_l$) at a time with finite $t_{rr'}$ only intra-layer (orange lines) or between neighboring layers (green lines). For clarity, we only show hopping concerning a single site (dashed circle).} \label{fig:FQNN}
\end{figure}

Here, we consider a fermion neural network (FNN) that possesses the benefits of both the classical and quantum worlds. Being a fermion model over a network of sites (neurons), its model parameters, such as inter-site hopping amplitudes and onsite potentials, grant extensive degrees of freedom. The input, regardless of quantum or classical, is incorporated as a part of the model system, whose resulting physical properties, such as local density of states (LDOS) or conditional conductance (CC), exhibit rich expressions and serve as the FNN outputs. Like back-propagation in classical ANNs, we establish efficient optimization for FNNs following a layered architecture (Fig. \ref{fig:FQNN}). We demonstrate machine-learning applications of FNNs on classical MNIST benchmarks \cite{MNIST1998} with excellent efficiency and accuracy.

Importantly, FNNs can apply directly to quantum data and systems and offer in-situ analysis without bridging or preprocessing. Such compatibility is valuable for studying quantum matters whose defining signature is unknown or unavailable or whose physics is too difficult to analyze, such as strongly correlated systems. Indeed, we showcase successful characterizations of topological phases and emergent charge orders with FNN machine learning. Here, we analyze interacting FNN-joint models with Matsubara Green's functions and dynamical mean-field theory (DMFT) \cite{DMFT1996, Metzner1989, Vollhardt2012, VollhardtBook2012}, especially suitable with the FNN's large coordination number and compatible with our efficient optimization.

The FNN's quantum nature also entitles a unique perspective: quantum correlations generalize FNNs' architecture and, like the residual networks \cite{He2016resnet}, address the vanishing gradient problem \cite{VanGrad2001, Han1995sigmoid}; quantum entanglement \cite{Amico2008, Eisert2010} delivers interesting practices of interpretable machine learning \cite{Zhang2020, Ribeiro2016, Arrieta2020, Arnold2022}, including analysis of training dynamics, logic flow, and generative criteria.

\section{FNN for machine learning}\label{sec:fnnintro}

Without loss of generality, our FNN model takes the following form:
\begin{equation}
\hat H = \sum_{rr'} t_{rr'} c^\dagger_{r} c_{r'} + \sum_{r} \mu_{r} c^\dagger_{r} c_{r}, \label{eq:ham}
\end{equation}
where $c_{r}$ is an electron annihilation operator at site $r$. The hopping amplitudes $t_{rr'} = t^{*}_{r'r}$ \footnote{We may generalize Eq. \ref{eq:ham} to non-Hermitian FNNs with $t_{rr'}\neq t_{r'r}^*$ \cite{nhermi2020, nhermi2021} and closer resemblance to classical ANNs with unidirectional information flow.} and the onsite potentials $\mu_{r}$ are model parameters up for machine learning, analogous to the weights and biases of a classical ANN. We label each site (neuron) with $r=(l, m)$, where $l \in [0, L]$ and $m$ denote the layer and the intra-layer coordinate, respectively. We allow finite $t_{rr'}$ only for $|l-l'| \le 1$, including intra-layer hopping as in Hopfield networks \cite{Hopfield1982}; see an illustration of the FNN architecture in Fig. \ref{fig:FQNN}.

The input data or quantum system ($l=0$) connects to the FNN's first layer ($l'=1$) via the hopping terms $t_{(0,m),(1,m')}$, after which the entire quantum model's specific physical properties represent the outputs. For example, we may use the LDOS on the neurons in the last layer ($l=L$, $m=1, \cdots, M_L$):
\begin{equation}
    y_m = -\frac1{\pi}\Im\left[\hat{\bf G}_{L, L}^{(L)}\right]_{m, m}, \label{eq:gfLDOS}
\end{equation}
as categorical outputs - the neuron with the largest LDOS stands for the FNN's decision in a classification problem. Here, $[\hat{\bf G}_{l, l'}^{(L)}]_{m, m'}$ is the matrix element between the $(l, m)$ and $(l' ,m')$ sites of Green's functions $\hat{\bf G}=(z-\hat{H})^{-1}$ for an overall system $\hat{H}$ including $l=0,1,2,\cdots,L$ layers. Likewise, we may use the FNN's CC, i.e., the localization property across its depth:
\begin{equation}
y=\sum_{m}\left|\left[\hat{\bf G}_{0, L}^{(L)}\right]_{m, 1}\right|^2, \label{eq:gfCC}
\end{equation}
as a binary output for an FNN with $M_L=1$. In supervised machine learning, we optimize the model parameters so that the outputs $\bm y$ approach the desired $\bm y_{tar}$ for a training set. Thus, we define a loss function $\mathcal{L}({\bm y, \bm y_{tar}})$ as the divergence between $\bm y$ and $\bm y_{tar}$, and employ stochastic gradient descent with its model-parameter derivatives:
\begin{equation}
\Delta t_{rr'} = -\eta\frac{\partial\mathcal{L}}{\partial t_{rr'}},\
\Delta \mu_{r} = -\eta\frac{\partial\mathcal{L}}{\partial \mu_{r}}, \label{eq:gradesc}
\end{equation}
where $\eta$ is the learning rate. Once the training converges, we can use the FNNs to analyze test inputs.

Importantly, we can evaluate the outputs (Eq. \ref{eq:gfLDOS} and Eq. \ref{eq:gfCC}) and the gradients (Eq. \ref{eq:gradesc}) efficiently via layer-by-layer updates of $\hat {\bf G}$ \cite{RGF2006}:
\begin{eqnarray}
 \hat{\bf G}_{l, l}^{(l)}&=&\left[z\hat{I}-\hat{H}_{l}-\hat{T}_{l-1}^\dagger\hat{\bf G}_{l-1,l-1}^{(l-1)}\hat{T}_{l-1}\right]^{-1}, \nonumber \\
\hat{\bf G}_{i, l}^{(l)}&=&\hat{\bf G}_{i,l-1}^{(l-1)}\hat{T}_{l-1}\hat{\bf G}_{l, l}^{(l)}, \label{eq:RGF}
\end{eqnarray}
where we arrange $\hat{H} = \hat{H}_0+\sum_{l=1}^{L}\hat{H}_l+\hat{T}_{l-1}+\hat{T}^\dagger_{l-1}$ in Eq. \ref{eq:ham} into components $\hat{H}_{l}$ within the $l^{th}$ layer and hopping $\hat{T}_l$ from the $(l+1)^{th}$ to the $l^{th}$ layer. For non-interacting systems, we consider retarded Green's functions in $z=E+\ti\gamma$ at the Fermi energy $E$ plus a small imaginary part $\gamma>0$; for interacting quantum systems, we resort to Matsubara Green's functions in imaginary frequencies $z=\ti\omega_n$. Since the functions mapping each iteration's outputs to the next are highly nonlinear, FNNs possess powerful expressions and chain-rule gradient solutions - merits that made ANN's name in machine learning. Also, Eq. \ref{eq:RGF}'s time complexity is polynomial in the number of neurons $M_l$ in each layer, similar to ANNs. For regularization, we include a weight decay $\lambda$ \cite{wd1991} on both $t_{rr'}$ and $\mu_r$. Further details are in Appendices \ref{app:regularization} and \ref{app:chainrule}.

\section{Classical examples}\label{sec:classical}

First, we apply FNN machine learning on MNIST and encode each image $x_{\vec m} \in [0, 1.0]$, $\vec{m}=(m_x, m_y)$ as onsite potentials of a model:
\begin{equation}
\hat{H}_{0} = \sum_{\vec{m}} x_{\vec{m}} c^\dagger_{\vec{m}} c_{\vec{m}}, \label{eq:encode_ham}
\end{equation}
which is incorporated with the FNN as its $l=0$ layer.

\begin{figure}
\includegraphics[width=0.95\linewidth]{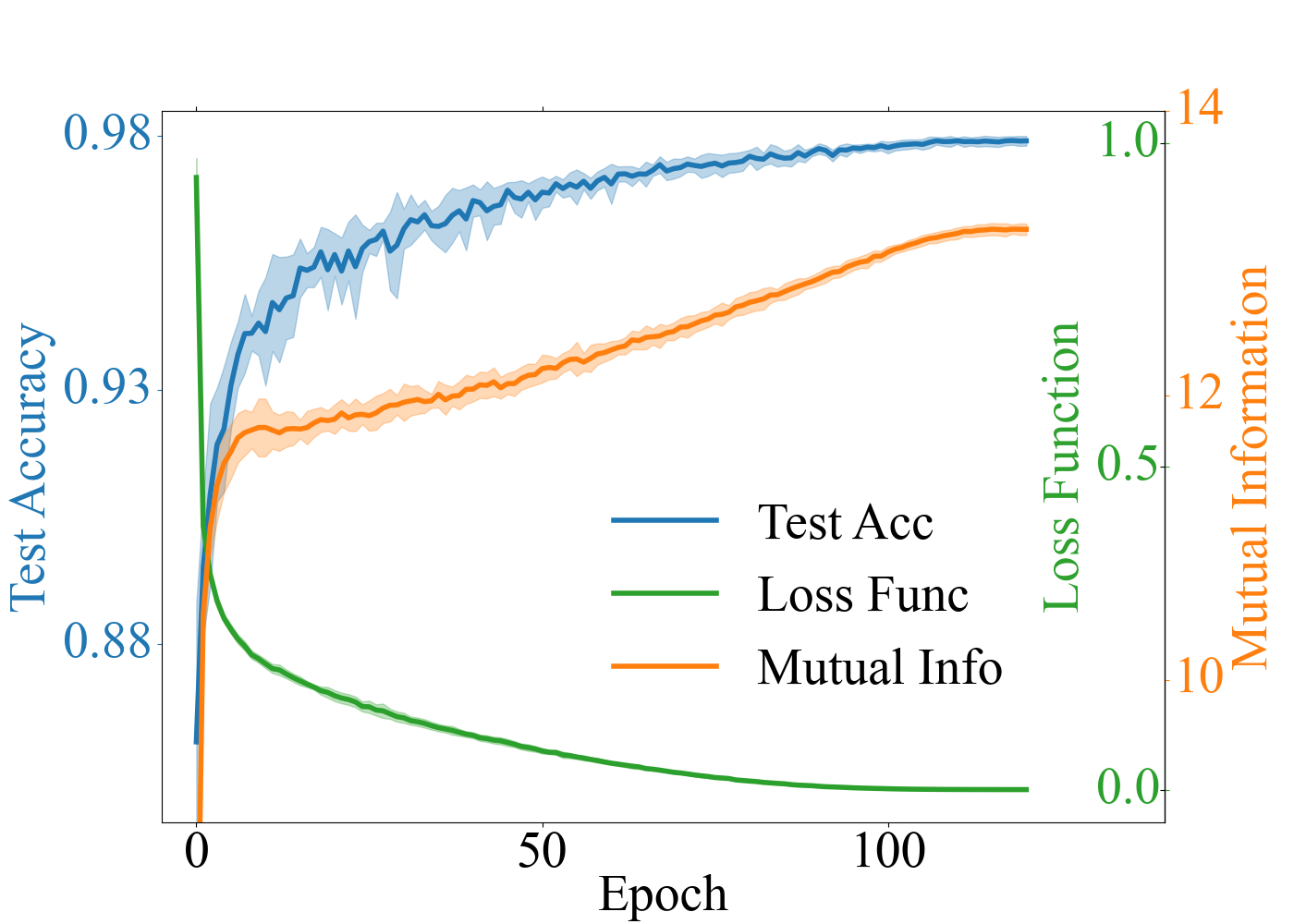}
\caption{The training loss function, the test accuracy, and the mutual information $I_{0L}$ between the input ($l=0$) and the output ($l=L$) layers consistently and dynamically depict the FNN's convergence on MNIST during supervised machine learning. The shades represent standard deviation over ten trials.}
\label{fig:MI}
\end{figure}

We compare different FNN architectures: the first FNN consists of $[100, 64, 10]$ neurons in three layers, with $t_{rr'}$ fully connecting each neuron to every neuron in its own and adjacent layers (Fig. \ref{fig:FQNN}); we also consider a more locally-connected FNN with $[13\times 13, 6 \times 6, 10]$ neurons and local hopping among its $l=1, 2$ layers' 2D lattices; in both FNNs, the $l=3$ layer remains fully-connected, whose LDOS represents the ten MNIST categories. We set $\eta=0.005$, $\gamma=0.005$, and $\lambda=0.001$. Further details and examples are in Appendix \ref{app:architecture}.

With our efficient optimization, supervised machine learning achieves high-quality convergence on MNIST. With a fully-connected FNN, we reach 98.13\% accuracy (Fig. \ref{fig:MI}) \footnote{The state-of-the-art 99.87\% accuracy is achieved by ensemble learning with homogeneous vector capsules and convolutional neural networks, please refer to \url{https://paperswithcode.com/sota/image-classification-on-mnist} for details. Likewise, FNNs have much room for faster convergence and better accuracy, achievable with proper hyper-parameter settings and architecture expansions; see Appendix \ref{app:architecture}}, comparable with fully-connected classical ANNs with approximately the same size ($\sim98.47\%$ \footnote{See benchmark ANN accuracy at \url{http://yann.lecun.com/exdb/mnist/}.}). Interestingly, we still reach a satisfactory 97.36\% accuracy with a locally-connected FNN. We attribute such ease of architectural constraints in FNNs to quantum correlation, which tends beyond the range of hopping $t_{rr'}$ via perturbation series.

We emphasize the diverse possibilities over FNN input and output formalisms. For instance, instead of Eq. \ref{eq:encode_ham}, we may encode an image as an external LDOS field $\Im (\hat{\bf G}_{00})_{\vec m\vec m} \propto x(\vec m)$ coupled to the FNN. The FNN achieves a higher 98.54\% accuracy on MNIST, where we set $\eta=0.001$, $\gamma=0.001$, and $\lambda=0.001$; see further details in Appendix \ref{sec:extLDOS}.

\section{Examples: topological insulators}\label{sec:chern}

Next, we train FNNs to distinguish Chern insulators (Chern number $C=1$) and normal insulators ($C=0$). For example, we consider the non-interacting Hamiltonian \cite{qlt2016}:
\begin{eqnarray}
\hat{H}_0(\kappa)&=&\sum_{\vec{r}}(-1)^y c^{\dagger}_{\vec{r}+\hat{x}}c_{\vec{r}}+\left[1+(-1)^y(1-\kappa)\right]c^{\dagger}_{\vec{r}+\hat{y}}c_{\vec{r}}  \label{eq:Ham_ins}  \nonumber \\
&+&(-1)^y\frac{\text{i}\kappa}{2}\left[c^{\dagger}_{\vec{r}+\hat{x}+\hat{y}}c_{\vec{r}}-c^{\dagger}_{\vec{r}+\hat{x}-\hat{y}}c_{\vec{r}}\right]+\text{h.c.}, \label{eq:Ham_ins}
\end{eqnarray}
on a $12\times 12$ square lattice $\vec{r}=(x,y)$, representing a Chern insulator if $\kappa>\kappa_C=0.5$ and a normal insulator otherwise at Fermi energy $E=0$. We also add quenched disorder for a more diverse training set and confirm the disordered models' topological categories via the real-space Kubo formula \cite{Kitaev2006, qlt2016}; see details in Appendix \ref{app:disChern}. Such topological phases require quantum portrayals hard for classical ANNs, whose machine-learning successes usually relied on presumption-based preprocessing of entanglement \cite{vanNieuwenburg2017}, edge states \cite{Ohtsuki2016, Ohtsuki2017}, or quantum operators \cite{qlt2016, FrankMLZ2}, unavailable for general topological phases.

The FNN directly incorporates each sample model as its $l=0$ layer to offer in-situ analysis via designated physical properties. We first employ an FNN with $[100, [64] \times 3, 2]$ neurons in five layers. The LDOS on its two neurons in the $l=L$ layer represents whether $\hat{H}_0(\kappa)$ is a Chern insulator. For such a binary output, we may also use the CC, i.e., we want an FNN that localizes (with vanishing Green's functions) across its depth if $\hat H_0(\kappa)$ is a normal insulator, and vice versa. This FNN has $[100, [64] \times 5, 1]$ neurons among seven layers. We set $\eta=0.001 \sim 0.005$, $\gamma=0.01$, and $\lambda=0.001$.

\begin{figure}
    \includegraphics[width=1.0\linewidth]{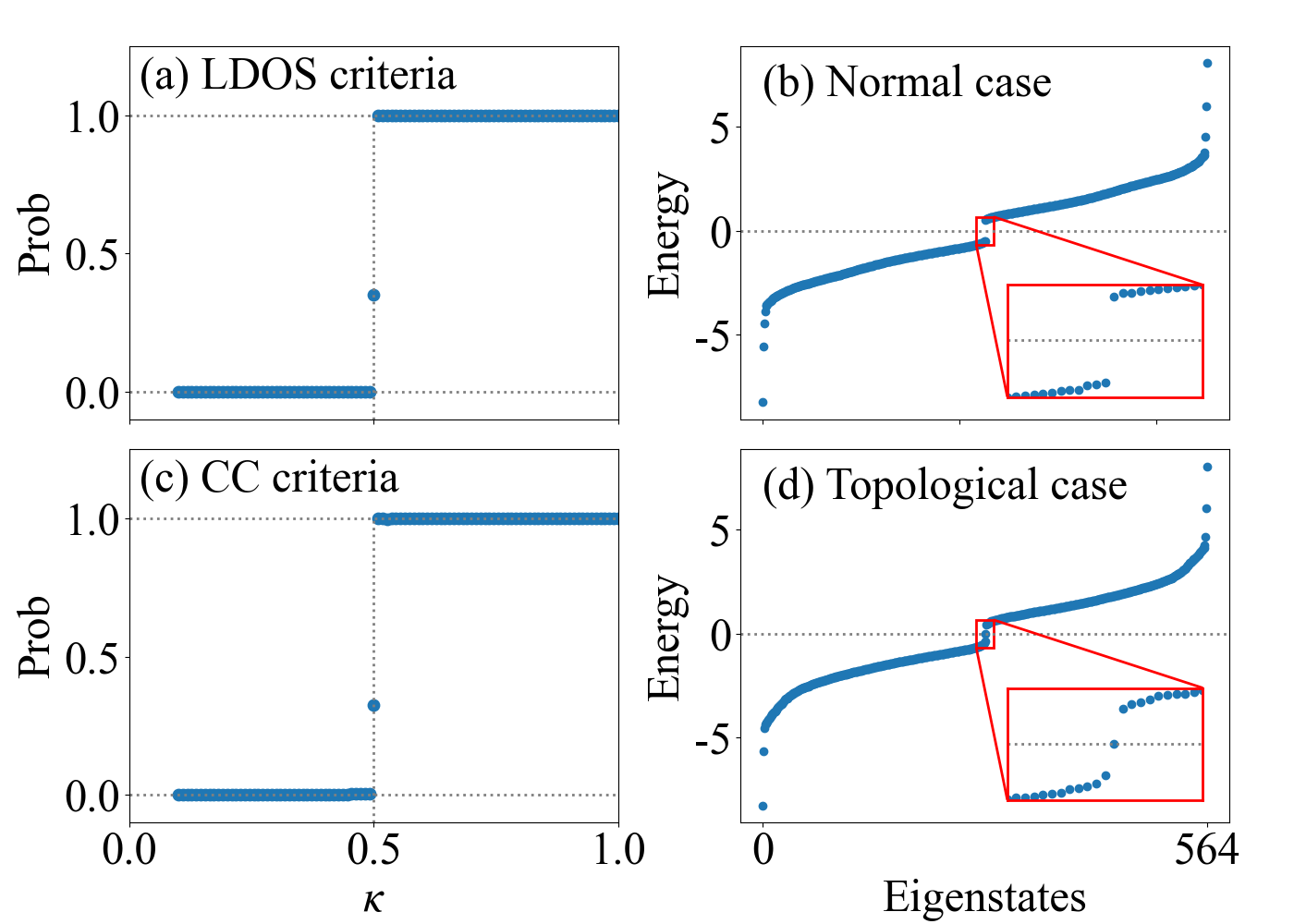}
	\caption{(a) and (c): With FNN's LDOS and CC as outputs, respectively, the ratios of Chern-insulator responses for the clean model in Eq. \ref{eq:Ham_ins} over $\kappa \in [0.1, 1.0]$ are consistent with the phase diagram with $\kappa_C=0.5$. (b) and (d): With FNN's CC as outputs, the energy spectra of the entire system (combining the original insulator and the FNN) differ by the absence or presence of an in-gap state at $E\approx 0$ if the input $\hat H_0$ is a normal or a Chern insulator. All results are after successful supervised machine learning. } \label{fig:PD}
\end{figure}

After convergent supervised machine learning, we apply the FNNs to infer the topological phase of various models. The phase diagram of clean models achieves $100\%$ accuracy and a consistent topological transition at $\kappa=0.5$; see Fig. \ref{fig:PD}(a)(c). Even for models with moderate disorders, the FNNs with LDOS (CC) output achieves 99.83\% (99.51\%) accuracy and 100\% (99.99\%) test area under the receiver operating characteristic curve (AUROC) \cite{AUC1997, ROC2006}. Interestingly, accompanying the change of CC, we observe a sharp contrast in the entire system's energy spectra: while a localization gap around $E\approx 0$ commonly exists (Fig. \ref{fig:PD}(b)) when the trained FNN combines a normal insulator, a state responsible for Green's functions' extended behaviors emerges at $E\approx 0$ (Fig. \ref{fig:PD}(d)) when the incorporated input model is a Chern insulator. Here in Fig. \ref{fig:PD}(b)(d), we have chosen a different FNN with 98.1\% accuracy for clearer signatures. Such physics is useful for FNN interpretability, as we will demonstrate.

\section{Examples: strongly correlated systems}\label{sec:scs}

Strongly correlated systems pose challenges to conventional analysis and, in turn, bridging or preprocessing for classical machine learning. However, we can analyze target properties and phases by FNN machine learning with direct quantum input. Here, we consider the Falicov-Kimbal (FK) model on a 2D square lattice \cite{Hettler1998, FK2003, FKBrandt1989, FKBrandt1990, FKModelQMC2006}:
\begin{eqnarray}
\hat{H}_0 &=& t\sum_{\langle ij\rangle} c^\dagger_i c_j +t'\sum_{\llangle ik\rrangle} c^\dagger_i c_k + \text{h.c.}\nonumber\\
& &  - \mu \sum_i c^\dagger_i c_i + E_f\sum_i n^f_{i} + U\sum_i c^\dagger_i c_i n^f_{i}, \label{eq:fkmodel}
\end{eqnarray}
where $n^f_{i}=\{0, 1\}$ is the number of localized $f$ electrons at site $i$. We focus on scenarios where the $f$ ($c$) electrons are exactly (nearly) at half-filling via adjusting $E_f$ ($\mu$). We set $t=1$ as our unit of energy. At temperature $T=0$, a checkerboard (stripe) charge order emerges at small (large) next-nearest neighbor hopping, separated by a phase transition at $t'_C/t \sim 0.7$ and half-filling; at larger $T$, the critical point broadens into an intermediate region, and the insulating charge orders may give way to a disordered metal \cite{FKModelQMC2006}. We further summarize and discuss the phase diagrams of the Falicov-Kimbal model in Appendices \ref{sec:fkpd} and \ref{sec:fkdmft}.

We apply FNN to infer a phase diagram versus $t'/t$ at low $T=0.005$. First, we incorporate $\hat{H}_0$ in Eq. \ref{eq:fkmodel} as the $l=0$ layer of the FNN, which retains the same architecture as the topological-insulator case - the last layer's LDOS signals the checkerboard and stripe charge orders, respectively. In training set, we include diverse models with variable $t'/t$ and $\mu$ close to half-filling and deep in the respective charge orders. After supervised machine learning, we apply the FNN to probe the charge ordering of models with varying $t'/t$ traversing phase transitions; see Fig. \ref{fig:FKmodelPD} inset for the parameter settings. In Appendix \ref{sec:FNNMLFK2}, we also study a metal-insulator phase diagram versus the interacting strength $U$ at relatively higher temperatures (e.g., $T=0.11$) for the pristine FK model ($t'=0$).

\begin{figure}
	\includegraphics[width=1.0\linewidth]{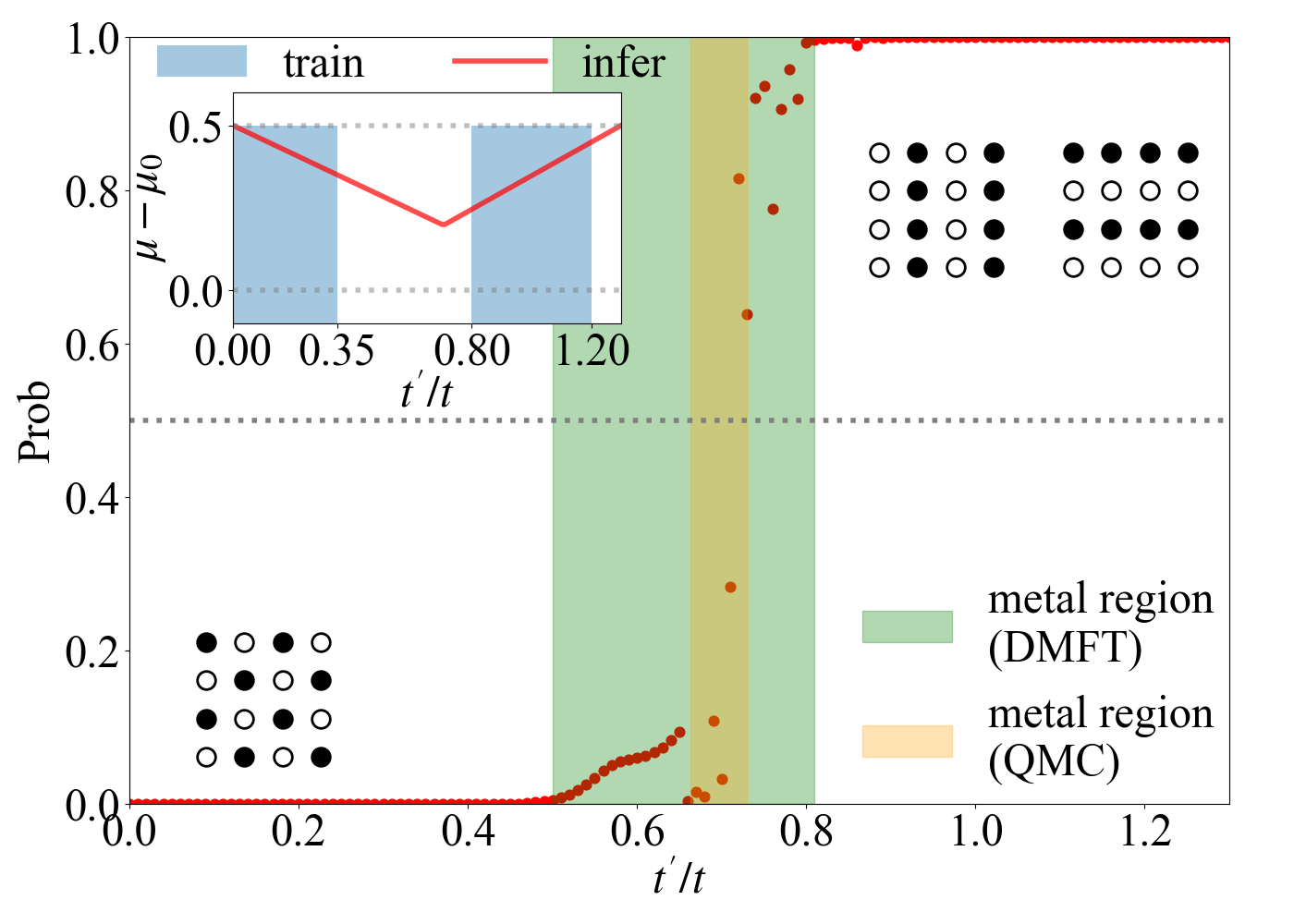}
	\caption{Using LDOS as outputs, the FNN's responses on the FK models in Eq. \ref{eq:fkmodel} indicate a transition from a checkerboard to a stripe charge order. $T=0.005$ and $U=1.0$. The yellow and green shadows are the disordered transition regions upon the same model space, as determined by DMFT in Appendix \ref{sec:fkdmft} and QMC approaches \cite{FKModelQMC2006}. The upper-left inset shows the training set in the ($t'/t$, $\mu$) parameter space (shaded blue region) and the choice of $\mu$ when inferring the phase diagram (red line). $\mu_0$ is the half-filling value of $\mu$ at $U=E_f=0$.}
	\label{fig:FKmodelPD}
\end{figure}

With a finite interaction $U$ present in the $l=0$ layer, the entire FNN system becomes an interacting quantum problem. Therefore, instead of Green's functions at the Fermi energy $E$, we analyze and optimize the FNN via Matsubara Green's functions $\hat{\bf G}(\ti\omega_n)$ and self-energies $\Sigma_r(\ti\omega_n)$ in imaginary frequencies $\omega_n = 2\pi T(n+1/2)$, $n\in \mathbb{Z}$. Such formalism would also work if interactions were present in the FNN itself ($l=1,2,\cdots$). Such interactions also help generalize explicitly beyond the constraints of local self-energies introduced by DMFT routines. Since the recursions in Eq. \ref{eq:RGF} remain valid, and the LDOS at site $r$ takes the form \cite{Nandini1995}:
\begin{equation}
\rho(r) = -\frac{1}{\pi T}G\left(r, \tau=\frac{1}{2T}\right) = \frac{i}{\pi}\sum_n (-1)^n \hat{\bf G}(\ti\omega_n)_{rr},
\end{equation}
we can efficiently trace the loss function to FNN parameters via the chain rule across multiple $n\in [-n_0, n_0-1]$ in parallel. We set $n_0 = 20$ for demonstrations. Further algorithmic details are in Appendix \ref{app:mats_alg}.

Meanwhile, we resort to DMFT for self-energies $\Sigma_r(\ti\omega_n)$, mapping each site $r$ onto a local impurity problem with a dynamical environment \cite{DMFT1996, Metzner1989, Vollhardt2012, VollhardtBook2012}. DMFT may have limited control over the original quantum systems, especially in low dimensions. Indeed, comparisons with controlled quantum Monte Carlo (QMC) calculations \cite{FKModelQMC2006} show that DMFT overestimates the small yet finite transition window between the two charge orders, a reflection of its approximate nature; see Appendix \ref{sec:fkdmft}. However, after coupling with an FNN, especially a fully-connected one, the entire system obtains a large coordination number that makes DMFT more suitable \footnote{The number of nearest neighbors in the original 2D plane remains limited even after coupling to FNNs; therefore, we may generalize DMFT to local clusters, i.e., dynamical cluster approximation \cite{Hettler1998}, for better-controlled analysis.}. In practice, we implement machine learning by alternating between the gradient-descent optimizations (FNN parameters) in real space and DMFT calculations (self-energies) in imaginary-frequency space and establishing consistency between them; see details in Appendix \ref{app:mats_alg}. Such a quantum many-body algorithm does not resemble any classical ANNs.

A well-trained FNN can deliver consistent and confident results on FK models in either the checkerboard or stripe charge-ordering phase, achieving both 100\% test accuracy and AUROC. We summarize the results in Fig. \ref{fig:FKmodelPD}. We emphasize that our analysis generally differs from direct implementation of DMFT upon the target models (Fig. \ref{fig:FKmodelPD} and Appendix \ref{sec:fkdmft}), neither does the FNN make decisions by the (residue) order parameters presented in its $l=0$ layer, which essentially vanishes and leaves no trace of the suggested phase diagram - once incorporated into an FNN, the physics of the quantum model as a part of the overall system differs drastically from its independent self. Instead, we establish direct connections between the target physics and FNN outputs via supervised machine learning, an unprecedented perspective for strongly correlated systems, where QMC is commonly unavailable due to the sign problem and ED and DMRG are limited to small systems. In addition, due to competing orders and domains, real-space implementations of DMFT are sometimes unstable and hard to converge and evaluate; FNN machine learning fully circumvents such difficulties.

\section{Physical insights and interpretability from FNN machine learning}\label{sec:intp}

\begin{figure}
\includegraphics[width=0.95\linewidth]{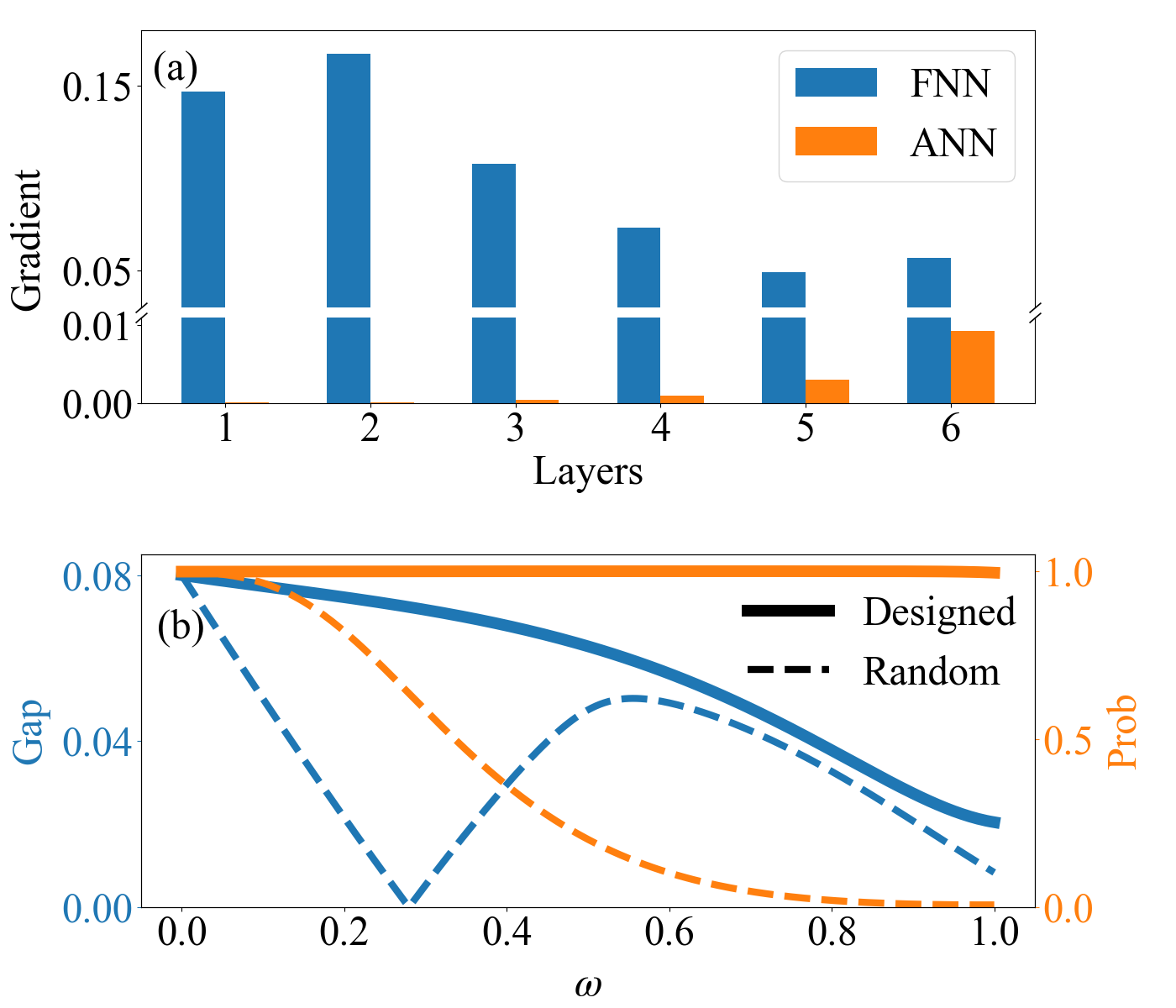}
\caption{(a): The (modulus) gradient distributions among ANN and FNN (with CC output) show the former's vanishing gradient problem and the latter's more even distribution, which achieves similar effects as residual networks despite sequential architectures. The results are averaged over the first three epochs and normalized with the last layer's values (not shown). We employ identical architecture, initialization, and models for ANN and FNN and preprocess the former's dataset through quantum operators \cite{qlt2016}. (b): While random perturbations topple a near-critical Chern insulator $\kappa=0.52 \gtrsim \kappa_C$ quickly, perturbations following the Chern-insulator criteria generalize the topological phase until much larger strength $\omega$. We may track the transition with the spectral gap or FNN output. } \label{fig:adv}
\end{figure}

FNN also offers physics insights into the vanishing-gradient problem - decreasing gradients on layers farther from outputs, which had plagued deep ANNs until convolutional neural networks \cite{Fukushima1980, MNIST1998} and residual blocks \cite{He2016resnet}. Similar vanishing gradients known as barren plateaus also afflict quantum-circuit optimization \cite{McClean2018, Cerezo2021BP, Wang2021}. From a response-theory perspective, the correlation between outputs and model parameters naturally decays with their distances on FNNs respecting locality over the layers. However, as the CC's contributions distribute across the entire depth, such outputs depend more evenly on parameters from all layers. Indeed, as shown in Fig. \ref{fig:adv}(a), a sharp contrast emerges between the gradient distributions among FNNs (with CC output) and classical ANNs during supervised machine learning. We discuss the mechanism behind the physics and differences from residual blocks in Appendix \ref{app:chainrule}.

In addition, FNNs allow novel quantum perspectives for interpretable machine learning. For example, mutual information measures the entanglement between subregions $A$ and $B$ \cite{Adami1997, Groisman2005}:
\begin{equation}
I_{AB}=S_A+S_B-S_{AB}, \label{eq:mi}
\end{equation}
where $S_A=-\text{tr}(\hat \rho_A \log \hat \rho_A)$ and $\hat \rho_A=\text{tr}_{\bar A}(\hat \rho)$ are the entanglement entropy and (reduced) density operator on $A$. Thus, the mutual information $I_{0L}$ between the $l=0$ and $l=L$ layers evaluates an FNN's capacity to orderly transform inputs into outputs, an alternative measure of machine-learning progress; see Fig. \ref{fig:MI} for such mid-training dynamics. Similarly, given a target input, we can use mutual information to simplify and visualize the logic flow along FNNs, as we demonstrate in Appendix \ref{app:logicflow}. 

FNNs with CC outputs (Figs. \ref{fig:PD}(b) and (d)) also support physics-guided interpretive and generative machine learning. For example, we can formulate Chern-insulator criteria from trained FNNs. As we have discussed in Sec. \ref{sec:chern}, the target CC behavior requires an emergent state at $E\eqsim 0$ (Fig. \ref{fig:PD}(d)), which suggests the Hamiltonian of the overall system possesses a zero eigenvalue:
\begin{equation}
\det\begin{pmatrix} \hat H_0 & \hat T_0 \\ \hat T_0^\dagger & \hat H_{l \ge 1} \end{pmatrix} = 0, \label{eq:det}
\end{equation}
when the input model $\hat H_0$ is a Chern insulator. Here, $\hat{T}_0$ is the hopping from the FNN first layer ($l=1$) to the input model ($l=0$), and vice versa for $\hat{T}^\dagger_0$. $\hat H_{l \ge 1}$ is the Hamiltonian of the $l=1, 2, \cdots, L$ layers of the FNN. Vital information condensed from supervised machine learning is encoded in $\hat H_{l \ge 1}$, $\hat{T}_0$, and $\hat{T}_0^\dagger$.

To see what kind of $\hat H_0$ satisfies Eq. \ref{eq:det}, we note that for a given Chern insulator $\hat H_{ch}$ that satisfies Eq. \ref{eq:det}, the models:
\begin{equation}
\hat H_{0}  = \hat H_{ch} + \sum_i \omega_i \hat{h}_{\delta i} + \omega^{*}_i \hat{h}^\dagger_{\delta i}, \label{eq:Htcriteria}
\end{equation}
also satisfies Eq. \ref{eq:det} for small $\omega_i$, and should also be determined by the FNN as a Chern-insulator. Here, $\hat h_{i}$ $(i=1, 2, \cdots)$ are normalized columns linearly dependent on the overall Hamiltonian in Eq. \ref{eq:det}. We also require the single-column matrices $\hat h_{\delta i}$ are physically local. Interpretive criteria such as Eq. \ref{eq:Htcriteria} empower us to employ an FNN, trained for discriminative objectives, as a generative model. We can add perturbations following such criteria to generate new Chern-insulating models, like image AI generates new pictures with same contents but modified styles \cite{StableDiffusion_2022}.

To verify, we start with a Chern insulator $\hat H_{ch}$ close to the topological transition, e.g., Eq. \ref{eq:Ham_ins} with $\kappa=0.52$, and evaluate the consequences of perturbations either random or directed by the criteria in Eq. \ref{eq:Htcriteria}. While random perturbations generally harm the Chern insulator and topple it relatively easily, the designed perturbations keep the model steadily in the topological phase until the amplitude of the perturbation becomes fairly large; see Fig. \ref{fig:adv}(b).

\section{Discussion}

We have shown that FNN inherits classical ANNs' efficient optimization and powerful expression while exhibiting applicability in both classical and quantum worlds. By direct quantum input and offering in-situ machine learning, FNNs not only retain a potentially advantageous quantum perspective on quantum problems \cite{Harrow2017, Shor1997, Terhal2004, Aaronson2013, Aaronson2017, Bravyi2018, Liu2021, Huang2021QA, Huang2022QA, Arute2019, Zhong2020, Wu2021}, but also forego the costly and knowledge-reliant preprocessing, such as projective and response measurements on the original quantum systems. Also, we can choose physical measurables with readily available response measurements as FNN outputs \cite{Tersoff1985, Crommie1993, Wees1988}. By implementing FNN machine learning in quantum experiments and simulations, we make identifying physical properties and phases, sometimes complex with current instruments and routines, much more straightforward. For example, such modules will be convenient for analyzing quantum neural networks and circuits \cite{biamonte2017, Cerezo2021, ParaShift2018, ParaShift2019, Cong2019, Bittel2021, Abbas2021, Cerezo2022, Wang2023}, whose otherwise projective measurements are challenging and require post-processing. 

With non-dynamical and local setups, FNNs are viable in analog quantum simulations \cite{Buluta2009, Georgescu2014}, which have recently exercised precise controls \cite{Barthelemy2013, Browaeys2020} and reached hundreds of sites \cite{Scholl2021, Ebadi2022, Bluvstein2021}, especially in quantum-dot \cite{Hanson2007, Barthelemy2013, Byrnes2008, Braakman2013, Hensgens2017, Diepen2021} and Rydberg-atom arrays \cite{Saffman2010, Weimer2010, Omran2019, Browaeys2020, Moreno2021, Scholl2021, Bluvstein2021, Ebadi2022, Cuadra2023}. Finally, FNN machine learning offers an efficient avenue for designing novel functions, such as CC, in quantum systems.

\section*{Acknowledgments} 
We thank Zhenduo Wang for insightful discussions, and support from the National Key R\&D Program of China (Grant No. 2021YFA1401900) and the National Natural Science Foundation of China (Grant No. 12174008 and Grant No. 92270102). The computation was supported by the High-performance Computing Platform of Peking University. The source code is available in Ref. \footnote{\url{https://github.com/PeilinZHENG/FQNN}}.

\appendix

\section{FNN regularization}\label{app:regularization}

\begin{figure}
    \includegraphics[width=1.0\linewidth]{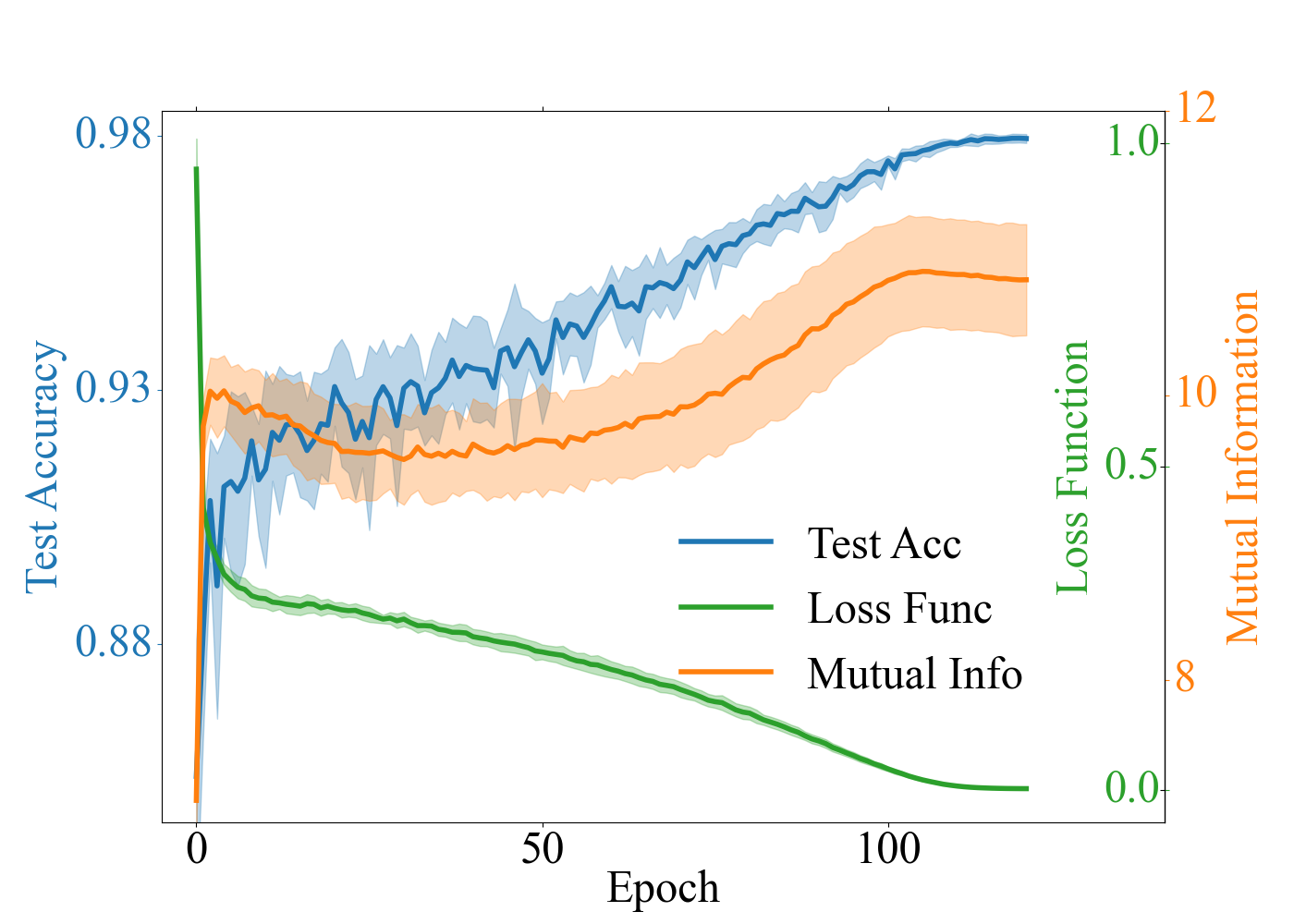}
	\includegraphics[width=1.02\linewidth]{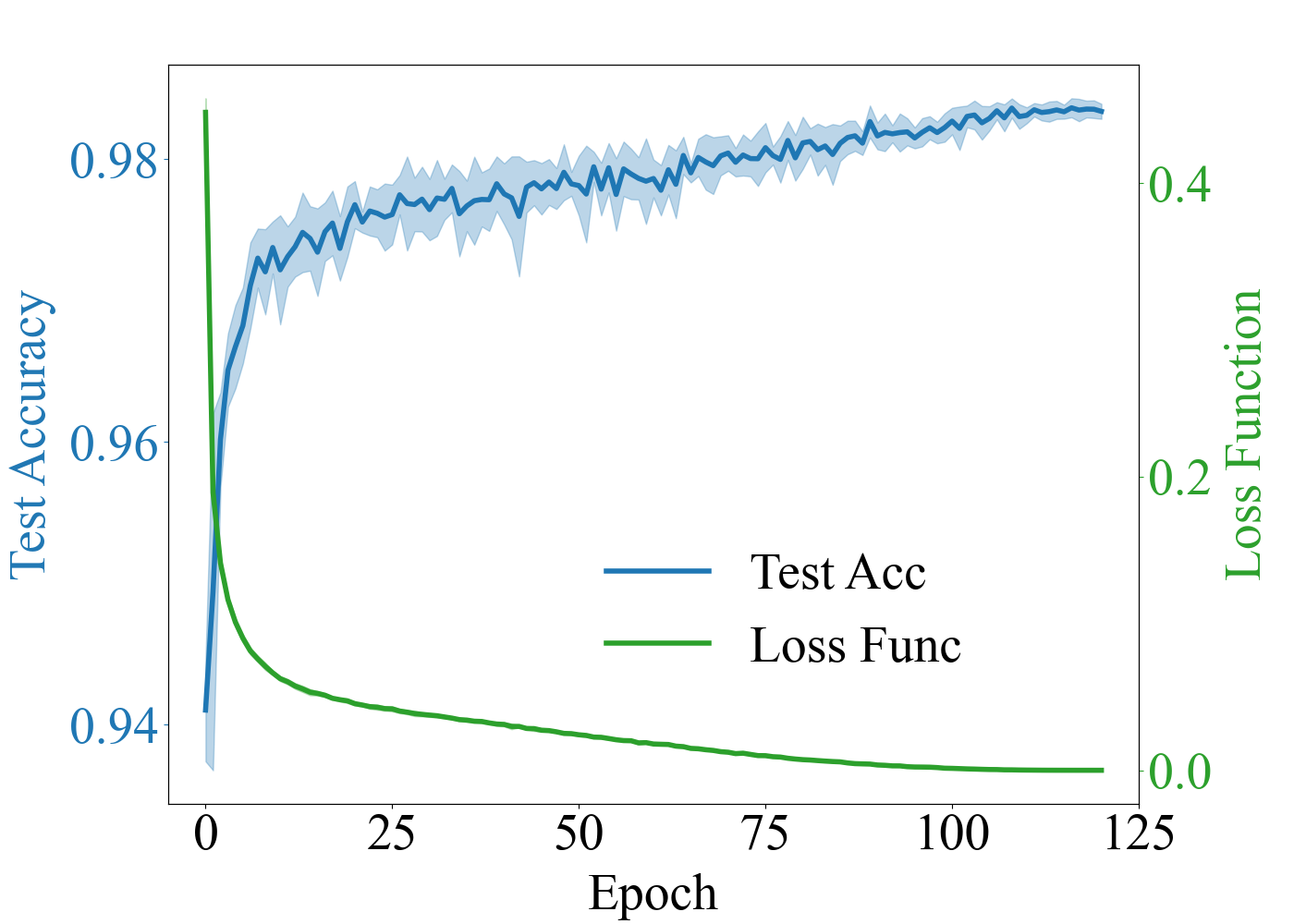}
	\caption{The FNN performance on the MNIST dataset receives an additional boost by including a weight decay $\lambda=0.001$ as regularization. In comparison to Fig. \ref{fig:MI} (image input as onsite potentials, the top) and Fig. \ref{fig:LA} (image input as an external LDOS field, the bottom), the FNNs trained with $\lambda=0.001$ are generally capable of achieving higher test accuracy at convergence. } \label{fig:reg}
\end{figure}

Regularization plays a vital role in reducing over-fitting and reaching optimal performance in supervised machine learning. Common regularization approaches for classical ANNs are weight decay \cite{wd1991} and dropout \cite{dropout2014}. Noting the similarity between FNN's $\hat{T}_{l-1}$ and $\hat{H}_l$ with classical ANN's weights and biases, we have studied carrying over a weight decay during FNNs' supervised machine learning. Indeed, we have observed improved results by including a small weight decay of $\lambda=0.001$ in various scenarios; see Fig. \ref{fig:reg} for examples. In addition, most of the optimal results in Tabs. \ref{tab:MNIST} and \ref{tab:MNIST_LDOS} are obtained with the weight decay. As a caveat, we also observe increased fluctuations in mid-training performances with such regularization.

Another regularization is the small imaginary part $\gamma$ added to the energy $E$. Physically, such $\gamma$ introduces a finite level width, reduces singularities, and may originate from the averaged effect of random quenched disorder. Like dropout introduces effective averaging over multiple descendants of a classical ANN, a proper value of $\gamma$ may represent an average over various random disorder configurations over the FNN, thus reducing sensitivity towards details and enhancing generality. A closer comparison with dropout for classical ANNs would suggest training with variable disorder configurations, which are then replaced with a corresponding $\gamma$ for tests and applications. We leave such endeavors and more throughout investigations on FNN regularization to future studies.

\section{Gradient chain rule from recursive Green's functions}\label{app:chainrule}

To characterize the quantum model in Eq. \ref{eq:ham}, we can evaluate the expectation values $\langle \Phi |\hat O|\Phi \rangle$ for the many-fermion ground state $|\Phi\rangle =\prod_{\epsilon_n \le E} c^\dagger_{n} |0\rangle$, where $c^\dagger_{n}$ corresponds to the energy eigenstates $\hat H = \sum_n \epsilon_n c^\dagger_{n} c_{n}$ and we set the Fermi energy $E=0$. Alternatively, we may resort to Green's functions:
\begin{equation}
    \hat G(z)=\left(z\hat I-\hat H\right)^{-1}, \label{eq:greens}
\end{equation}
where $z=E+\ti\gamma$ possesses a small imaginary part $\gamma>0$ that attributes a finite level width and avoids singularities for retarded Green's functions, or $z=\ti\omega_n$ for Matsubara Green's functions.

The recursive Green's function method \cite{RGF2006} derives Eq. \ref{eq:greens} via recursion. Starting from the system's first layer, the recursive expressions ($i, j<N$):
\begin{eqnarray}
\hat{\bf G}_{N, N}^{(N)}&=&\left[z\hat{I}-\hat{H}_{N}-\hat{T}_{N-1}^\dagger\hat{\bf G}_{N-1,N-1}^{(N-1)}\hat{T}_{N-1}\right]^{-1}\ , \nonumber \\
\hat{\bf G}_{i, N}^{(N)}&=&\hat{\bf G}_{i,N-1}^{(N-1)}\hat{T}_{N-1}\hat{\bf G}_{N, N}^{(N)}\ , \nonumber \\
\hat{\bf G}_{N, j}^{(N)}&=&\hat{\bf G}_{N, N}^{(N)}\hat{T}_{N-1}^\dagger\hat{\bf G}_{N-1,j}^{(N-1)}\ ,\label{eq:RGF4} \\
\hat{\bf G}_{i,j}^{(N)}&=&\hat{\bf G}_{i,j}^{(N-1)} +\hat{\bf G}_{i,N-1}^{(N-1)}\hat{T}_{N-1}\hat{\bf G}_{N, N}^{(N)}\hat{T}_{N-1}^\dagger\hat{\bf G}_{N-1,j}^{(N-1)}\ , \nonumber
\end{eqnarray}
yield the Green's functions of a system with the first $N$ layers:
\begin{equation}
    \begin{split}
        &\hat{\bf G}^{(N)}=
        \begin{pmatrix}
        \hat{\bf G}_{0,0}^{(N)} & \cdots & \hat{\bf G}_{0,N-1}^{(N)} & \hat{\bf G}_{0,N}^{(N)}\\
        \vdots & \ddots & \vdots & \vdots \\
        \hat{\bf G}_{N-1,0}^{(N)} & \cdots & \hat{\bf G}_{N-1,N-1}^{(N)} & \hat{\bf G}_{N-1,N}^{(N)} \\
        \hat{\bf G}_{N,0}^{(N)} & \cdots & \hat{\bf G}_{N,N-1}^{(N)} & \hat{\bf G}_{N,N}^{(N)}
        \end{pmatrix},
    \end{split}\label{eq:G^(N+1)}
\end{equation}
given those of a system with the first $N-1$ layers,
\begin{equation}
    \begin{split}
        &\hat{\bf G}^{(N-1)}=
        \begin{pmatrix}
        \hat{\bf G}_{0,0}^{(N-1)} & \cdots &  \hat{\bf G}_{0,N-1}^{(N-1)}\\
        \vdots & \ddots & \vdots \\
        \hat{\bf G}_{N-1,0}^{(N-1)} & \cdots & \hat{\bf G}_{N-1,N-1}^{(N-1)}
        \end{pmatrix}, \\
    \end{split}\label{eq:G^(N)}
\end{equation}
adding one layer at each iteration until we reach the entire system ($N=L$), whose physical properties correspond to the FNN outputs. In practice, one first solves for $\hat{\bf G}_{N, N}^{(N)}$ in the first line in Eq. \ref{eq:RGF4}, which is then plugged into the second to last lines.

Such iterative dependencies also allow us to utilize the chain rule and collectively obtain the differentials of the loss function $\mathcal L$ with respect to the FNN parameters implicitly in  $\hat{T}$ and $\hat{H}$, whose definitions are below Eq. \ref{eq:RGF}. For convenience, we define the following real matrices \cite{ComMat1974}:
\begin{eqnarray}
    \bm{T}^{(N-1)}&=&
    \begin{pmatrix}
    \Re\left(\hat{T}_{N-1}\right) & \Im\left(\hat{T}_{N-1}\right) \\
    -\Im\left(\hat{T}_{N-1}\right) & \Re\left(\hat{T}_{N-1}\right)
    \end{pmatrix}, \nonumber \\
    \bm{H}^{(N)}&=&
    \begin{pmatrix}
    \Re\left(\hat{H}_{N}\right) & \Im\left(\hat{H}_{N}\right) \\
    -\Im\left(\hat{H}_{N}\right) & \Re\left(\hat{H}_{N}\right)
    \end{pmatrix},\\
    \bm{O}^{(N)}&=&
    \begin{pmatrix}
    \Re\left(\hat{\bf G}_{N,N}^{(N)}\right) & \Im\left(\hat{\bf G}_{N,N}^{(N)}\right)\\
    -\Im\left(\hat{\bf G}_{N,N}^{(N)}\right) & \Re\left(\hat{\bf G}_{N,N}^{(N)}\right)
    \end{pmatrix}=\left(\bm{X}^{(N)}\right)^{-1}, \nonumber\\
    \bm{X}^{(N)}&=&z\hat{I}-\bm{H}^{(N)}-\left(\bm{T}^{(N-1)}\right)^T\bm{O}^{(N-1)}\bm{T}^{(N-1)}\ , \nonumber \\
    \bm{Y}^{(N)}&=&
    \begin{pmatrix}
    \Re\left(\hat{G}_{0,N}^{(N)}\right) & \Im\left(\hat{G}_{0,N}^{(N)}\right)\\
    -\Im\left(\hat{G}_{0,N}^{(N)}\right) & \Re\left(\hat{G}_{0,N}^{(N)}\right)
    \end{pmatrix} =\bm{Y}^{(N-1)}\bm{T}^{(N-1)}\bm{O}^{(N)}\ , \nonumber
\end{eqnarray}
where $\bm{X}^{(N)}$, $\bm{Y}^{(N)}$ and $\bm{O}^{(N)}$ are the input and output of the $N^{th}$ iteration (first two lines in Eq. \ref{eq:RGF4}), with parametric dependence on the $N^{th}$ layer.

First, we consider the case where the FNN outputs $\bm{y}$ are the LDOS over the $M_L$ neurons in the last layer $N=L$; see Eq. \ref{eq:gfLDOS}. $\mathcal L$ depends explicitly on $\bm{y}$, e.g., a mean-square or cross-entropy error loss function. With the chain rule, we have:
\begin{eqnarray}
    \frac{\partial\mathcal{L}}{\partial T_{ij}^{(N-1)}}&=&\frac{\partial\mathcal{L}}{\partial O_{mn}^{(N)}}\frac{\partial O_{mn}^{(N)}}{\partial T_{ij}^{(N-1)}}=\epsilon_{mn}^{(N)}\alpha_{mn,ij}^{(N)}, \nonumber\\
    \frac{\partial\mathcal{L}}{\partial H_{ij}^{(N)}}&=&\frac{\partial\mathcal{L}}{\partial O_{mn}^{(N)}}\frac{\partial O_{mn}^{(N)}}{\partial H_{ij}^{(N)}}=\epsilon_{mn}^{(N)}\beta_{mn,ij}^{(N)},\label{eq:pH}
\end{eqnarray}
where:
\begin{equation}
    \begin{split}
       \epsilon_{mn}^{(L)}&=\sum_{k=1}^{M_L}\frac{\partial\mathcal{L}}{\partial y_k}\frac{\partial y_k}{\partial O_{mn}^{(L)}}\\
       &=\sum_{k=1}^{M_L}\frac{\partial\mathcal{L}}{\partial y_k}\frac{\delta_{k+M_L,m}\delta_{kn}-\delta_{km}\delta_{k+M_L,n}}{\pi}\\
       &=\frac{\partial\mathcal{L}}{\partial y_n}\frac{\delta_{n+M_L,m}-\delta_{m+M_L,n}}{\pi}\ ,
    \end{split}\label{eq:eM}
\end{equation}
and the rest ($N<L$) are obtainable via a second (backward) recursion:
\begin{equation}
    \begin{split}
        \epsilon_{mn}^{(N)}&=\frac{\partial\mathcal{L}}{\partial O_{mn}^{(N)}}=\sum_{kl}\epsilon_{kl}^{(N+1)}\frac{\partial O_{kl}^{(N+1)}}{\partial O_{mn}^{(N)}}\\
        &=\sum_{kl}\epsilon_{kl}^{(N+1)}\gamma_{kl,mn}^{(N+1)},
    \end{split}\label{eq:epsilon}
\end{equation}

We note that the values of $\alpha$, $\beta$, and $\gamma$ necessary for \ref{eq:pH} and \ref{eq:epsilon} are fully determined in the previous recursion towards the FNN outputs:
\begin{eqnarray}
    \alpha_{mn,ij}^{(N)}&=&\frac{\partial O_{mn}^{(N)}}{\partial T_{ij}^{(N-1)}}  \nonumber \\ &=& -\sum_{p}O_{mp}^{(N)}O_{jn}^{(N)}\left[\left(T^{(N-1)}\right)^TO^{(N-1)}\right]_{pi}\nonumber\\
    & & \indent \quad +O_{mj}^{(N)}O_{pn}^{(N)}\left[O^{(N-1)}T^{(N-1)}\right]_{ip}\ , \nonumber \\
    \beta_{mn,ij}^{(N)}&=&\frac{\partial O_{mn}^{(N)}}{\partial H_{ij}^{(N)}}=O_{mi}^{(N)}O_{jn}^{(N)}\ ,\label{eq:beta}\\
    \gamma_{kl,mn}^{(N+1)}&=&\frac{\partial O_{kl}^{(N+1)}}{\partial O_{mn}^{(N)}}=\sum_{pq}O_{kp}^{(N+1)}O_{ql}^{(N+1)}\left(T^{(N)}\right)^T_{pm}T^{(N)}_{nq}\ ,\nonumber
\end{eqnarray}
where we have employed the following formulas \cite{MatCook2012}:
\begin{eqnarray}
    \frac{\partial O_{mn}^{(N)}}{\partial X_{pq}^{(N)}}&=&-O_{mp}^{(N)}O_{qn}^{(N)}\ ,\nonumber \\
    \frac{\partial X_{pq}^{(N)}}{\partial T_{ij}^{(N-1)}}&=&\delta_{qj}\left[\left(T^{(N-1)}\right)^TO^{(N-1)}\right]_{pi}\nonumber\\
    & &+\delta_{pj}\left[O^{(N-1)}T^{(N-1)}\right]_{iq}\ ,\nonumber\\
    \frac{\partial X_{pq}^{(N)}}{\partial H_{ij}^{(N)}}&=&-\delta_{pi}\delta_{qj}\ ,\\
    \frac{\partial X_{pq}^{(N)}}{\partial O_{mn}^{(N-1)}}&=&-\left(T^{(N-1)}\right)^T_{pm}T^{(N-1)}_{nq}\ . \nonumber
\end{eqnarray}

\begin{figure}
	\includegraphics[width=0.8\linewidth]{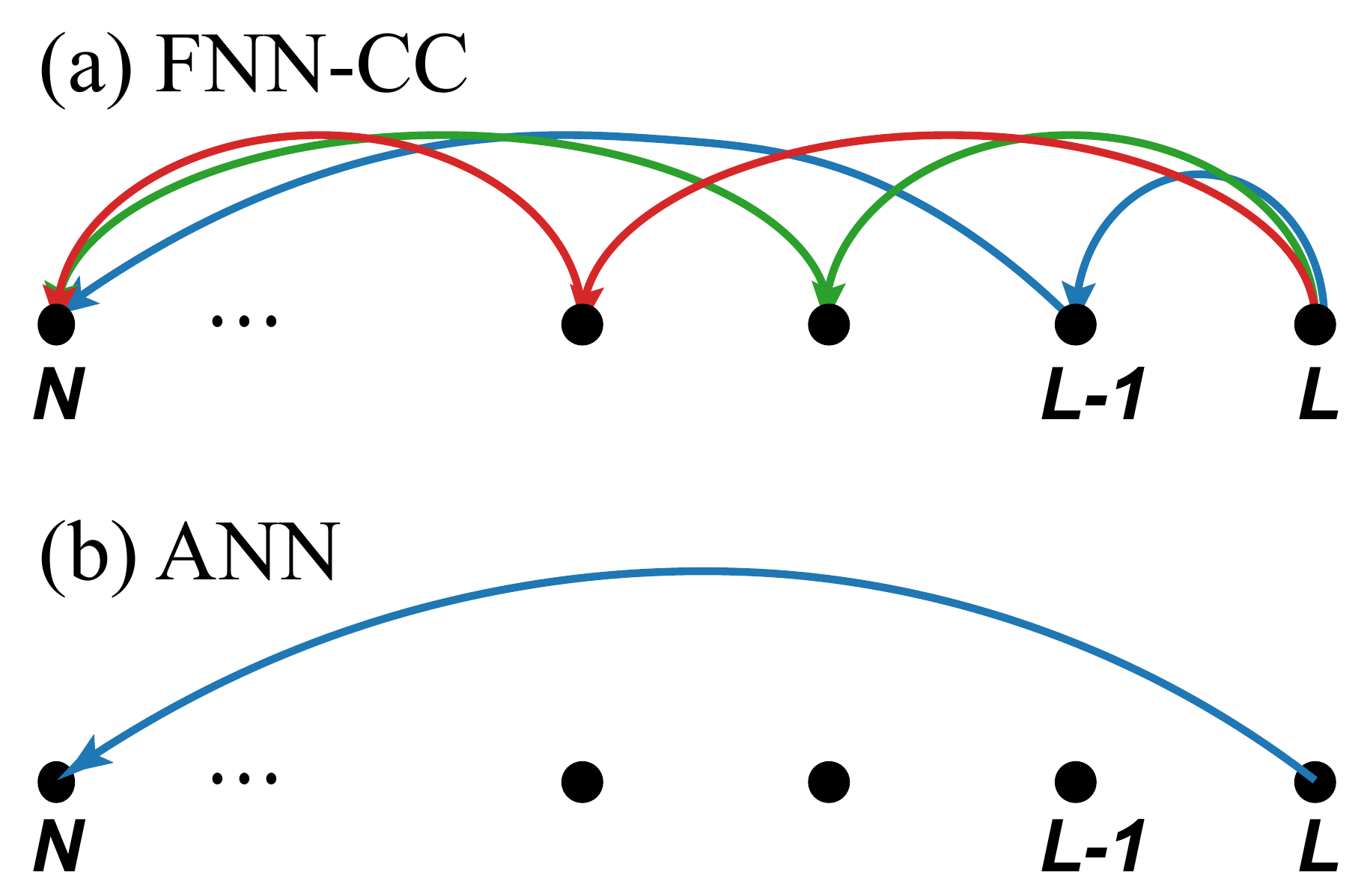}
	\caption{Compared to (b) a conventional deep ANN, where the gradients pass to earlier layers via a single channel and thus inevitably decay over distances, (a) an FNN with CC output possesses multiple channels thanks to the summation over $N'$ in Eq. \ref{eq:pH1}.} \label{fig:AP3_grad}
\end{figure}

Next, we consider the case where the FNN output takes the form of the CC, Eq. \ref{eq:gfCC}. Once again, we apply the chain rule to get:
\begin{eqnarray}
    \frac{\partial\mathcal{L}}{\partial T_{ij}^{(N-1)}}&=& \frac{\partial\mathcal{L}}{\partial y}\sum_{kl}\frac{\partial y}{\partial Y_{kl}^{(L)}}\Bigg[\frac{\partial Y_{kl}^{(L)}}{\partial Y_{uv}^{(N)}}\frac{\partial Y_{uv}^{(N)}}{\partial T_{ij}^{(N-1)}} \nonumber\\
    & &+\sum_{N'=N}^{L}\frac{\partial Y_{kl}^{(L)}}{\partial Y_{uv}^{(N')}}\frac{\partial Y_{uv}^{(N')}}{\partial O_{mn}^{(N)}}\frac{\partial O_{mn}^{(N)}}{\partial T_{ij}^{(N-1)}}\Bigg] \nonumber\\
    &=&\frac{\partial\mathcal{L}}{\partial y}\sum_{kl}Y_{kl}^{(L)}\Bigg[\theta_{kl,uv}^{(N)}Y_{ui}^{(N-1)}O_{jv}^{(N)} \label{eq:pH1} \\
    & &+\sum_{N'=N}^{L}\theta_{kl,uv}^{(N')}\phi_{uv,mn}^{(N',N)}\alpha_{mn,ij}^{(N)}\Bigg], \nonumber\\
    \frac{\partial\mathcal{L}}{\partial H_{ij}^{(N)}}&=& \frac{\partial\mathcal{L}}{\partial y}\sum_{kl}\frac{\partial y}{\partial Y_{kl}^{(L)}}\left[\sum_{N'=N}^{L}\frac{\partial Y_{kl}^{(L)}}{\partial Y_{uv}^{(N')}}\frac{\partial Y_{uv}^{(N')}}{\partial O_{mn}^{(N)}}\frac{\partial O_{mn}^{(N)}}{\partial H_{ij}^{(N)}}\right]\nonumber\\
    &=&\frac{\partial\mathcal{L}}{\partial y}\sum_{kl}Y_{kl}^{(L)}\left[\sum_{N'=N}^{L}\theta_{kl,uv}^{(N')}\phi_{uv,mn}^{(N',N)}\beta_{mn,ij}^{(N)}\right], \nonumber
\end{eqnarray}
where:
\begin{eqnarray}
    \theta_{kl,uv}^{(L)}&=&\frac{\partial Y_{kl}^{(L)}}{\partial Y_{uv}^{(L)}}=\delta_{ku}\delta_{lv}\ ,  \\
    \phi_{uv,mn}^{(N',N')}&=&\frac{\partial Y_{uv}^{(N')}}{\partial O_{mn}^{(N')}}=\delta_{vn}\sum_{r}Y_{ur}^{(N'-1)}T_{rm}^{(N'-1)}\ ,\nonumber
\end{eqnarray}
and the rest ($N < L$ for $\theta$ and $N < N'$ for $\phi$) are obtainable via (back-ward) recursions:
\begin{eqnarray}
    \theta_{kl,uv}^{(N')}&=&\frac{\partial Y_{kl}^{(L)}}{\partial Y_{uv}^{(N')}}=\sum_{pq}\theta_{kl,pq}^{(N'+1)}\frac{\partial Y_{pq}^{(N'+1)}}{\partial Y_{uv}^{(N')}} \nonumber\\
        &=& \sum_{qr}\theta_{kl,uq}^{(N'+1)}T_{vr}^{(N')}O_{rq}^{(N'+1)}\ , \\
    \phi_{uv,mn}^{(N',N)}&=&\frac{\partial Y_{uv}^{(N')}}{\partial O_{mn}^{(N)}}=\sum_{pq}\phi_{uv,pq}^{(N',N+1)}\gamma_{pq,mn}^{(N+1)}. \nonumber
\end{eqnarray}

Interestingly, due to the extra summation of $N'$ from $N$ to $L$ in Eq. \ref{eq:pH1}, an earlier layer receives a concentration and boost in its corresponding differentials from multiple channels contributed by all later layers (Fig. \ref{fig:AP3_grad}), thus circumventing the gradient decay like in classical ANNs; see Fig. \ref{fig:adv}. Such multi-channels of gradients towards the earlier layers are also present in residual neural networks \cite{He2016resnet} via additional connectivity in ANN architecture. In comparison, despite simple FNN architecture with local connectivity, we may still achieve additional gradient contributions, emerging not as a consequence of long-range connectivity but as the CC's characteristic global dependence.

\section{FNN architectures}\label{app:architecture}

We study more FNN architectures to compare the effect of connectivity and locality on machine-learning performance. We consider FNNs with the neurons in each layer forming a 2D square lattice and either no intra-layer hopping (unconnected), nearest neighbor intra-layer hopping, or nearest and next-nearest neighbor intra-layer hopping. We also consider two inter-layer connection formalisms: first, we consider FNNs with $[14\times 14, 7 \times 7, 10]$ neurons per layer, and tree-type inter-layer hopping - each neuron in the $1^{st}$ ($2^{nd}$) layer connects with $2\times 2$ neurons in the $0^{th}$ ($1^{st}$) layer and without overlap; then, we also consider FNNs with $[13\times 13, 6 \times 6, 10]$ neurons per layer, where the neighboring neurons' inter-layer partners share an overlap - inter-layer $t_{rr'}$ connects neuron at $(m_1, m_2)$ in the $1^{st}$ ($2^{nd}$) layer and neurons at $(m'_1, m'_2)$ in the $0^{th}$ ($1^{st}$) layer, $m'_{j}\in [2m_{j}-1, 2m_{j}+2]$ ($m'_{j}\in [2m_{j}-1, 2m_{j}+1]$). In all scenarios, the $l=3$ layer consists of fully-connected neurons whose LDOS represents FNNs' classification outputs.

\begin{table}[!htbp]
    \centering
    \begin{tabular}{|c|c|c|}
    \hline
    \diagbox{Intra-layer}{Inter-layer} & Tree & Overlapping \\
    \hline
    Unconnected & 93.43\% & 97.17\% \\
    \hline
    Upto nearest neighbor & 95.61\% & 97.26\% \\
    \hline
    Upto next nearest neighbor & 96.19\% & 97.36\% \\
    \hline
    \end{tabular}
    \caption{FNNs' accuracy on the MNIST test dataset shows relatively insensitive dependence on the architecture - the degree of locality and connectivity. The image's pixel grayscale is encoded as the zeroth layer model's onsite potentials; see Eq. \ref{eq:encode_ham}.}
    \label{tab:MNIST}
\end{table}

\begin{table}[!htbp]
    \centering
    \begin{tabular}{|c|c|c|}
    \hline
    \diagbox{Intra-layer}{Inter-layer} & Tree & Overlapping \\
    \hline
    Unconnected & 93.40\% & 97.84\% \\
    \hline
    Upto nearest neighbor & 96.27\% & 97.68\% \\
    \hline
    Upto next nearest neighbor & 96.16\% & 97.72\% \\
    \hline
    \end{tabular}
    \caption{FNNs' accuracy on the MNIST test dataset shows relatively insensitive dependence on the architecture - the degree of locality and connectivity. The image's pixel grayscale is encoded as an external, static LDOS field coupled to the FNN's first layer; see Appendix \ref{sec:extLDOS}.}
    \label{tab:MNIST_LDOS}
\end{table}

Applying such FNNs to supervised machine learning on the MNIST dataset, encoded as onsite potentials as Eq. \ref{eq:encode_ham} an external LDOS field in Appendix \ref{sec:extLDOS}, we summarize the optimal performances in Tabs. \ref{tab:MNIST} and \ref{tab:MNIST_LDOS}. We note that poor connectivity, e.g., the tree architecture (`tree column' and `unconnected' row), may adversely impact the FNN capacity - such locality is too strict to allow sufficient correlations between neurons in the first layers. Fortunately, unlike classical ANNs, FNNs allow intra-layer connections in addition to inter-layer hopping, which we observe bears more significant contributions. Given sufficient connectivity above the tree architecture, FNNs' performances increase quickly and saturate to the level of fully-connected FNNs. Further reduction of locality tends to receive a diminishing margin.

\section{FNN inputs: encoding classical data as external LDOS} \label{sec:extLDOS}

Given a classical training set, there are diverse encoding methods for FNNs' inputs. For example, in addition to the onsite potential $\mu_{\vec m}$ of a model, we may also regard each sample as $\Im (\hat{\bf G}_{00})_{\vec m\vec m}=-\pi\cdot x(\vec m)$, an external and static LDOS field coupled to the FNN's first layer ($l=1$) via $T_0$ and $T^{\dagger}_0$, the hopping between the $l=0$ and $l=1$ layers. As such inputs do not specify an explicit $\hat{H}_0$, the mutual information $I_{0L}$ between the input and output layers is no longer directly available.

\begin{figure}
	\includegraphics[width=1.02\linewidth]{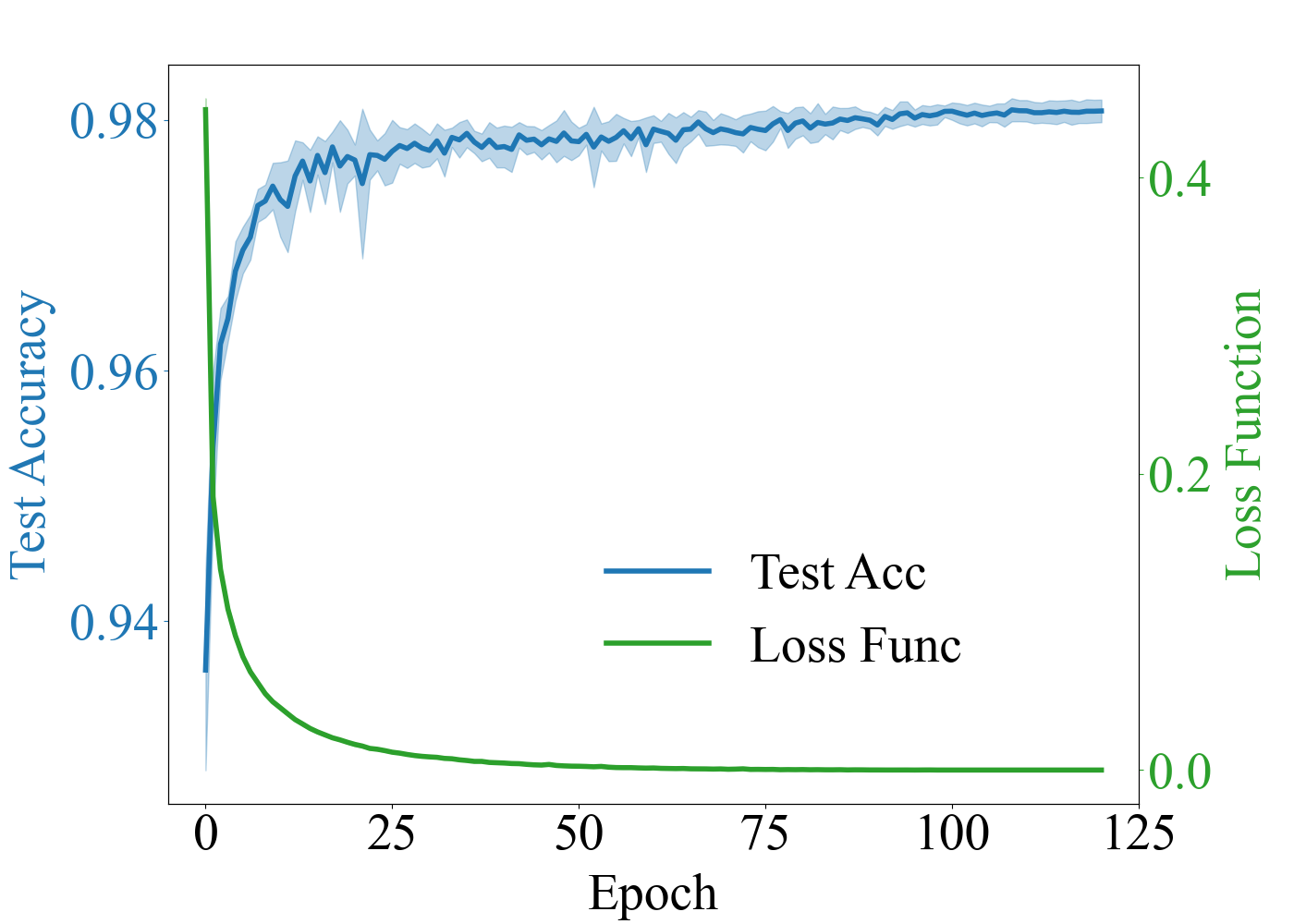}
	\caption{The loss function on the training set and the accuracy over the test set consistently demonstrate the convergence of a fully-connected FNN in supervised machine learning on the MNIST dataset encoded as external LDOS. The shades illustrate typical standard deviation over 10 trails. } \label{fig:LA}
\end{figure}

Applying such encoding to the classical MNIST dataset, we carry out supervised machine learning with FNNs and settings similar to Sec. \ref{sec:classical}. The LDOS over the ten neurons in the last layer serves as outputs. We summarize the performance on training and convergence in Fig. \ref{fig:LA}. With adequate tuning and training, FNNs can achieve 98.54\% accuracy on the MNIST dataset, fully comparable to classical ANNs with similar scale and complexity.

We have also investigated adding nearest neighbor hopping to the onsite potential Hamiltonian in Eq. \ref{eq:encode_ham}. The performance remains on par with the scenario without such hopping.

\section{Topological phases of disordered models}\label{app:disChern}

To increase the diversity of the insulator models in Eq. \ref{eq:Ham_ins}, we also include in their Hamiltonians the following terms on random quenched disorder:\begin{eqnarray}
\hat{H}_{dis}&=&\sum_{\vec{r}}w^{(1)}_{\vec{r}}c^{\dagger}_{\vec{r}+\hat{x}}c_{\vec{r}}+w^{(-1)}_{\vec{r}}c^{\dagger}_{\vec{r}+\hat{y}}c_{\vec{r}} + w^{(2)}_{\vec{r}}c^{\dagger}_{\vec{r}+\hat{x}+\hat{y}}c_{\vec{r}}
\nonumber \\
& & + w^{(-2)}_{\vec{r}}c^{\dagger}_{\vec{r}+\hat{x}-\hat{y}}c_{\vec{r}}+\text{h.c.} + w^{(0)} _{\vec{r}} c^{\dagger}_{\vec{r}}c_{\vec{r}}, \label{eq:Ham_ins_dis}
\end{eqnarray}
where $w^{(0)}_{\vec{r}}\in [-W_0, W_0],\ w^{(\pm 1)}_{\vec{r}} \in [-W_1, W_1],\ w^{(\pm 2)}_{\vec{r}} \in [-W_2, W_2]$ are perturbations to the onsite potential, nearest neighbor hopping and next-nearest neighbor hopping, respectively. In practice, we set $W_0\in[1.0,3.0],\ W_1\in[0,1.0],\ W_2\in[0,0.5]$.

Since such disorder may modify the topological character of a model, it may be rash to determine whether samples with $\hat{H}_{dis}$ are Chern or normal insulators just with their $\kappa$ value in the clean limit. Instead, we employ the real-space Kubo formula \cite{Kitaev2006, qlt2016} to evaluate the transverse conductance in accord with the Chern number:
\begin{equation}
    C=\frac{4\pi \ti}{N}\sum_{jkl} P_{jk}P_{kl}P_{lj}S_{\triangle jkl}, \label{eq:cn}
\end{equation}
where $P_{jk}=\langle c^\dagger_j c_k\rangle$ is the two-point correlator between sites $i$ and $j$, $S_{\triangle jkl}$ is the signed area of triangle $jkl$, and $N$ is the total number of sites. Given the locality of the insulator models, we limit the summation in Eq. \ref{eq:cn} to triangles no larger than a cut-off length scale $d=3$ for simplicity. Finally, for model samples with an apparent spectral gap and $0.7 \le C \le 1.0$ or $0 \le C \le 0.3$ after the cut-off, we include them in our training set as Chern insulators and normal insulators, respectively.

\section{Phase diagrams of the Falicov-Kimbal model} \label{sec:fkpd}

The Falicov-Kimbal (FK) model has a long and successful history in the studies of correlated electron systems, with a broad range of applications in the context of metal-semiconductor transition, binary alloys, crystallization, etc. As a variant of the Hubbard model, the FK model in Eq. \ref{eq:fkmodel}:
\begin{eqnarray}
\hat{H}_0 &=& t\sum_{\langle ij\rangle} c^\dagger_i c_j +t'\sum_{\llangle ik\rrangle} c^\dagger_i c_k + \text{h.c.} \nonumber \\
& &  - \mu \sum_i c^\dagger_i c_i + E_f\sum_i n^f_{i} + U\sum_i c^\dagger_i c_i n^f_{i}, \label{eq:fkmodel2}
\end{eqnarray}
characterizes a quantum many-body system with two species of electrons - the itinerant $c$ electrons and the localized $f$ electrons - interacting with each other. Without loss of generality, we focus on the FK model at exactly half-filling on a 2D square lattice, where the number of $f$ electrons is equal to the number of $c$ electrons, and their sum is equal to the number of lattice sites.

\begin{figure}
    \includegraphics[width=0.98\linewidth]{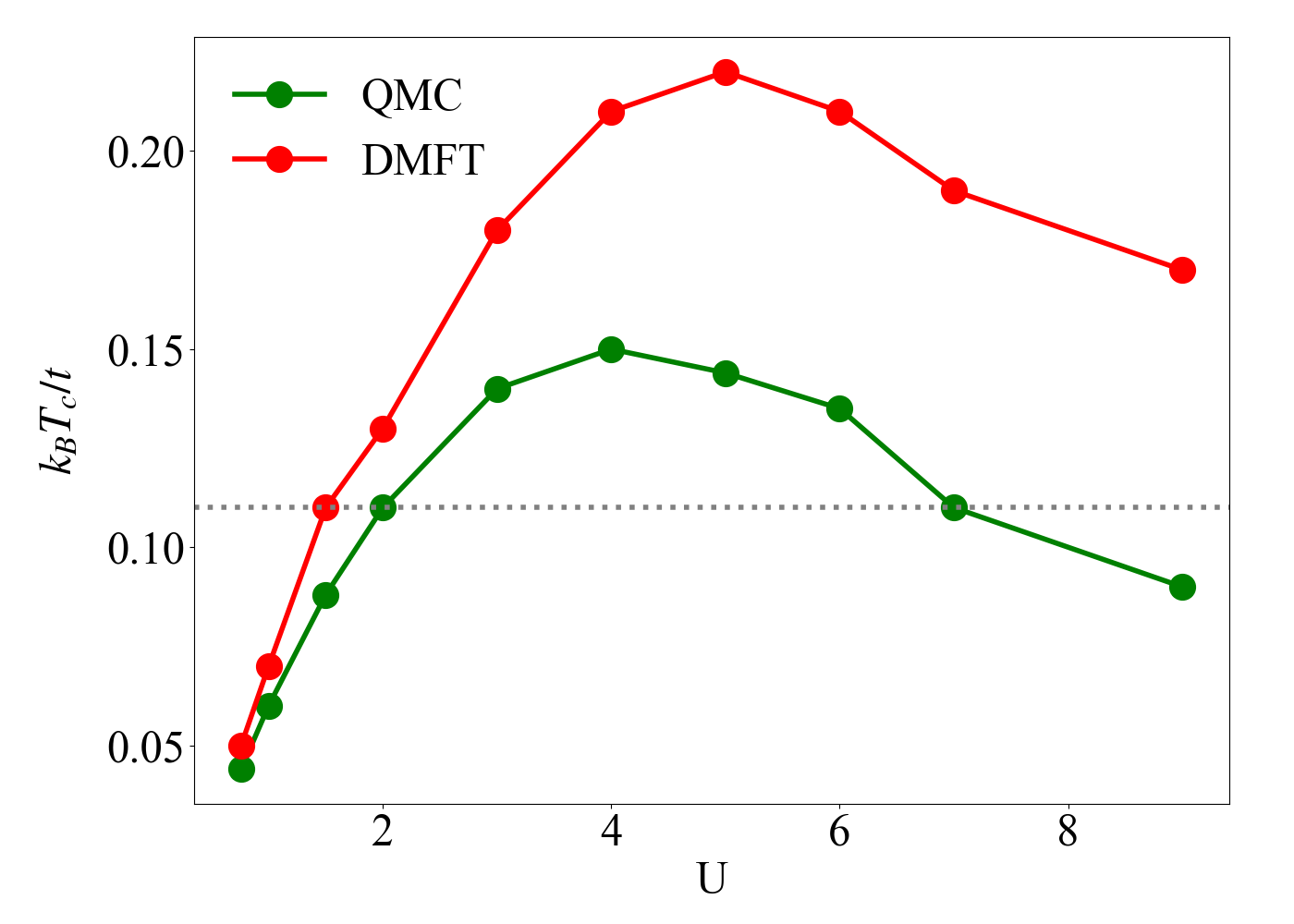}
    \includegraphics[width=0.94\linewidth]{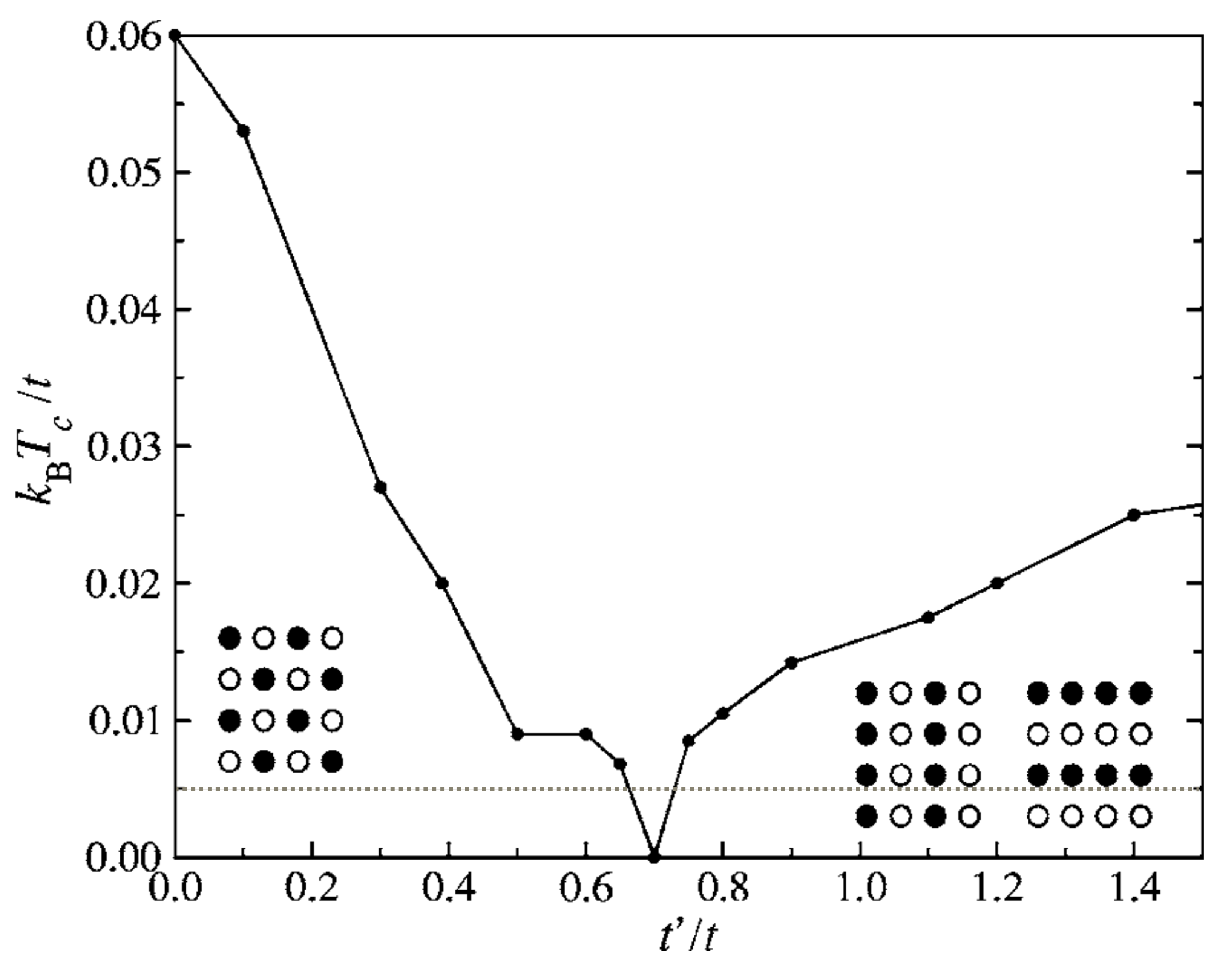}
	\caption{Top: The phase diagram of the pristine FK model ($t'=0$) in terms of the temperature $T$ and interaction strength $U$ consists of a disordered metal phase at higher temperatures and an insulating charge order at lower temperatures. The dashed line at $T=0.11$ corresponds to the phase space we study in Appendix \ref{sec:FNNMLFK2} using FNN machine learning. Bottom: The phase diagram of the FK model in terms of the temperature $T$ and the next nearest neighbor hopping $t'$ consists of two charge orders - checkerboard at small $t'/t$ and stripe at large $t'/t$ - separated by a critical point at $T=0$ that expands into a transition region at finite $T$. The dashed line corresponds to the temperature $T=0.005$, where we analyze the FK model's charge orders using FNN machine learning. The figures' data is partly obtained from Ref. \cite{FKModelQMC2006}.} \label{fig:FKPD}
\end{figure}

First, we focus on the pristine FK model with the next nearest neighbor hopping $t'=0$. At sufficiently low temperatures, the half-filled FK model possesses a long-range charge order, i.e., the electrons form a checkerboard pattern (Fig. \ref{fig:FKPD}(b) inset), the same as in the ground state. The system enters a disordered metal phase at higher temperatures through a metal-insulator transition. The transition temperature $T_C$ depends on the interaction strength $U$. We set $t=1$ as our unit of energy, like in Sec. \ref{sec:scs}. A general solution to the FK model is not known. Fortunately, a controlled analysis of the FK model on a 2D square lattice is available within the quantum Monte Carlo framework - after integrating out the $c$ electrons, we can sample the $f$-electron configurations without the sign problem \cite{FKModelQMC2006}. According to the presence and absence of the order parameter, the established phase diagram is in Fig. \ref{fig:FKPD}(a) and compared with the approximate DMFT results, as discussed in Appendix \ref{sec:fkdmft}.

It is also interesting to take the next nearest neighbor hopping $t'$ into account in addition to the nearest neighbor hopping $t$. In the limit of small $t'/t$, we expect that the checkerboard charge order still dominates in the ground state or at low temperatures. In the opposite limit, however, the ground state will have the form of vertical or horizontal stripes. We set $U/t=1$ as in Sec. \ref{sec:scs}, and the QMC phase diagram \cite{FKModelQMC2006} is in Fig. \ref{fig:FKPD}(b): the two distinctive charge orders are separated by a phase transition near $t_C/t\sim 0.7$ at $T\rightarrow 0$; at higher temperatures, the critical point expands into a transition region, where the system evolves into a disordered metal. In practice, we strictly enforce the half-filling condition for the localized $f$ electrons via adjusting $E_f$, while relaxing the half-filling condition for the itinerant $c$ electrons to a degree by varying $\mu$ a bit (inset of Fig. \ref{fig:FKmodelPD}), as long as the charge orders remain stable. Such variations yield more diverse models for the training set.

\section{DMFT results of the Falicov-Kimbal model} \label{sec:fkdmft}

Dynamical mean-field theory (DMFT) maps a strongly-correlated lattice model, intractable in general, to a (series of) local impurity problems, such as the Anderson impurity model, solvable through various schemes. In doing so, we assume the lattice self-energy is a (momentum-independent) local quantity, which becomes exact in the limit of large coordination numbers.

In practice, the self-consistent solution of DMFT commonly consists of the following iterations until convergence: (1) given the self-energies $\Sigma$, solve Matsubara Green's functions of the lattice model $G_{loc}$; (2) solve each local impurity model's Green's functions $G_{imp}$; determine the self-energies $\Sigma$ for the next iteration. For instance, we illustrate our DMFT routine for the half-filled FK model in Eq. \ref{eq:fkmodel2} in Algorithm \ref{alg:DMFT}. Here, we determine the parameter $E_f$ by setting half-filling $\bar n_f=1/2$ across all local impurity models:
\begin{equation}
\langle n^f\rangle=\left\{1+\exp{\left[ \frac{E-\mu}{T}-\sum_n \ln{\left(1- \frac{U}{ \mathcal{G}_0^{-1}(\ti\omega_n)}\right)}\right]}\right\}^{-1},
\end{equation}
where the summation is over imaginary frequencies $\omega_n = 2\pi T(n+1/2)$, $n\in \mathbb{Z}$ and $n\in[-n_0,n_0 - 1]$.

\begin{algorithm}
	\caption{DMFT for the half-filled FK model}
	\label{alg:DMFT}
	\begin{algorithmic}[1]
	    \Require
	    \Statex Lattice problem $\hat{H}_0$ with local interaction $U$ as local impurity models over $M_0$ respective sites, temperature $T$, Fermi energy $\mu$, update rate $\lambda$, and convergence criteria $\epsilon$
	    \Ensure
	    \Statex self-energies $\bm{\Sigma}(\ti\omega_n)$ and Green's functions $G_{loc}(\ti\omega_n)$
	    \State Initialize $\bm{\Sigma}(\ti\omega_n)$ for each of the $M_0$ sites;
		\Repeat
		    \State $G_{loc}(\ti\omega_n)=\left[\ti\omega_n+\mu-\hat{H}_0-\mbox{diag}\left(\bm{\Sigma}(\ti\omega_n)\right)\right]^{-1}_{ii}$;
            \State $\mathcal{G}_0^{-1}(\ti\omega_n)=G_{loc}^{-1}(\ti\omega_n)+\bm{\Sigma}(\ti\omega_n)$ for each local impurity;
            \State Determine $E_f$ for $\langle \bar n^f\rangle=1/2$ via binary search;
            \State $G_{imp}(\ti\omega_n)= \frac{n^f}{\mathcal{G}_0^{-1}(\ti\omega_n)-U}+\frac{1-n^f}{\mathcal{G}_0^{-1}(\ti\omega_n)}$ for each impurity;
            \State $\bm{\Sigma}(\ti\omega_n) \leftarrow (1-\lambda)\bm{\Sigma}(\ti\omega_n)+ \lambda \left[\mathcal{G}_0^{-1}(\ti\omega_n)-G_{imp}^{-1}(\ti\omega_n)\right]$;
		\Until{$\sum_n\left\|G_{imp}(\ti\omega_n)-G_{loc}(\ti\omega_n)\right\|_2<\epsilon$}
	\end{algorithmic}
\end{algorithm}

\begin{figure}
	\includegraphics[width=1.0\linewidth]{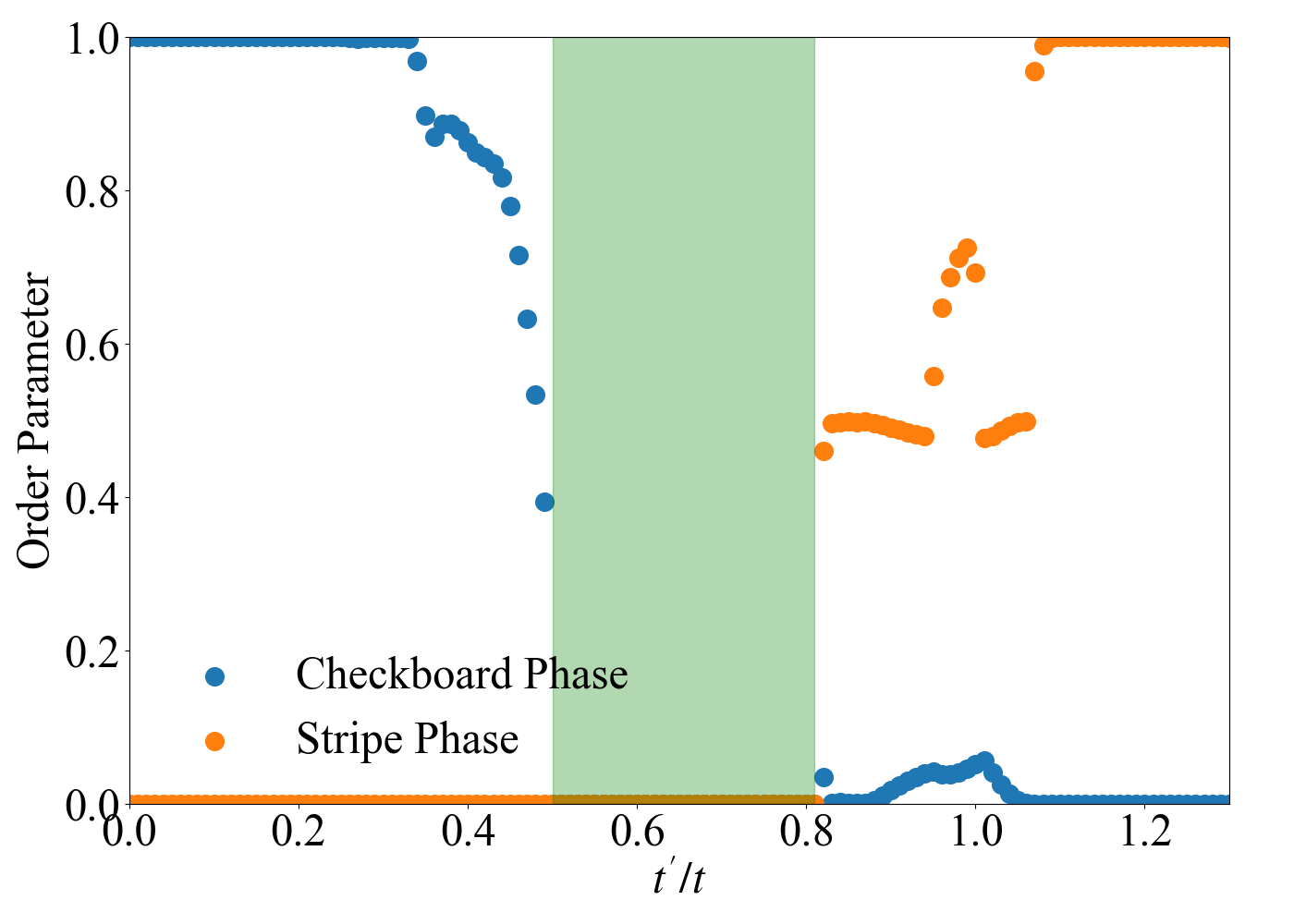}
	\caption{We determine the phase diagram of the FK model in Eq. \ref{eq:fkmodel} according to the order parameters of the checkerboard order and the stripe order via momentum-space DMFT. Both order parameters vanish for $t'/t$ between 0.5 and 0.81 (shaded green region), indicating an intermediate disordered metal phase at this finite $T=0.005$. The values of $\mu$ follow the red line in Fig. \ref{fig:FKmodelPD}.}
	\label{fig:DMFTop}
\end{figure}

In particular, we base the DMFT formalism in the momentum space with a $2\times2$ unit cell, sufficient for the emergence of the checkerboard and stripe orders. The resulting order parameters, $|\overline{n^f (-1)^{x+y}}|$ for the checkerboard order and $|\overline{n^f(-1)^{x}}|$ ($|\overline{n^f (-1)^{y}|}$) for the stripe order, are summarized in Fig. \ref{fig:DMFTop}, indicating a finite intermediate region for a disordered metallic phase. Though qualitatively consistent, the DMFT outcomes are approximate for such a strongly-correlated system in 2D, especially around its transitions, resulting in a broader range of intermediate disordered phases than quantum Monte Carlo calculations (Fig. \ref{fig:FKmodelPD} and Ref. \cite{FKModelQMC2006}). Such discrepancy is also apparent in Fig. \ref{fig:FKPD}a, as DMFT overestimate the transition temperatures $T_C$.

In addition, we carry out DMFT calculations fully in real space, with qualitatively consistent yet more unstable results. The difficulty is mainly in self-consistent convergence and order-parameter detection due to intertwined domains upon the extensive degrees of freedom.

In comparison, when we employ machine learning on such a strongly-correlated system by coupling it to an FNN, we carry out the process entirely in real space without resorting to any unit cells, which require presumed knowledge, e.g., symmetry-breaking order parameters. Indeed, the FNN does not cling to such order parameters, which are vanishingly small and meaningless in the coupled system, but instead to the original model's intrinsic quantum properties. In addition, the strongly-correlated system's link to the fully-connected FNN vastly increased the overall coordination number and effective dimensionality, which makes DMFT a much better and controlled approximation. Therefore, despite the DMFT and real-space nature of our FNN machine learning, we obtain results with quality and accuracy surpassing direct DMFT application on target systems (Fig. \ref{fig:FKmodelPD}).

\section{Supervised machine learning algorithm for interacting FNN} \label{app:mats_alg}

As discussed in Sec. \ref{sec:fnnintro}, we analyze and optimize interacting FNNs with their Matsubara Green's functions. Generalizing the recursive Green's functions in Eq. \ref{eq:RGF} and the corresponding descending-gradient chain rule in Appendix \ref{app:chainrule} to imaginary frequencies is straightforward. On the other hand, it is essential to derive the self-energies, dynamically dependent on the FNN model parameters, in a self-consistent fashion. Therefore, for supervised machine learning of an interacting FNN, we employ the following steps in each iteration: first, we fix the FNN model parameters and calculate the self-energies as a series of local impurity problems within the DMFT formalism; then, we incorporate the self-energies and evaluate Matsubara Green's functions, FNN outputs, and cost functions, etc., via the efficient recursive Green's function approach; next, we calculate the descending gradients through the chain rule and optimize the FNN via gradient descent, i.e., Eq. \ref{eq:gradesc}. For applying such an FNN, we only require a single iteration with the first two steps. In practice, we carry out a given number of iterations and keep the FNN with the best test accuracy in the process; see Algorithm \ref{alg:FNNwithDMFT} for details.

\begin{algorithm}
	\caption{Supervised machine learning for interacting FNN}
	\label{alg:FNNwithDMFT}
	\begin{algorithmic}[1]
	    \Require
	    \Statex FNN architecture, training and test sets, MAX$\_$EPOCHS
	    \Ensure
	    \Statex Trained FNN
	    \State Randomly initialize the FNN parameters $\{t_{rr'}, \mu_r\}$;
        \State MAX$\_$ACCUARCY=0;
		\For{$i=1$ to MAX$\_$EPOCHS}:
            \For{batch in training set}:
                \For{$\hat{H}_0$ in batch}:
                    \State Run Algorithm \ref{alg:DMFT} for self-energies $\bm{\Sigma}(\ti\omega_n)$;
                    \State Evaluate $\mathcal{L}[y_\text{FNN}(\hat{H}_0+\bm{\Sigma}(\ti\omega_n)), y_{tar}]$;
                    \State Calculate gradients $\{\partial \mathcal{L}/\partial t_{rr'}, \partial \mathcal{L}/\partial \mu_r\}$;
                \EndFor
                \State Update $\{t_{rr'}, \mu_r\}$ following batch gradient;
            \EndFor
            \State Correct=0;
            \For{$\hat{H}_0$ in test set}:
                \State Run Algorithm \ref{alg:DMFT} for self-energies $\bm{\Sigma}(\ti\omega_n)$;
                \State $\text{Prob}=\text{Softmax}[y_\text{FNN}(\hat{H}_0+\bm{\Sigma}(\ti\omega_n))]$;
                \State Correct += ($\arg\max{(\text{Prob})} == y_{tar}$);
            \EndFor
            \State Accuarcy=Correct/len(test set);
            \If{Accuarcy$>$MAX$\_$ACCUARCY}:
                \State MAX$\_$ACCUARCY=Accuarcy;
                \State Save FNN;
            \EndIf
		\EndFor
	\end{algorithmic}
\end{algorithm}

We note that self-energies depict interactions within an FNN. For the settings within our examples, where the interactions are contributed by and limited to the input models, it suffices to carry out the first step (self-energy calculations) upon only the FNN's $l=0$ layer. Likewise, we may regard supervised machine learning of non-interacting FNNs as Algorithm \ref{alg:FNNwithDMFT} with the first step (self-energy calculations) short-circuited (in addition to Green's functions in real frequency/energy).

\section{Machine learning metal-insulator transition in the Falicov-Kimbal model} \label{sec:FNNMLFK2}

\begin{figure}
	\includegraphics[width=1.0\linewidth]{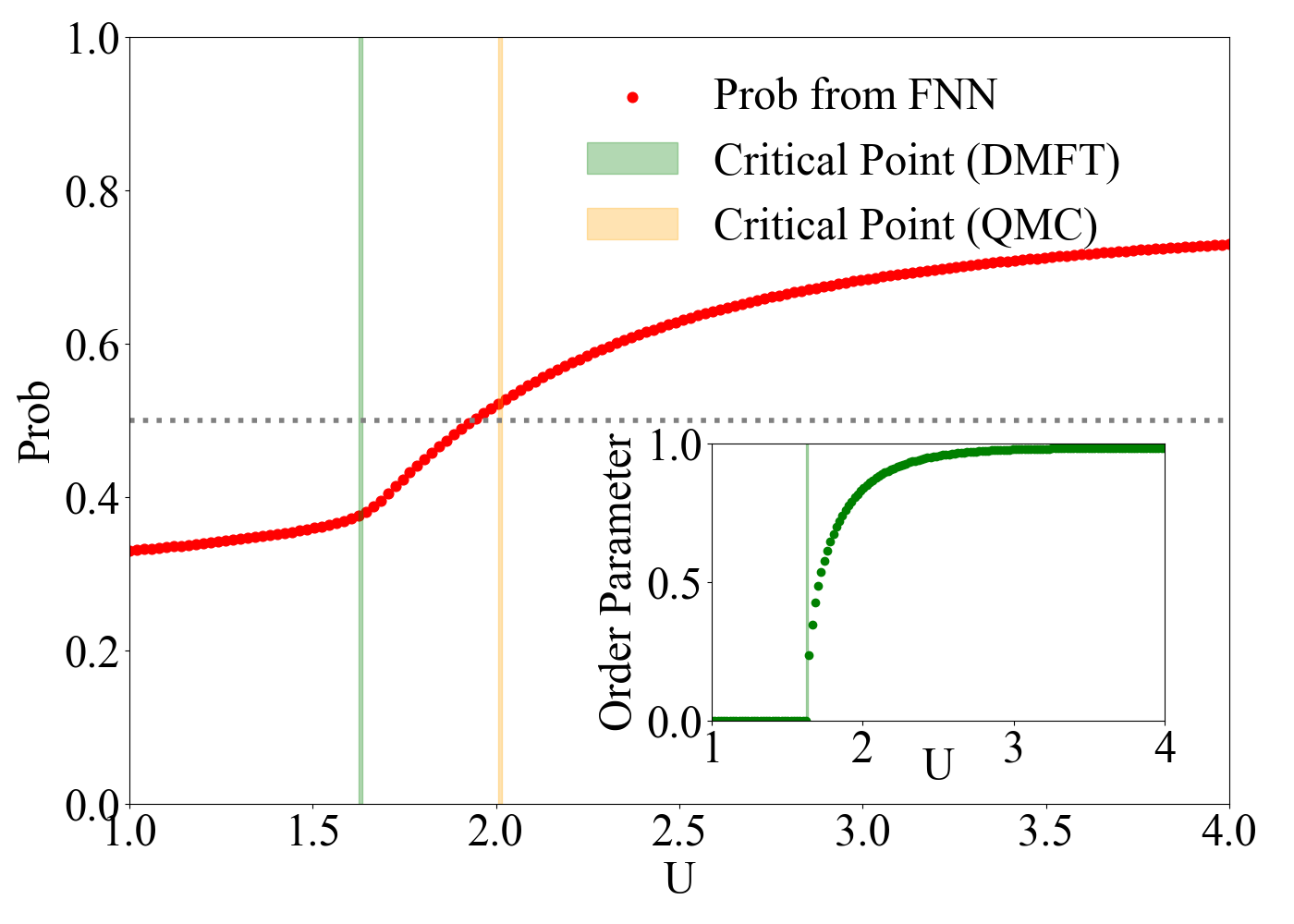}
	\caption{Using the last layer's LDOS as outputs, the FNN's insulator responses on the FK model $\hat{H}_0$ at exactly half-filled in Eq. \ref{eq:fkmit} indicate a metal-insulator transition at finite $T=0.11$. The yellow and green vertical lines are the critical point $U_C$ via QMC \cite{FKModelQMC2006} and DMFT (Appendix \ref{sec:fkdmft}) approaches upon the same parameter space. The inset shows the checkerboard order parameter of the insulating phase obtained in DMFT. }
	\label{fig:MIT}
\end{figure}

As another example, we study the metal-insulator transition of the FK model at higher temperature $T$ and exactly half-filling for both the $f$ and $c$ electrons. The Hamiltonian is \cite{Huang2017}:
\begin{eqnarray}
\hat{H}_0 &=& t\sum_{\langle ij\rangle} c^\dagger_i c_j + \text{h.c.} + U\sum_i\left(c^\dagger_i c_i-\frac{1}{2}\right)\left(n^f_{i}-\frac{1}{2}\right), \nonumber\\ \label{eq:fkmit}
\end{eqnarray}
corresponding to Eq. \ref{eq:fkmodel} with $t'=0$, $\mu=U/2$, and $E_f=-\mu$. Eq. \ref{eq:fkmit}'s particle-hole symmetry guarantees the half-filling condition. As we discuss in Appendix \ref{sec:fkpd}, at finite temperatures, the model prefers a disordered metallic phase (insulating checkerboard charge order) at small (large) $U$, see Fig. \ref{fig:MIT} inset and Ref. \cite{FKModelQMC2006}.

Similar to the Sec. \ref{sec:scs}, we couple the strongly-correlated model $\hat{H}_0$ in Eq. \ref{eq:fkmit} to an FNN and employ supervised machine learning. For the training set, we randomly sample models within the parameter region $T\in[0.1, 0.2]$ and $U\in[1.0, 4.0]$ away from the phase transitions. Once the training converges, we apply the FNN towards models with $U\in [1.0, 4.0]$ at $T=0.11$, whose results are in Fig. \ref{fig:MIT}. We note that the accuracy reaches 99.7\%, with a transition point between the numerical values obtained via quantum Monte Carlo (QMC) and DMFT. However, the FNN confidence is not very high, which is typical for machine learning of metal-insulator transitions, comparing gapless versus gapped phases and finite versus zero order parameters on finite-size systems. We also note the existence of a clear kink in the FNN output near the transition point indicated in DMFT.

\section{Mutual information and local unitary transformations for FNN logic flow} \label{app:logicflow}

For a post-learning FNN, we can employ mutual information to track the input-to-out logic flow through its full depth.

First, given a target input, we can establish mutual information between every neuron and the output neurons, whose LDOS holds the FNN's decision. Given a fermion tight-binding model, e.g., the FNN Hamiltonian in Eq. \ref{eq:ham}, we can analyze its ground-state entanglement entropy \cite{Amico2008, Eisert2010} and mutual information \cite{Adami1997, Groisman2005} starting from its two-point correlators $C_{ij}=\langle c_i^\dagger c_j\rangle$. With the correlation matrix $\bm{C}_A$ defined in a subsystem $A$, the corresponding entanglement Hamiltonian  $\bm{h}_A$ is:
\begin{equation}
    \bm{h}_A=\left\{\ln{\left[\left(1-\bm{C}_A\right)/\bm{C}_A\right]}\right\}^T\ , \label{eq:hC}
\end{equation}
which gives the reduced density matrix for the subsystem $A$ \cite{Peschel2009}:
\begin{equation}
    \rho_A=\frac1{Z_A}\exp{\left[-\sum_{i,j\in A}(h_A)_{ij}c_i^\dagger c_j\right]}.
\end{equation}

Since the partition function takes the form:
\begin{eqnarray}
    Z_A&=&\text{tr}\left(e^{-\bm{h}_A}\right)=\prod_k \left(1+e^{-\varepsilon^A_k}\right)\ , \nonumber \\
    \varepsilon^A_k &=&\ln{\left[\left(1-\zeta^A_k\right)/\zeta^A_k\right]}\ ,\label{eq:varepsilonzeta}
\end{eqnarray}
where $\varepsilon^A_k$ and $\zeta^A_k$ are the respective eigenvalues of $\bm{h}_A$ and $\bm{C}_A$ \cite{Peschel2003, Turner2010}, we can straightforwardly derive the Von Neumann entanglement entropy:
\begin{equation}
\begin{split}
    S_A&=-\text{tr}\left(\rho_A\ln{\rho_A}\right)\\
    &=\sum_k\left[\frac{\varepsilon^A_k}{e^{\varepsilon^A_k}+1}+\ln{\left(1+e^{-\varepsilon^A_k}\right)}\right]\ ,
\end{split}
\end{equation}
which we can calculate from first $C_{ij}$ and then $\zeta^A_k$. As a side note, we can also evaluate the Renyi entanglement entropy as:
\begin{equation}
\begin{split}
    (S_A)_\alpha&=\frac1{1-\alpha}\ln{\left[\text{tr}\left(\rho_A^\alpha\right)\right]}\\
    &=\frac1{1-\alpha}\sum_k\left[\ln{\left(1+e^{-\alpha\varepsilon^A_k}\right)}-\alpha\ln{\left(1+e^{-\varepsilon^A_k}\right)}\right]\ .
\end{split}\label{eq:renyi}
\end{equation}

For the FNN in Eq. \ref{eq:ham}, we first diagonalize the Hamiltonian:
\begin{equation}
\hat H = (c^\dagger Q)\Lambda(Q^\dagger c)=\alpha^\dagger\Lambda\alpha, \label{eq:ham_supp}
\end{equation}
where $\Lambda = \mbox{diag}(\{\epsilon_n\})$ are the energy eigenvalues and $\alpha_n^\dagger=\sum_{m}c_m^\dagger Q_{mn}$ are the corresponding eigenstates. Then, since the ground state fills (leaves empty) all states with $\epsilon_n\le0$ ($\epsilon_n>0$):
\begin{equation}
    |\Phi\rangle=\prod_{\epsilon_n\le0}\alpha_n^\dagger|0\rangle,
\end{equation}
where $|0\rangle$ is the vacuum state, we can obtain the correlation matrix:
\begin{equation}
    \bm{C}=\left(Q\tilde{\Lambda}Q^\dagger\right)^T,
\end{equation}
where $\tilde{\Lambda}$ is the occupation matrix with $1$'s in the diagonal for all $\epsilon_n\le0$.

In order to unambiguously locate the logic flow, we can perform a unitary transformation $U_l$ on each layer so that the mutual information in Eq. \ref{eq:mi} vanishes between the last layer and as many neurons in that layer as possible, simplifying the logic flows. We note that the mutual information $I_{AB}=0$ between two subsystems $A$ and $B$ if and only if the correlation matrix is block-diagonal:
\begin{equation}
    \bm{C}_{AB}=
        \begin{pmatrix}
        \bm{C}_{A}  &  \bm{0}\\
        \bm{0}  & \bm{C}_{B}
        \end{pmatrix},
\end{equation}
and so does the entanglement Hamiltonian $\bm{h}_{AB}$, yielding eigenvalues and entanglement entropy fully canceled by those from the separate subsystems $\bm{h}_A$ and $\bm{h}_B$. Physically, the subsystems $A$ and $B$ behave decoupled.

To preserve the layered FNN architecture, we considered layer-wise unitary transformations, which take a block-diagonal form:
\begin{equation}
    U=
    \begin{pmatrix}
        U_0 & \bm{0} & \cdots & \cdots & \cdots & \bm{0}\\
        \bm{0} & \ddots & \ddots & \ddots & \ddots & \vdots \\
        \vdots & \ddots & U_l & \bm{0} & \cdots & \bm{0} \\
        \vdots & \ddots & \bm{0} & \ddots & \ddots & \vdots \\
        \vdots & \ddots & \vdots & \ddots & U_{L-1} & \bm{0} \\
        \bm{0} & \cdots & \bm{0} & \cdots & \bm{0} & U_L
        \end{pmatrix}\ , \label{eq:U}
\end{equation}
where $U_l$ is a unitary matrix of $M_l\times M_l$ acting on the $l^{th}$ layer. $U$ transforms the overall correlation matrix (and simultaneously the FNN model) as follows:
\begin{equation}
    \begin{split}
        \bm{C}'&=U^\dagger\bm{C}U\\
        &=\begin{pmatrix}
        \bm{C}'_{0,0} & \cdots & \cdots & \cdots & \cdots & \bm{C}'_{0,L}\\
        \vdots & \ddots & \ddots & \ddots & \ddots & \vdots\\
        \vdots & \ddots & \bm{C}'_{l,l} & \cdots & \cdots & \bm{C}'_{l, L} \\
        \vdots & \ddots & \vdots & \ddots & \ddots & \vdots \\
        \vdots & \ddots & \vdots & \ddots & \bm{C}'_{L-1, L-1} & \bm{C}'_{L-1,L} \\
        \bm{C}'_{L,0} & \cdots & \bm{C}'_{L,l} & \cdots & \bm{C}'_{L,L-1} & \bm{C}'_{L,L}
        \end{pmatrix}\ ,
    \end{split} \label{eq:C'}
\end{equation}
where $\bm{C}'_{l, l'}=(U_l)^\dagger\bm{C}_{l, l'}U_{l'}$. Importantly, we wish to zero as many rows from the $M_l\times M_L$ matrix $\bm{C}'_{l, L}$ as possible so that these neurons in the $l^{th}$ layer do not correlate with the last layer, diminishing their mutual information. Commonly, we have more hidden neurons than output neurons, $M_l>M_L$. For such purpose, we can transform $\bm{C}'_{l, L}$ into an upper triangular matrix:
\begin{equation}
\bm{C}'_{l, L}=
    \begin{pmatrix}
    a_{11} & a_{12} & \cdots & a_{1,M_L} \\
    0 & a_{22} & \cdots & a_{2,M_L} \\
    \vdots & \ddots & \ddots & \vdots \\
    0 & \cdots & 0 & a_{M_L, M_L} \\
    0 & \cdots & \cdots & 0 \\
    \vdots & \ddots & \ddots & \vdots \\
    0 & \cdots & \cdots & 0
    \end{pmatrix}_{M_l\times M_L}, \label{eq:C'_iL}
\end{equation}
by making $U_L = I$ an identity so that the output neurons retain their meanings and obtaining $U_l^\dagger$ via the Gram-Schmidt orthogonalization. In the new FNN after the overall unitary transformation $U$, only the first $M_L$ neurons in a layer may possess finite mutual information with the last layer.

\begin{figure}
	\includegraphics[width=1.0\linewidth]{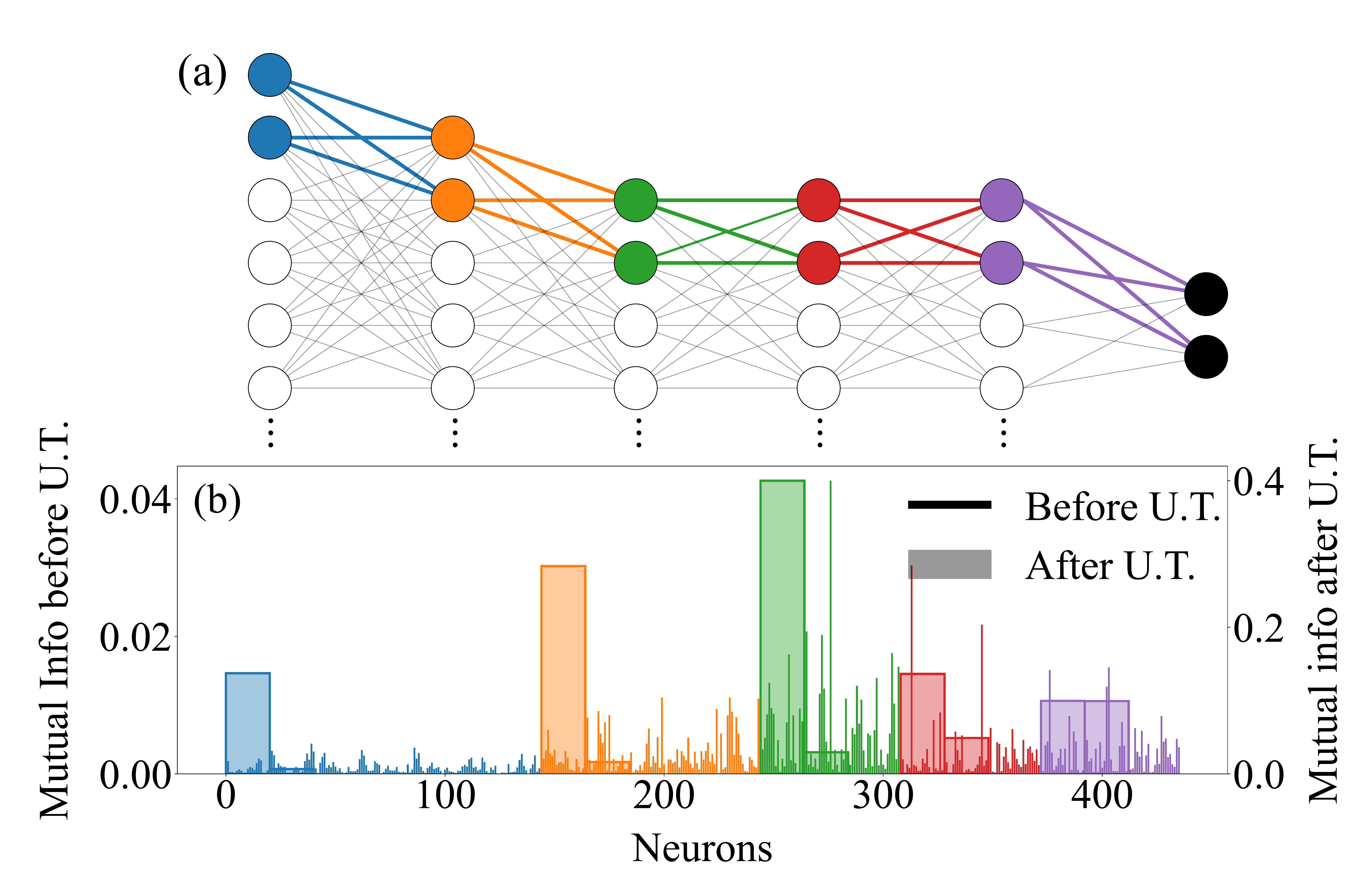}
	\caption{After proper unitary transformations (UTs), the mutual information of every single neutron and the output neurons exhibits a highly concentrated pattern, allowing us to trace the logic flows from any input sample - a Chern insulator model in this case - through the full depth of the FNN to its outputs accurately without information loss, as schematically shown in (a). (b) shows the mutual information distributions before and after the layer-wise unitary transformations.}
	\label{fig:logic_flow}
\end{figure}

For example, for the FNN in Fig. \ref{fig:logic_flow}(a), a logic flow example given a Chern insulator input and the mutual information distributions before and after the unitary transformation are in Fig. \ref{fig:logic_flow}(a)(b). We note that such unitary transformations are controlled, invertible, and sample-wise applicable - no transformation of this kind is available to classical ANNs. While L1 regularization may eliminate some weights in classical ANNs, its effect is stochastic, uncontrolled, and at the cost of ANNs' complexity and power; it also applies only to the ANN level instead of specific samples.

The analogous idea of network simplification and logic-flow analysis also exists in classical ANNs \cite{Frankle2018, Blalock2020}. One can keep the ANN performance while pruning most model parameters, resulting in simpler architecture and better interpretability. Compared to a general compression technique such as pruning, however, our logic-flow analysis works for individual FNN inputs. Naturally, different inputs incur different logic flows, for which we may simply employ different local unitary transformations.

\bibliography{refs}

\begin{thebibliography}{124}%
\makeatletter
\providecommand \@ifxundefined [1]{%
 \@ifx{#1\undefined}
}%
\providecommand \@ifnum [1]{%
 \ifnum #1\expandafter \@firstoftwo
 \else \expandafter \@secondoftwo
 \fi
}%
\providecommand \@ifx [1]{%
 \ifx #1\expandafter \@firstoftwo
 \else \expandafter \@secondoftwo
 \fi
}%
\providecommand \natexlab [1]{#1}%
\providecommand \enquote  [1]{``#1''}%
\providecommand \bibnamefont  [1]{#1}%
\providecommand \bibfnamefont [1]{#1}%
\providecommand \citenamefont [1]{#1}%
\providecommand \href@noop [0]{\@secondoftwo}%
\providecommand \href [0]{\begingroup \@sanitize@url \@href}%
\providecommand \@href[1]{\@@startlink{#1}\@@href}%
\providecommand \@@href[1]{\endgroup#1\@@endlink}%
\providecommand \@sanitize@url [0]{\catcode `\\12\catcode `\$12\catcode
  `\&12\catcode `\#12\catcode `\^12\catcode `\_12\catcode `\%12\relax}%
\providecommand \@@startlink[1]{}%
\providecommand \@@endlink[0]{}%
\providecommand \url  [0]{\begingroup\@sanitize@url \@url }%
\providecommand \@url [1]{\endgroup\@href {#1}{\urlprefix }}%
\providecommand \urlprefix  [0]{URL }%
\providecommand \Eprint [0]{\href }%
\providecommand \doibase [0]{https://doi.org/}%
\providecommand \selectlanguage [0]{\@gobble}%
\providecommand \bibinfo  [0]{\@secondoftwo}%
\providecommand \bibfield  [0]{\@secondoftwo}%
\providecommand \translation [1]{[#1]}%
\providecommand \BibitemOpen [0]{}%
\providecommand \bibitemStop [0]{}%
\providecommand \bibitemNoStop [0]{.\EOS\space}%
\providecommand \EOS [0]{\spacefactor3000\relax}%
\providecommand \BibitemShut  [1]{\csname bibitem#1\endcsname}%
\let\auto@bib@innerbib\@empty
\bibitem [{\citenamefont {LeCun}\ \emph {et~al.}(2015)\citenamefont {LeCun},
  \citenamefont {Bengio},\ and\ \citenamefont {Hinton}}]{Lecun2015Deep}%
  \BibitemOpen
  \bibfield  {author} {\bibinfo {author} {\bibfnamefont {Y.}~\bibnamefont
  {LeCun}}, \bibinfo {author} {\bibfnamefont {Y.}~\bibnamefont {Bengio}},\ and\
  \bibinfo {author} {\bibfnamefont {G.}~\bibnamefont {Hinton}},\ }\bibfield
  {title} {\bibinfo {title} {Deep learning},\ }\href
  {https://doi.org/10.1038/nature14539} {\bibfield  {journal} {\bibinfo
  {journal} {Nature}\ }\textbf {\bibinfo {volume} {521}},\ \bibinfo {pages}
  {436} (\bibinfo {year} {2015})}\BibitemShut {NoStop}%
\bibitem [{\citenamefont {Jordan}\ and\ \citenamefont
  {Mitchell}(2015)}]{Jordan2015Machine}%
  \BibitemOpen
  \bibfield  {author} {\bibinfo {author} {\bibfnamefont {M.}~\bibnamefont
  {Jordan}}\ and\ \bibinfo {author} {\bibfnamefont {T.}~\bibnamefont
  {Mitchell}},\ }\bibfield  {title} {\bibinfo {title} {Machine learning:
  Trends, perspectives, and prospects},\ }\href
  {https://doi.org/10.1126/science.aaa8415} {\bibfield  {journal} {\bibinfo
  {journal} {Science}\ }\textbf {\bibinfo {volume} {349}},\ \bibinfo {pages}
  {255} (\bibinfo {year} {2015})}\BibitemShut {NoStop}%
\bibitem [{\citenamefont {{Michael A. Nielsen}}(2015)}]{MLbook}%
  \BibitemOpen
  \bibfield  {author} {\bibinfo {author} {\bibnamefont {{Michael A.
  Nielsen}}},\ }\href@noop {} {\emph {\bibinfo {title} {{Neural Networks and
  Deep Learning}}}}\ (\bibinfo  {publisher} {{Determination Press}},\ \bibinfo
  {year} {2015})\BibitemShut {NoStop}%
\bibitem [{\citenamefont {Silver}\ \emph {et~al.}(2016)\citenamefont {Silver},
  \citenamefont {Huang}, \citenamefont {Maddison}, \citenamefont {Guez},
  \citenamefont {Sifre}, \citenamefont {van~den Driessche}, \citenamefont
  {Schrittwieser}, \citenamefont {Antonoglou}, \citenamefont {Panneershelvam},
  \citenamefont {Lanctot}, \citenamefont {Dieleman}, \citenamefont {Grewe},
  \citenamefont {Nham}, \citenamefont {Kalchbrenner}, \citenamefont
  {Sutskever}, \citenamefont {Lillicrap}, \citenamefont {Leach}, \citenamefont
  {Kavukcuoglu}, \citenamefont {Graepel},\ and\ \citenamefont
  {Hassabis}}]{Silver2016}%
  \BibitemOpen
  \bibfield  {author} {\bibinfo {author} {\bibfnamefont {D.}~\bibnamefont
  {Silver}}, \bibinfo {author} {\bibfnamefont {A.}~\bibnamefont {Huang}},
  \bibinfo {author} {\bibfnamefont {C.~J.}\ \bibnamefont {Maddison}}, \bibinfo
  {author} {\bibfnamefont {A.}~\bibnamefont {Guez}}, \bibinfo {author}
  {\bibfnamefont {L.}~\bibnamefont {Sifre}}, \bibinfo {author} {\bibfnamefont
  {G.}~\bibnamefont {van~den Driessche}}, \bibinfo {author} {\bibfnamefont
  {J.}~\bibnamefont {Schrittwieser}}, \bibinfo {author} {\bibfnamefont
  {I.}~\bibnamefont {Antonoglou}}, \bibinfo {author} {\bibfnamefont
  {V.}~\bibnamefont {Panneershelvam}}, \bibinfo {author} {\bibfnamefont
  {M.}~\bibnamefont {Lanctot}}, \bibinfo {author} {\bibfnamefont
  {S.}~\bibnamefont {Dieleman}}, \bibinfo {author} {\bibfnamefont
  {D.}~\bibnamefont {Grewe}}, \bibinfo {author} {\bibfnamefont
  {J.}~\bibnamefont {Nham}}, \bibinfo {author} {\bibfnamefont {N.}~\bibnamefont
  {Kalchbrenner}}, \bibinfo {author} {\bibfnamefont {I.}~\bibnamefont
  {Sutskever}}, \bibinfo {author} {\bibfnamefont {T.}~\bibnamefont
  {Lillicrap}}, \bibinfo {author} {\bibfnamefont {M.}~\bibnamefont {Leach}},
  \bibinfo {author} {\bibfnamefont {K.}~\bibnamefont {Kavukcuoglu}}, \bibinfo
  {author} {\bibfnamefont {T.}~\bibnamefont {Graepel}},\ and\ \bibinfo {author}
  {\bibfnamefont {D.}~\bibnamefont {Hassabis}},\ }\bibfield  {title} {\bibinfo
  {title} {Mastering the game of go with deep neural networks and tree
  search},\ }\href {https://doi.org/10.1038/nature16961} {\bibfield  {journal}
  {\bibinfo  {journal} {Nature}\ }\textbf {\bibinfo {volume} {529}},\ \bibinfo
  {pages} {484} (\bibinfo {year} {2016})}\BibitemShut {NoStop}%
\bibitem [{\citenamefont {Silver}\ \emph {et~al.}(2017)\citenamefont {Silver},
  \citenamefont {Schrittwieser}, \citenamefont {Simonyan}, \citenamefont
  {Antonoglou}, \citenamefont {Huang}, \citenamefont {Guez}, \citenamefont
  {Hubert}, \citenamefont {Baker}, \citenamefont {Lai}, \citenamefont {Bolton},
  \citenamefont {Chen}, \citenamefont {Lillicrap}, \citenamefont {Hui},
  \citenamefont {Sifre}, \citenamefont {van~den Driessche}, \citenamefont
  {Graepel},\ and\ \citenamefont {Hassabis}}]{Silver2017}%
  \BibitemOpen
  \bibfield  {author} {\bibinfo {author} {\bibfnamefont {D.}~\bibnamefont
  {Silver}}, \bibinfo {author} {\bibfnamefont {J.}~\bibnamefont
  {Schrittwieser}}, \bibinfo {author} {\bibfnamefont {K.}~\bibnamefont
  {Simonyan}}, \bibinfo {author} {\bibfnamefont {I.}~\bibnamefont
  {Antonoglou}}, \bibinfo {author} {\bibfnamefont {A.}~\bibnamefont {Huang}},
  \bibinfo {author} {\bibfnamefont {A.}~\bibnamefont {Guez}}, \bibinfo {author}
  {\bibfnamefont {T.}~\bibnamefont {Hubert}}, \bibinfo {author} {\bibfnamefont
  {L.}~\bibnamefont {Baker}}, \bibinfo {author} {\bibfnamefont
  {M.}~\bibnamefont {Lai}}, \bibinfo {author} {\bibfnamefont {A.}~\bibnamefont
  {Bolton}}, \bibinfo {author} {\bibfnamefont {Y.}~\bibnamefont {Chen}},
  \bibinfo {author} {\bibfnamefont {T.}~\bibnamefont {Lillicrap}}, \bibinfo
  {author} {\bibfnamefont {F.}~\bibnamefont {Hui}}, \bibinfo {author}
  {\bibfnamefont {L.}~\bibnamefont {Sifre}}, \bibinfo {author} {\bibfnamefont
  {G.}~\bibnamefont {van~den Driessche}}, \bibinfo {author} {\bibfnamefont
  {T.}~\bibnamefont {Graepel}},\ and\ \bibinfo {author} {\bibfnamefont
  {D.}~\bibnamefont {Hassabis}},\ }\bibfield  {title} {\bibinfo {title}
  {Mastering the game of go without human knowledge},\ }\href
  {https://doi.org/10.1038/nature24270} {\bibfield  {journal} {\bibinfo
  {journal} {Nature}\ }\textbf {\bibinfo {volume} {550}},\ \bibinfo {pages}
  {354} (\bibinfo {year} {2017})}\BibitemShut {NoStop}%
\bibitem [{\citenamefont {Jumper}\ \emph {et~al.}(2021)\citenamefont {Jumper},
  \citenamefont {Evans}, \citenamefont {Pritzel}, \citenamefont {Green},
  \citenamefont {Figurnov}, \citenamefont {Ronneberger}, \citenamefont
  {Tunyasuvunakool}, \citenamefont {Bates}, \citenamefont {{\v{Z}}{\'i}dek},
  \citenamefont {Potapenko}, \citenamefont {Bridgland}, \citenamefont {Meyer},
  \citenamefont {Kohl}, \citenamefont {Ballard}, \citenamefont {Cowie},
  \citenamefont {Romera-Paredes}, \citenamefont {Nikolov}, \citenamefont
  {Jain}, \citenamefont {Adler}, \citenamefont {Back}, \citenamefont
  {Petersen}, \citenamefont {Reiman}, \citenamefont {Clancy}, \citenamefont
  {Zielinski}, \citenamefont {Steinegger}, \citenamefont {Pacholska},
  \citenamefont {Berghammer}, \citenamefont {Bodenstein}, \citenamefont
  {Silver}, \citenamefont {Vinyals}, \citenamefont {Senior}, \citenamefont
  {Kavukcuoglu}, \citenamefont {Kohli},\ and\ \citenamefont
  {Hassabis}}]{Jumper2021}%
  \BibitemOpen
  \bibfield  {author} {\bibinfo {author} {\bibfnamefont {J.}~\bibnamefont
  {Jumper}}, \bibinfo {author} {\bibfnamefont {R.}~\bibnamefont {Evans}},
  \bibinfo {author} {\bibfnamefont {A.}~\bibnamefont {Pritzel}}, \bibinfo
  {author} {\bibfnamefont {T.}~\bibnamefont {Green}}, \bibinfo {author}
  {\bibfnamefont {M.}~\bibnamefont {Figurnov}}, \bibinfo {author}
  {\bibfnamefont {O.}~\bibnamefont {Ronneberger}}, \bibinfo {author}
  {\bibfnamefont {K.}~\bibnamefont {Tunyasuvunakool}}, \bibinfo {author}
  {\bibfnamefont {R.}~\bibnamefont {Bates}}, \bibinfo {author} {\bibfnamefont
  {A.}~\bibnamefont {{\v{Z}}{\'i}dek}}, \bibinfo {author} {\bibfnamefont
  {A.}~\bibnamefont {Potapenko}}, \bibinfo {author} {\bibfnamefont
  {A.}~\bibnamefont {Bridgland}}, \bibinfo {author} {\bibfnamefont
  {C.}~\bibnamefont {Meyer}}, \bibinfo {author} {\bibfnamefont {S.~A.~A.}\
  \bibnamefont {Kohl}}, \bibinfo {author} {\bibfnamefont {A.~J.}\ \bibnamefont
  {Ballard}}, \bibinfo {author} {\bibfnamefont {A.}~\bibnamefont {Cowie}},
  \bibinfo {author} {\bibfnamefont {B.}~\bibnamefont {Romera-Paredes}},
  \bibinfo {author} {\bibfnamefont {S.}~\bibnamefont {Nikolov}}, \bibinfo
  {author} {\bibfnamefont {R.}~\bibnamefont {Jain}}, \bibinfo {author}
  {\bibfnamefont {J.}~\bibnamefont {Adler}}, \bibinfo {author} {\bibfnamefont
  {T.}~\bibnamefont {Back}}, \bibinfo {author} {\bibfnamefont {S.}~\bibnamefont
  {Petersen}}, \bibinfo {author} {\bibfnamefont {D.}~\bibnamefont {Reiman}},
  \bibinfo {author} {\bibfnamefont {E.}~\bibnamefont {Clancy}}, \bibinfo
  {author} {\bibfnamefont {M.}~\bibnamefont {Zielinski}}, \bibinfo {author}
  {\bibfnamefont {M.}~\bibnamefont {Steinegger}}, \bibinfo {author}
  {\bibfnamefont {M.}~\bibnamefont {Pacholska}}, \bibinfo {author}
  {\bibfnamefont {T.}~\bibnamefont {Berghammer}}, \bibinfo {author}
  {\bibfnamefont {S.}~\bibnamefont {Bodenstein}}, \bibinfo {author}
  {\bibfnamefont {D.}~\bibnamefont {Silver}}, \bibinfo {author} {\bibfnamefont
  {O.}~\bibnamefont {Vinyals}}, \bibinfo {author} {\bibfnamefont {A.~W.}\
  \bibnamefont {Senior}}, \bibinfo {author} {\bibfnamefont {K.}~\bibnamefont
  {Kavukcuoglu}}, \bibinfo {author} {\bibfnamefont {P.}~\bibnamefont {Kohli}},\
  and\ \bibinfo {author} {\bibfnamefont {D.}~\bibnamefont {Hassabis}},\
  }\bibfield  {title} {\bibinfo {title} {Highly accurate protein structure
  prediction with alphafold},\ }\href
  {https://doi.org/10.1038/s41586-021-03819-2} {\bibfield  {journal} {\bibinfo
  {journal} {Nature}\ }\textbf {\bibinfo {volume} {596}},\ \bibinfo {pages}
  {583} (\bibinfo {year} {2021})}\BibitemShut {NoStop}%
\bibitem [{\citenamefont {Davies}\ \emph {et~al.}(2021)\citenamefont {Davies},
  \citenamefont {Veli{\v{c}}kovi{\'{c}}}, \citenamefont {Buesing},
  \citenamefont {Blackwell}, \citenamefont {Zheng}, \citenamefont
  {Toma{\v{s}}ev}, \citenamefont {Tanburn}, \citenamefont {Battaglia},
  \citenamefont {Blundell}, \citenamefont {Juh{\'a}sz}, \citenamefont
  {Lackenby}, \citenamefont {Williamson}, \citenamefont {Hassabis},\ and\
  \citenamefont {Kohli}}]{Davies2021}%
  \BibitemOpen
  \bibfield  {author} {\bibinfo {author} {\bibfnamefont {A.}~\bibnamefont
  {Davies}}, \bibinfo {author} {\bibfnamefont {P.}~\bibnamefont
  {Veli{\v{c}}kovi{\'{c}}}}, \bibinfo {author} {\bibfnamefont {L.}~\bibnamefont
  {Buesing}}, \bibinfo {author} {\bibfnamefont {S.}~\bibnamefont {Blackwell}},
  \bibinfo {author} {\bibfnamefont {D.}~\bibnamefont {Zheng}}, \bibinfo
  {author} {\bibfnamefont {N.}~\bibnamefont {Toma{\v{s}}ev}}, \bibinfo {author}
  {\bibfnamefont {R.}~\bibnamefont {Tanburn}}, \bibinfo {author} {\bibfnamefont
  {P.}~\bibnamefont {Battaglia}}, \bibinfo {author} {\bibfnamefont
  {C.}~\bibnamefont {Blundell}}, \bibinfo {author} {\bibfnamefont
  {A.}~\bibnamefont {Juh{\'a}sz}}, \bibinfo {author} {\bibfnamefont
  {M.}~\bibnamefont {Lackenby}}, \bibinfo {author} {\bibfnamefont
  {G.}~\bibnamefont {Williamson}}, \bibinfo {author} {\bibfnamefont
  {D.}~\bibnamefont {Hassabis}},\ and\ \bibinfo {author} {\bibfnamefont
  {P.}~\bibnamefont {Kohli}},\ }\bibfield  {title} {\bibinfo {title} {Advancing
  mathematics by guiding human intuition with ai},\ }\href
  {https://doi.org/10.1038/s41586-021-04086-x} {\bibfield  {journal} {\bibinfo
  {journal} {Nature}\ }\textbf {\bibinfo {volume} {600}},\ \bibinfo {pages}
  {70} (\bibinfo {year} {2021})}\BibitemShut {NoStop}%
\bibitem [{\citenamefont {Agostinelli}\ \emph {et~al.}(2019)\citenamefont
  {Agostinelli}, \citenamefont {McAleer}, \citenamefont {Shmakov},\ and\
  \citenamefont {Baldi}}]{agostinelli2019solving}%
  \BibitemOpen
  \bibfield  {author} {\bibinfo {author} {\bibfnamefont {F.}~\bibnamefont
  {Agostinelli}}, \bibinfo {author} {\bibfnamefont {S.}~\bibnamefont
  {McAleer}}, \bibinfo {author} {\bibfnamefont {A.}~\bibnamefont {Shmakov}},\
  and\ \bibinfo {author} {\bibfnamefont {P.}~\bibnamefont {Baldi}},\ }\bibfield
   {title} {\bibinfo {title} {Solving the rubik’s cube with deep
  reinforcement learning and search},\ }\href
  {https://www.nature.com/articles/s42256-019-0070-z} {\bibfield  {journal}
  {\bibinfo  {journal} {Nature Machine Intelligence}\ }\textbf {\bibinfo
  {volume} {1}},\ \bibinfo {pages} {356} (\bibinfo {year} {2019})}\BibitemShut
  {NoStop}%
\bibitem [{\citenamefont {Carleo}\ \emph {et~al.}(2019)\citenamefont {Carleo},
  \citenamefont {Cirac}, \citenamefont {Cranmer}, \citenamefont {Daudet},
  \citenamefont {Schuld}, \citenamefont {Tishby}, \citenamefont
  {Vogt-Maranto},\ and\ \citenamefont {Zdeborov\'a}}]{Carleo2019}%
  \BibitemOpen
  \bibfield  {author} {\bibinfo {author} {\bibfnamefont {G.}~\bibnamefont
  {Carleo}}, \bibinfo {author} {\bibfnamefont {I.}~\bibnamefont {Cirac}},
  \bibinfo {author} {\bibfnamefont {K.}~\bibnamefont {Cranmer}}, \bibinfo
  {author} {\bibfnamefont {L.}~\bibnamefont {Daudet}}, \bibinfo {author}
  {\bibfnamefont {M.}~\bibnamefont {Schuld}}, \bibinfo {author} {\bibfnamefont
  {N.}~\bibnamefont {Tishby}}, \bibinfo {author} {\bibfnamefont
  {L.}~\bibnamefont {Vogt-Maranto}},\ and\ \bibinfo {author} {\bibfnamefont
  {L.}~\bibnamefont {Zdeborov\'a}},\ }\bibfield  {title} {\bibinfo {title}
  {Machine learning and the physical sciences},\ }\href
  {https://doi.org/10.1103/RevModPhys.91.045002} {\bibfield  {journal}
  {\bibinfo  {journal} {Rev. Mod. Phys.}\ }\textbf {\bibinfo {volume} {91}},\
  \bibinfo {pages} {045002} (\bibinfo {year} {2019})}\BibitemShut {NoStop}%
\bibitem [{\citenamefont {Carleo}\ and\ \citenamefont
  {Troyer}(2017)}]{Carleo2016}%
  \BibitemOpen
  \bibfield  {author} {\bibinfo {author} {\bibfnamefont {G.}~\bibnamefont
  {Carleo}}\ and\ \bibinfo {author} {\bibfnamefont {M.}~\bibnamefont
  {Troyer}},\ }\bibfield  {title} {\bibinfo {title} {{Solving the quantum
  many-body problem with artificial neural networks}},\ }\href
  {https://doi.org/10.1126/science.aag2302} {\bibfield  {journal} {\bibinfo
  {journal} {Science}\ }\textbf {\bibinfo {volume} {355}},\ \bibinfo {pages}
  {602} (\bibinfo {year} {2017})}\BibitemShut {NoStop}%
\bibitem [{\citenamefont {Deng}\ \emph {et~al.}(2017)\citenamefont {Deng},
  \citenamefont {Li},\ and\ \citenamefont {Das~Sarma}}]{Deng2016}%
  \BibitemOpen
  \bibfield  {author} {\bibinfo {author} {\bibfnamefont {D.-L.}\ \bibnamefont
  {Deng}}, \bibinfo {author} {\bibfnamefont {X.}~\bibnamefont {Li}},\ and\
  \bibinfo {author} {\bibfnamefont {S.}~\bibnamefont {Das~Sarma}},\ }\bibfield
  {title} {\bibinfo {title} {Machine learning topological states},\ }\href
  {https://doi.org/10.1103/PhysRevB.96.195145} {\bibfield  {journal} {\bibinfo
  {journal} {Phys. Rev. B}\ }\textbf {\bibinfo {volume} {96}},\ \bibinfo
  {pages} {195145} (\bibinfo {year} {2017})}\BibitemShut {NoStop}%
\bibitem [{\citenamefont {Torlai}\ \emph {et~al.}(2018)\citenamefont {Torlai},
  \citenamefont {Mazzola}, \citenamefont {Carrasquilla}, \citenamefont
  {Troyer}, \citenamefont {Melko},\ and\ \citenamefont {Carleo}}]{Torlai2018}%
  \BibitemOpen
  \bibfield  {author} {\bibinfo {author} {\bibfnamefont {G.}~\bibnamefont
  {Torlai}}, \bibinfo {author} {\bibfnamefont {G.}~\bibnamefont {Mazzola}},
  \bibinfo {author} {\bibfnamefont {J.}~\bibnamefont {Carrasquilla}}, \bibinfo
  {author} {\bibfnamefont {M.}~\bibnamefont {Troyer}}, \bibinfo {author}
  {\bibfnamefont {R.}~\bibnamefont {Melko}},\ and\ \bibinfo {author}
  {\bibfnamefont {G.}~\bibnamefont {Carleo}},\ }\bibfield  {title} {\bibinfo
  {title} {Neural-network quantum state tomography},\ }\href
  {https://doi.org/10.1038/s41567-018-0048-5} {\bibfield  {journal} {\bibinfo
  {journal} {Nature Physics}\ }\textbf {\bibinfo {volume} {14}},\ \bibinfo
  {pages} {447} (\bibinfo {year} {2018})}\BibitemShut {NoStop}%
\bibitem [{\citenamefont {Melnikov}\ \emph {et~al.}(2018)\citenamefont
  {Melnikov}, \citenamefont {Poulsen~Nautrup}, \citenamefont {Krenn},
  \citenamefont {Dunjko}, \citenamefont {Tiersch}, \citenamefont {Zeilinger},\
  and\ \citenamefont {Briegel}}]{Melnikov2018}%
  \BibitemOpen
  \bibfield  {author} {\bibinfo {author} {\bibfnamefont {A.~A.}\ \bibnamefont
  {Melnikov}}, \bibinfo {author} {\bibfnamefont {H.}~\bibnamefont
  {Poulsen~Nautrup}}, \bibinfo {author} {\bibfnamefont {M.}~\bibnamefont
  {Krenn}}, \bibinfo {author} {\bibfnamefont {V.}~\bibnamefont {Dunjko}},
  \bibinfo {author} {\bibfnamefont {M.}~\bibnamefont {Tiersch}}, \bibinfo
  {author} {\bibfnamefont {A.}~\bibnamefont {Zeilinger}},\ and\ \bibinfo
  {author} {\bibfnamefont {H.~J.}\ \bibnamefont {Briegel}},\ }\bibfield
  {title} {\bibinfo {title} {Active learning machine learns to create new
  quantum experiments},\ }\href {https://doi.org/10.1073/pnas.1714936115}
  {\bibfield  {journal} {\bibinfo  {journal} {Proceedings of the National
  Academy of Sciences}\ }\textbf {\bibinfo {volume} {115}},\ \bibinfo {pages}
  {1221} (\bibinfo {year} {2018})}\BibitemShut {NoStop}%
\bibitem [{\citenamefont {Luo}\ and\ \citenamefont {Clark}(2019)}]{Luo2019}%
  \BibitemOpen
  \bibfield  {author} {\bibinfo {author} {\bibfnamefont {D.}~\bibnamefont
  {Luo}}\ and\ \bibinfo {author} {\bibfnamefont {B.~K.}\ \bibnamefont
  {Clark}},\ }\bibfield  {title} {\bibinfo {title} {Backflow transformations
  via neural networks for quantum many-body wave functions},\ }\href
  {https://doi.org/10.1103/PhysRevLett.122.226401} {\bibfield  {journal}
  {\bibinfo  {journal} {Phys. Rev. Lett.}\ }\textbf {\bibinfo {volume} {122}},\
  \bibinfo {pages} {226401} (\bibinfo {year} {2019})}\BibitemShut {NoStop}%
\bibitem [{\citenamefont {Zhang}\ \emph
  {et~al.}(2020{\natexlab{a}})\citenamefont {Zhang}, \citenamefont {Zheng},
  \citenamefont {Zhang},\ and\ \citenamefont {Deng}}]{Zhang2020TQC}%
  \BibitemOpen
  \bibfield  {author} {\bibinfo {author} {\bibfnamefont {Y.-H.}\ \bibnamefont
  {Zhang}}, \bibinfo {author} {\bibfnamefont {P.-L.}\ \bibnamefont {Zheng}},
  \bibinfo {author} {\bibfnamefont {Y.}~\bibnamefont {Zhang}},\ and\ \bibinfo
  {author} {\bibfnamefont {D.-L.}\ \bibnamefont {Deng}},\ }\bibfield  {title}
  {\bibinfo {title} {Topological quantum compiling with reinforcement
  learning},\ }\href {https://doi.org/10.1103/PhysRevLett.125.170501}
  {\bibfield  {journal} {\bibinfo  {journal} {Phys. Rev. Lett.}\ }\textbf
  {\bibinfo {volume} {125}},\ \bibinfo {pages} {170501} (\bibinfo {year}
  {2020}{\natexlab{a}})}\BibitemShut {NoStop}%
\bibitem [{\citenamefont {Fournier}\ \emph {et~al.}(2020)\citenamefont
  {Fournier}, \citenamefont {Wang}, \citenamefont {Yazyev},\ and\ \citenamefont
  {Wu}}]{Fournier2020}%
  \BibitemOpen
  \bibfield  {author} {\bibinfo {author} {\bibfnamefont {R.}~\bibnamefont
  {Fournier}}, \bibinfo {author} {\bibfnamefont {L.}~\bibnamefont {Wang}},
  \bibinfo {author} {\bibfnamefont {O.~V.}\ \bibnamefont {Yazyev}},\ and\
  \bibinfo {author} {\bibfnamefont {Q.}~\bibnamefont {Wu}},\ }\bibfield
  {title} {\bibinfo {title} {Artificial neural network approach to the analytic
  continuation problem},\ }\href
  {https://doi.org/10.1103/PhysRevLett.124.056401} {\bibfield  {journal}
  {\bibinfo  {journal} {Phys. Rev. Lett.}\ }\textbf {\bibinfo {volume} {124}},\
  \bibinfo {pages} {056401} (\bibinfo {year} {2020})}\BibitemShut {NoStop}%
\bibitem [{\citenamefont {Huang}\ \emph
  {et~al.}(2022{\natexlab{a}})\citenamefont {Huang}, \citenamefont {Kueng},
  \citenamefont {Torlai}, \citenamefont {Albert},\ and\ \citenamefont
  {Preskill}}]{Huang2022ML}%
  \BibitemOpen
  \bibfield  {author} {\bibinfo {author} {\bibfnamefont {H.-Y.}\ \bibnamefont
  {Huang}}, \bibinfo {author} {\bibfnamefont {R.}~\bibnamefont {Kueng}},
  \bibinfo {author} {\bibfnamefont {G.}~\bibnamefont {Torlai}}, \bibinfo
  {author} {\bibfnamefont {V.~V.}\ \bibnamefont {Albert}},\ and\ \bibinfo
  {author} {\bibfnamefont {J.}~\bibnamefont {Preskill}},\ }\bibfield  {title}
  {\bibinfo {title} {Provably efficient machine learning for quantum many-body
  problems},\ }\href {https://doi.org/10.1126/science.abk3333} {\bibfield
  {journal} {\bibinfo  {journal} {Science}\ }\textbf {\bibinfo {volume}
  {377}},\ \bibinfo {pages} {eabk3333} (\bibinfo {year}
  {2022}{\natexlab{a}})}\BibitemShut {NoStop}%
\bibitem [{\citenamefont {Zhang}\ and\ \citenamefont {Kim}(2017)}]{qlt2016}%
  \BibitemOpen
  \bibfield  {author} {\bibinfo {author} {\bibfnamefont {Y.}~\bibnamefont
  {Zhang}}\ and\ \bibinfo {author} {\bibfnamefont {E.-A.}\ \bibnamefont
  {Kim}},\ }\bibfield  {title} {\bibinfo {title} {{Quantum Loop Topography for
  Machine Learning}},\ }\href {https://doi.org/10.1103/PhysRevLett.118.216401}
  {\bibfield  {journal} {\bibinfo  {journal} {Phys. Rev. Lett.}\ }\textbf
  {\bibinfo {volume} {118}},\ \bibinfo {pages} {216401} (\bibinfo {year}
  {2017})}\BibitemShut {NoStop}%
\bibitem [{\citenamefont {Carrasquilla}\ and\ \citenamefont
  {Melko}(2017)}]{Melko20161}%
  \BibitemOpen
  \bibfield  {author} {\bibinfo {author} {\bibfnamefont {J.}~\bibnamefont
  {Carrasquilla}}\ and\ \bibinfo {author} {\bibfnamefont {R.~G.}\ \bibnamefont
  {Melko}},\ }\bibfield  {title} {\bibinfo {title} {Machine learning phases of
  matter},\ }\href {http://dx.doi.org/10.1038/nphys4035} {\bibfield  {journal}
  {\bibinfo  {journal} {Nature Physics}\ }\textbf {\bibinfo {volume} {13}},\
  \bibinfo {pages} {431} (\bibinfo {year} {2017})}\BibitemShut {NoStop}%
\bibitem [{\citenamefont {Ohtsuki}\ and\ \citenamefont
  {Ohtsuki}(2016)}]{Ohtsuki2016}%
  \BibitemOpen
  \bibfield  {author} {\bibinfo {author} {\bibfnamefont {T.}~\bibnamefont
  {Ohtsuki}}\ and\ \bibinfo {author} {\bibfnamefont {T.}~\bibnamefont
  {Ohtsuki}},\ }\bibfield  {title} {\bibinfo {title} {{Deep Learning the
  Quantum Phase Transitions in Random Two-Dimensional Electron Systems}},\
  }\href {https://doi.org/10.7566/JPSJ.85.123706} {\bibfield  {journal}
  {\bibinfo  {journal} {Journal of the Physical Society of Japan}\ }\textbf
  {\bibinfo {volume} {85}},\ \bibinfo {pages} {123706} (\bibinfo {year}
  {2016})}\BibitemShut {NoStop}%
\bibitem [{\citenamefont {Ohtsuki}\ and\ \citenamefont
  {Ohtsuki}(2017)}]{Ohtsuki2017}%
  \BibitemOpen
  \bibfield  {author} {\bibinfo {author} {\bibfnamefont {T.}~\bibnamefont
  {Ohtsuki}}\ and\ \bibinfo {author} {\bibfnamefont {T.}~\bibnamefont
  {Ohtsuki}},\ }\bibfield  {title} {\bibinfo {title} {Deep learning the quantum
  phase transitions in random electron systems: Applications to three
  dimensions},\ }\href {https://doi.org/10.7566/JPSJ.86.044708} {\bibfield
  {journal} {\bibinfo  {journal} {Journal of the Physical Society of Japan}\
  }\textbf {\bibinfo {volume} {86}},\ \bibinfo {pages} {044708} (\bibinfo
  {year} {2017})}\BibitemShut {NoStop}%
\bibitem [{\citenamefont {van Nieuwenburg}\ \emph {et~al.}(2017)\citenamefont
  {van Nieuwenburg}, \citenamefont {Liu},\ and\ \citenamefont
  {Huber}}]{vanNieuwenburg2017}%
  \BibitemOpen
  \bibfield  {author} {\bibinfo {author} {\bibfnamefont {E.}~\bibnamefont {van
  Nieuwenburg}}, \bibinfo {author} {\bibfnamefont {Y.-H.}\ \bibnamefont
  {Liu}},\ and\ \bibinfo {author} {\bibfnamefont {S.}~\bibnamefont {Huber}},\
  }\bibfield  {title} {\bibinfo {title} {Learning phase transitions by
  confusion},\ }\href {https://doi.org/10.1038/nphys4037} {\bibfield  {journal}
  {\bibinfo  {journal} {Nature Physics}\ }\textbf {\bibinfo {volume} {13}},\
  \bibinfo {pages} {435} (\bibinfo {year} {2017})}\BibitemShut {NoStop}%
\bibitem [{\citenamefont {Zhang}\ \emph {et~al.}(2018)\citenamefont {Zhang},
  \citenamefont {Shen},\ and\ \citenamefont {Zhai}}]{Zhaihui2018}%
  \BibitemOpen
  \bibfield  {author} {\bibinfo {author} {\bibfnamefont {P.}~\bibnamefont
  {Zhang}}, \bibinfo {author} {\bibfnamefont {H.}~\bibnamefont {Shen}},\ and\
  \bibinfo {author} {\bibfnamefont {H.}~\bibnamefont {Zhai}},\ }\bibfield
  {title} {\bibinfo {title} {Machine learning topological invariants with
  neural networks},\ }\href {https://doi.org/10.1103/PhysRevLett.120.066401}
  {\bibfield  {journal} {\bibinfo  {journal} {Phys. Rev. Lett.}\ }\textbf
  {\bibinfo {volume} {120}},\ \bibinfo {pages} {066401} (\bibinfo {year}
  {2018})}\BibitemShut {NoStop}%
\bibitem [{\citenamefont {Zhang}\ \emph {et~al.}(2017)\citenamefont {Zhang},
  \citenamefont {Melko},\ and\ \citenamefont {Kim}}]{FrankMLZ2}%
  \BibitemOpen
  \bibfield  {author} {\bibinfo {author} {\bibfnamefont {Y.}~\bibnamefont
  {Zhang}}, \bibinfo {author} {\bibfnamefont {R.~G.}\ \bibnamefont {Melko}},\
  and\ \bibinfo {author} {\bibfnamefont {E.-A.}\ \bibnamefont {Kim}},\
  }\bibfield  {title} {\bibinfo {title} {{Machine learning ${\mathbb{Z}}_{2}$
  quantum spin liquids with quasiparticle statistics}},\ }\href
  {https://doi.org/10.1103/PhysRevB.96.245119} {\bibfield  {journal} {\bibinfo
  {journal} {Phys. Rev. B}\ }\textbf {\bibinfo {volume} {96}},\ \bibinfo
  {pages} {245119} (\bibinfo {year} {2017})}\BibitemShut {NoStop}%
\bibitem [{\citenamefont {Lian}\ \emph {et~al.}(2019)\citenamefont {Lian},
  \citenamefont {Wang}, \citenamefont {Lu}, \citenamefont {Huang},
  \citenamefont {Wang}, \citenamefont {Yuan}, \citenamefont {Zhang},
  \citenamefont {Ouyang}, \citenamefont {Wang}, \citenamefont {Huang},
  \citenamefont {He}, \citenamefont {Chang}, \citenamefont {Deng},\ and\
  \citenamefont {Duan}}]{Lian2019}%
  \BibitemOpen
  \bibfield  {author} {\bibinfo {author} {\bibfnamefont {W.}~\bibnamefont
  {Lian}}, \bibinfo {author} {\bibfnamefont {S.-T.}\ \bibnamefont {Wang}},
  \bibinfo {author} {\bibfnamefont {S.}~\bibnamefont {Lu}}, \bibinfo {author}
  {\bibfnamefont {Y.}~\bibnamefont {Huang}}, \bibinfo {author} {\bibfnamefont
  {F.}~\bibnamefont {Wang}}, \bibinfo {author} {\bibfnamefont {X.}~\bibnamefont
  {Yuan}}, \bibinfo {author} {\bibfnamefont {W.}~\bibnamefont {Zhang}},
  \bibinfo {author} {\bibfnamefont {X.}~\bibnamefont {Ouyang}}, \bibinfo
  {author} {\bibfnamefont {X.}~\bibnamefont {Wang}}, \bibinfo {author}
  {\bibfnamefont {X.}~\bibnamefont {Huang}}, \bibinfo {author} {\bibfnamefont
  {L.}~\bibnamefont {He}}, \bibinfo {author} {\bibfnamefont {X.}~\bibnamefont
  {Chang}}, \bibinfo {author} {\bibfnamefont {D.-L.}\ \bibnamefont {Deng}},\
  and\ \bibinfo {author} {\bibfnamefont {L.}~\bibnamefont {Duan}},\ }\bibfield
  {title} {\bibinfo {title} {Machine learning topological phases with a
  solid-state quantum simulator},\ }\href
  {https://doi.org/10.1103/PhysRevLett.122.210503} {\bibfield  {journal}
  {\bibinfo  {journal} {Phys. Rev. Lett.}\ }\textbf {\bibinfo {volume} {122}},\
  \bibinfo {pages} {210503} (\bibinfo {year} {2019})}\BibitemShut {NoStop}%
\bibitem [{\citenamefont {Zhang}\ \emph {et~al.}(2019)\citenamefont {Zhang},
  \citenamefont {Mesaros}, \citenamefont {Fujita}, \citenamefont {Edkins},
  \citenamefont {Hamidian}, \citenamefont {Ch'ng}, \citenamefont {Eisaki},
  \citenamefont {Uchida}, \citenamefont {Davis}, \citenamefont {Khatami} \emph
  {et~al.}}]{mlstm2019}%
  \BibitemOpen
  \bibfield  {author} {\bibinfo {author} {\bibfnamefont {Y.}~\bibnamefont
  {Zhang}}, \bibinfo {author} {\bibfnamefont {A.}~\bibnamefont {Mesaros}},
  \bibinfo {author} {\bibfnamefont {K.}~\bibnamefont {Fujita}}, \bibinfo
  {author} {\bibfnamefont {S.}~\bibnamefont {Edkins}}, \bibinfo {author}
  {\bibfnamefont {M.}~\bibnamefont {Hamidian}}, \bibinfo {author}
  {\bibfnamefont {K.}~\bibnamefont {Ch'ng}}, \bibinfo {author} {\bibfnamefont
  {H.}~\bibnamefont {Eisaki}}, \bibinfo {author} {\bibfnamefont
  {S.}~\bibnamefont {Uchida}}, \bibinfo {author} {\bibfnamefont {J.~S.}\
  \bibnamefont {Davis}}, \bibinfo {author} {\bibfnamefont {E.}~\bibnamefont
  {Khatami}}, \emph {et~al.},\ }\bibfield  {title} {\bibinfo {title} {Machine
  learning in electronic-quantum-matter imaging experiments},\ }\href
  {https://doi.org/10.1038/s41586-019-1319-8} {\bibfield  {journal} {\bibinfo
  {journal} {Nature}\ }\textbf {\bibinfo {volume} {570}},\ \bibinfo {pages}
  {484} (\bibinfo {year} {2019})}\BibitemShut {NoStop}%
\bibitem [{\citenamefont {Bohrdt}\ \emph {et~al.}(2019)\citenamefont {Bohrdt},
  \citenamefont {Chiu}, \citenamefont {Ji}, \citenamefont {Xu}, \citenamefont
  {Greif}, \citenamefont {Greiner}, \citenamefont {Demler}, \citenamefont
  {Grusdt},\ and\ \citenamefont {Knap}}]{Bohrdt2019}%
  \BibitemOpen
  \bibfield  {author} {\bibinfo {author} {\bibfnamefont {A.}~\bibnamefont
  {Bohrdt}}, \bibinfo {author} {\bibfnamefont {C.~S.}\ \bibnamefont {Chiu}},
  \bibinfo {author} {\bibfnamefont {G.}~\bibnamefont {Ji}}, \bibinfo {author}
  {\bibfnamefont {M.}~\bibnamefont {Xu}}, \bibinfo {author} {\bibfnamefont
  {D.}~\bibnamefont {Greif}}, \bibinfo {author} {\bibfnamefont
  {M.}~\bibnamefont {Greiner}}, \bibinfo {author} {\bibfnamefont
  {E.}~\bibnamefont {Demler}}, \bibinfo {author} {\bibfnamefont
  {F.}~\bibnamefont {Grusdt}},\ and\ \bibinfo {author} {\bibfnamefont
  {M.}~\bibnamefont {Knap}},\ }\bibfield  {title} {\bibinfo {title}
  {Classifying snapshots of the doped hubbard model with machine learning},\
  }\href {https://doi.org/10.1038/s41567-019-0565-x} {\bibfield  {journal}
  {\bibinfo  {journal} {Nature Physics}\ }\textbf {\bibinfo {volume} {15}},\
  \bibinfo {pages} {921} (\bibinfo {year} {2019})}\BibitemShut {NoStop}%
\bibitem [{\citenamefont {Rem}\ \emph {et~al.}(2019)\citenamefont {Rem},
  \citenamefont {K{\"a}ming}, \citenamefont {Tarnowski}, \citenamefont
  {Asteria}, \citenamefont {Fl{\"a}schner}, \citenamefont {Becker},
  \citenamefont {Sengstock},\ and\ \citenamefont {Weitenberg}}]{Rem2019}%
  \BibitemOpen
  \bibfield  {author} {\bibinfo {author} {\bibfnamefont {B.~S.}\ \bibnamefont
  {Rem}}, \bibinfo {author} {\bibfnamefont {N.}~\bibnamefont {K{\"a}ming}},
  \bibinfo {author} {\bibfnamefont {M.}~\bibnamefont {Tarnowski}}, \bibinfo
  {author} {\bibfnamefont {L.}~\bibnamefont {Asteria}}, \bibinfo {author}
  {\bibfnamefont {N.}~\bibnamefont {Fl{\"a}schner}}, \bibinfo {author}
  {\bibfnamefont {C.}~\bibnamefont {Becker}}, \bibinfo {author} {\bibfnamefont
  {K.}~\bibnamefont {Sengstock}},\ and\ \bibinfo {author} {\bibfnamefont
  {C.}~\bibnamefont {Weitenberg}},\ }\bibfield  {title} {\bibinfo {title}
  {Identifying quantum phase transitions using artificial neural networks on
  experimental data},\ }\href {https://doi.org/10.1038/s41567-019-0554-0}
  {\bibfield  {journal} {\bibinfo  {journal} {Nature Physics}\ }\textbf
  {\bibinfo {volume} {15}},\ \bibinfo {pages} {917} (\bibinfo {year}
  {2019})}\BibitemShut {NoStop}%
\bibitem [{\citenamefont {Peano}\ \emph {et~al.}(2021)\citenamefont {Peano},
  \citenamefont {Sapper},\ and\ \citenamefont {Marquardt}}]{Peano2021}%
  \BibitemOpen
  \bibfield  {author} {\bibinfo {author} {\bibfnamefont {V.}~\bibnamefont
  {Peano}}, \bibinfo {author} {\bibfnamefont {F.}~\bibnamefont {Sapper}},\ and\
  \bibinfo {author} {\bibfnamefont {F.}~\bibnamefont {Marquardt}},\ }\bibfield
  {title} {\bibinfo {title} {Rapid exploration of topological band structures
  using deep learning},\ }\href {https://doi.org/10.1103/PhysRevX.11.021052}
  {\bibfield  {journal} {\bibinfo  {journal} {Phys. Rev. X}\ }\textbf {\bibinfo
  {volume} {11}},\ \bibinfo {pages} {021052} (\bibinfo {year}
  {2021})}\BibitemShut {NoStop}%
\bibitem [{\citenamefont {Arnold}\ and\ \citenamefont
  {Sch\"afer}(2022)}]{Arnold2022}%
  \BibitemOpen
  \bibfield  {author} {\bibinfo {author} {\bibfnamefont {J.}~\bibnamefont
  {Arnold}}\ and\ \bibinfo {author} {\bibfnamefont {F.}~\bibnamefont
  {Sch\"afer}},\ }\bibfield  {title} {\bibinfo {title} {Replacing neural
  networks by optimal analytical predictors for the detection of phase
  transitions},\ }\href {https://doi.org/10.1103/PhysRevX.12.031044} {\bibfield
   {journal} {\bibinfo  {journal} {Phys. Rev. X}\ }\textbf {\bibinfo {volume}
  {12}},\ \bibinfo {pages} {031044} (\bibinfo {year} {2022})}\BibitemShut
  {NoStop}%
\bibitem [{\citenamefont {Zheng}\ \emph {et~al.}(2022)\citenamefont {Zheng},
  \citenamefont {Du},\ and\ \citenamefont {Zhang}}]{Zheng2022}%
  \BibitemOpen
  \bibfield  {author} {\bibinfo {author} {\bibfnamefont {P.-L.}\ \bibnamefont
  {Zheng}}, \bibinfo {author} {\bibfnamefont {S.-J.}\ \bibnamefont {Du}},\ and\
  \bibinfo {author} {\bibfnamefont {Y.}~\bibnamefont {Zhang}},\ }\bibfield
  {title} {\bibinfo {title} {Ground-state properties via machine learning
  quantum constraints},\ }\href
  {https://doi.org/10.1103/PhysRevResearch.4.L032043} {\bibfield  {journal}
  {\bibinfo  {journal} {Phys. Rev. Research}\ }\textbf {\bibinfo {volume}
  {4}},\ \bibinfo {pages} {L032043} (\bibinfo {year} {2022})}\BibitemShut
  {NoStop}%
\bibitem [{\citenamefont {Koolstra}\ \emph {et~al.}(2022)\citenamefont
  {Koolstra}, \citenamefont {Stevenson}, \citenamefont {Barzili}, \citenamefont
  {Burns}, \citenamefont {Siva}, \citenamefont {Greenfield}, \citenamefont
  {Livingston}, \citenamefont {Hashim}, \citenamefont {Naik}, \citenamefont
  {Kreikebaum}, \citenamefont {O'Brien}, \citenamefont {Santiago},
  \citenamefont {Dressel},\ and\ \citenamefont {Siddiqi}}]{Koolstra2022}%
  \BibitemOpen
  \bibfield  {author} {\bibinfo {author} {\bibfnamefont {G.}~\bibnamefont
  {Koolstra}}, \bibinfo {author} {\bibfnamefont {N.}~\bibnamefont {Stevenson}},
  \bibinfo {author} {\bibfnamefont {S.}~\bibnamefont {Barzili}}, \bibinfo
  {author} {\bibfnamefont {L.}~\bibnamefont {Burns}}, \bibinfo {author}
  {\bibfnamefont {K.}~\bibnamefont {Siva}}, \bibinfo {author} {\bibfnamefont
  {S.}~\bibnamefont {Greenfield}}, \bibinfo {author} {\bibfnamefont
  {W.}~\bibnamefont {Livingston}}, \bibinfo {author} {\bibfnamefont
  {A.}~\bibnamefont {Hashim}}, \bibinfo {author} {\bibfnamefont {R.~K.}\
  \bibnamefont {Naik}}, \bibinfo {author} {\bibfnamefont {J.~M.}\ \bibnamefont
  {Kreikebaum}}, \bibinfo {author} {\bibfnamefont {K.~P.}\ \bibnamefont
  {O'Brien}}, \bibinfo {author} {\bibfnamefont {D.~I.}\ \bibnamefont
  {Santiago}}, \bibinfo {author} {\bibfnamefont {J.}~\bibnamefont {Dressel}},\
  and\ \bibinfo {author} {\bibfnamefont {I.}~\bibnamefont {Siddiqi}},\
  }\bibfield  {title} {\bibinfo {title} {Monitoring fast superconducting qubit
  dynamics using a neural network},\ }\href
  {https://doi.org/10.1103/PhysRevX.12.031017} {\bibfield  {journal} {\bibinfo
  {journal} {Phys. Rev. X}\ }\textbf {\bibinfo {volume} {12}},\ \bibinfo
  {pages} {031017} (\bibinfo {year} {2022})}\BibitemShut {NoStop}%
\bibitem [{\citenamefont {Tibaldi}\ \emph {et~al.}(2023)\citenamefont
  {Tibaldi}, \citenamefont {Magnifico}, \citenamefont {Vodola},\ and\
  \citenamefont {Ercolessi}}]{Tibaldi2202}%
  \BibitemOpen
  \bibfield  {author} {\bibinfo {author} {\bibfnamefont {S.}~\bibnamefont
  {Tibaldi}}, \bibinfo {author} {\bibfnamefont {G.}~\bibnamefont {Magnifico}},
  \bibinfo {author} {\bibfnamefont {D.}~\bibnamefont {Vodola}},\ and\ \bibinfo
  {author} {\bibfnamefont {E.}~\bibnamefont {Ercolessi}},\ }\bibfield  {title}
  {\bibinfo {title} {{Unsupervised and supervised learning of interacting
  topological phases from single-particle correlation functions}},\ }\href
  {https://doi.org/10.21468/SciPostPhys.14.1.005} {\bibfield  {journal}
  {\bibinfo  {journal} {SciPost Phys.}\ }\textbf {\bibinfo {volume} {14}},\
  \bibinfo {pages} {005} (\bibinfo {year} {2023})}\BibitemShut {NoStop}%
\bibitem [{\citenamefont {Chang}\ \emph {et~al.}(2023)\citenamefont {Chang},
  \citenamefont {O'Leary}, \citenamefont {Su}, \citenamefont {Jacobs},
  \citenamefont {Kahn}, \citenamefont {Zettl}, \citenamefont {Ciston},
  \citenamefont {Ercius},\ and\ \citenamefont {Miao}}]{Chang2023}%
  \BibitemOpen
  \bibfield  {author} {\bibinfo {author} {\bibfnamefont {D.~J.}\ \bibnamefont
  {Chang}}, \bibinfo {author} {\bibfnamefont {C.~M.}\ \bibnamefont {O'Leary}},
  \bibinfo {author} {\bibfnamefont {C.}~\bibnamefont {Su}}, \bibinfo {author}
  {\bibfnamefont {D.~A.}\ \bibnamefont {Jacobs}}, \bibinfo {author}
  {\bibfnamefont {S.}~\bibnamefont {Kahn}}, \bibinfo {author} {\bibfnamefont
  {A.}~\bibnamefont {Zettl}}, \bibinfo {author} {\bibfnamefont
  {J.}~\bibnamefont {Ciston}}, \bibinfo {author} {\bibfnamefont
  {P.}~\bibnamefont {Ercius}},\ and\ \bibinfo {author} {\bibfnamefont
  {J.}~\bibnamefont {Miao}},\ }\bibfield  {title} {\bibinfo {title}
  {Deep-learning electron diffractive imaging},\ }\href
  {https://doi.org/10.1103/PhysRevLett.130.016101} {\bibfield  {journal}
  {\bibinfo  {journal} {Phys. Rev. Lett.}\ }\textbf {\bibinfo {volume} {130}},\
  \bibinfo {pages} {016101} (\bibinfo {year} {2023})}\BibitemShut {NoStop}%
\bibitem [{\citenamefont {Link}\ \emph {et~al.}(2023)\citenamefont {Link},
  \citenamefont {Gao}, \citenamefont {Kell}, \citenamefont {Breyer},
  \citenamefont {Eberz}, \citenamefont {Rauf},\ and\ \citenamefont
  {K\"ohl}}]{Link2023}%
  \BibitemOpen
  \bibfield  {author} {\bibinfo {author} {\bibfnamefont {M.}~\bibnamefont
  {Link}}, \bibinfo {author} {\bibfnamefont {K.}~\bibnamefont {Gao}}, \bibinfo
  {author} {\bibfnamefont {A.}~\bibnamefont {Kell}}, \bibinfo {author}
  {\bibfnamefont {M.}~\bibnamefont {Breyer}}, \bibinfo {author} {\bibfnamefont
  {D.}~\bibnamefont {Eberz}}, \bibinfo {author} {\bibfnamefont
  {B.}~\bibnamefont {Rauf}},\ and\ \bibinfo {author} {\bibfnamefont
  {M.}~\bibnamefont {K\"ohl}},\ }\bibfield  {title} {\bibinfo {title} {Machine
  learning the phase diagram of a strongly interacting fermi gas},\ }\href
  {https://doi.org/10.1103/PhysRevLett.130.203401} {\bibfield  {journal}
  {\bibinfo  {journal} {Phys. Rev. Lett.}\ }\textbf {\bibinfo {volume} {130}},\
  \bibinfo {pages} {203401} (\bibinfo {year} {2023})}\BibitemShut {NoStop}%
\bibitem [{\citenamefont {Lecun}\ \emph {et~al.}(1998)\citenamefont {Lecun},
  \citenamefont {Bottou}, \citenamefont {Bengio},\ and\ \citenamefont
  {Haffner}}]{MNIST1998}%
  \BibitemOpen
  \bibfield  {author} {\bibinfo {author} {\bibfnamefont {Y.}~\bibnamefont
  {Lecun}}, \bibinfo {author} {\bibfnamefont {L.}~\bibnamefont {Bottou}},
  \bibinfo {author} {\bibfnamefont {Y.}~\bibnamefont {Bengio}},\ and\ \bibinfo
  {author} {\bibfnamefont {P.}~\bibnamefont {Haffner}},\ }\bibfield  {title}
  {\bibinfo {title} {Gradient-based learning applied to document recognition},\
  }\href {https://doi.org/10.1109/5.726791} {\bibfield  {journal} {\bibinfo
  {journal} {Proceedings of the IEEE}\ }\textbf {\bibinfo {volume} {86}},\
  \bibinfo {pages} {2278} (\bibinfo {year} {1998})}\BibitemShut {NoStop}%
\bibitem [{\citenamefont {Georges}\ \emph {et~al.}(1996)\citenamefont
  {Georges}, \citenamefont {Kotliar}, \citenamefont {Krauth},\ and\
  \citenamefont {Rozenberg}}]{DMFT1996}%
  \BibitemOpen
  \bibfield  {author} {\bibinfo {author} {\bibfnamefont {A.}~\bibnamefont
  {Georges}}, \bibinfo {author} {\bibfnamefont {G.}~\bibnamefont {Kotliar}},
  \bibinfo {author} {\bibfnamefont {W.}~\bibnamefont {Krauth}},\ and\ \bibinfo
  {author} {\bibfnamefont {M.~J.}\ \bibnamefont {Rozenberg}},\ }\bibfield
  {title} {\bibinfo {title} {Dynamical mean-field theory of strongly correlated
  fermion systems and the limit of infinite dimensions},\ }\href
  {https://doi.org/10.1103/RevModPhys.68.13} {\bibfield  {journal} {\bibinfo
  {journal} {Rev. Mod. Phys.}\ }\textbf {\bibinfo {volume} {68}},\ \bibinfo
  {pages} {13} (\bibinfo {year} {1996})}\BibitemShut {NoStop}%
\bibitem [{\citenamefont {Metzner}\ and\ \citenamefont
  {Vollhardt}(1989)}]{Metzner1989}%
  \BibitemOpen
  \bibfield  {author} {\bibinfo {author} {\bibfnamefont {W.}~\bibnamefont
  {Metzner}}\ and\ \bibinfo {author} {\bibfnamefont {D.}~\bibnamefont
  {Vollhardt}},\ }\bibfield  {title} {\bibinfo {title} {Correlated lattice
  fermions in $d=\ensuremath{\infty}$ dimensions},\ }\href
  {https://doi.org/10.1103/PhysRevLett.62.324} {\bibfield  {journal} {\bibinfo
  {journal} {Phys. Rev. Lett.}\ }\textbf {\bibinfo {volume} {62}},\ \bibinfo
  {pages} {324} (\bibinfo {year} {1989})}\BibitemShut {NoStop}%
\bibitem [{\citenamefont {Vollhardt}(2012)}]{Vollhardt2012}%
  \BibitemOpen
  \bibfield  {author} {\bibinfo {author} {\bibfnamefont {D.}~\bibnamefont
  {Vollhardt}},\ }\bibfield  {title} {\bibinfo {title} {Dynamical mean-field
  theory for correlated electrons},\ }\href
  {https://doi.org/https://doi.org/10.1002/andp.201100250} {\bibfield
  {journal} {\bibinfo  {journal} {Annalen der Physik}\ }\textbf {\bibinfo
  {volume} {524}},\ \bibinfo {pages} {1} (\bibinfo {year} {2012})}\BibitemShut
  {NoStop}%
\bibitem [{\citenamefont {Vollhardt}\ \emph {et~al.}(2012)\citenamefont
  {Vollhardt}, \citenamefont {Byczuk},\ and\ \citenamefont
  {Kollar}}]{VollhardtBook2012}%
  \BibitemOpen
  \bibfield  {author} {\bibinfo {author} {\bibfnamefont {D.}~\bibnamefont
  {Vollhardt}}, \bibinfo {author} {\bibfnamefont {K.}~\bibnamefont {Byczuk}},\
  and\ \bibinfo {author} {\bibfnamefont {M.}~\bibnamefont {Kollar}},\ }\bibinfo
  {title} {Dynamical mean-field theory},\ in\ \href
  {https://doi.org/10.1007/978-3-642-21831-6_7} {\emph {\bibinfo {booktitle}
  {Strongly Correlated Systems: Theoretical Methods}}},\ \bibinfo {editor}
  {edited by\ \bibinfo {editor} {\bibfnamefont {A.}~\bibnamefont {Avella}}\
  and\ \bibinfo {editor} {\bibfnamefont {F.}~\bibnamefont {Mancini}}}\
  (\bibinfo  {publisher} {Springer Berlin Heidelberg},\ \bibinfo {address}
  {Berlin, Heidelberg},\ \bibinfo {year} {2012})\ pp.\ \bibinfo {pages}
  {203--236}\BibitemShut {NoStop}%
\bibitem [{\citenamefont {He}\ \emph {et~al.}(2016)\citenamefont {He},
  \citenamefont {Zhang}, \citenamefont {Ren},\ and\ \citenamefont
  {Sun}}]{He2016resnet}%
  \BibitemOpen
  \bibfield  {author} {\bibinfo {author} {\bibfnamefont {K.}~\bibnamefont
  {He}}, \bibinfo {author} {\bibfnamefont {X.}~\bibnamefont {Zhang}}, \bibinfo
  {author} {\bibfnamefont {S.}~\bibnamefont {Ren}},\ and\ \bibinfo {author}
  {\bibfnamefont {J.}~\bibnamefont {Sun}},\ }\bibfield  {title} {\bibinfo
  {title} {Deep residual learning for image recognition},\ }in\ \href@noop {}
  {\emph {\bibinfo {booktitle} {Proceedings of the IEEE Conference on Computer
  Vision and Pattern Recognition (CVPR)}}}\ (\bibinfo {year}
  {2016})\BibitemShut {NoStop}%
\bibitem [{\citenamefont {Kolen}\ and\ \citenamefont
  {Kremer}(2001)}]{VanGrad2001}%
  \BibitemOpen
  \bibfield  {author} {\bibinfo {author} {\bibfnamefont {J.~F.}\ \bibnamefont
  {Kolen}}\ and\ \bibinfo {author} {\bibfnamefont {S.~C.}\ \bibnamefont
  {Kremer}},\ }\bibinfo {title} {Gradient flow in recurrent nets: The
  difficulty of learning longterm dependencies},\ in\ \href
  {https://doi.org/10.1109/9780470544037.ch14} {\emph {\bibinfo {booktitle} {A
  Field Guide to Dynamical Recurrent Networks}}}\ (\bibinfo {year} {2001})\
  pp.\ \bibinfo {pages} {237--243}\BibitemShut {NoStop}%
\bibitem [{\citenamefont {Han}\ and\ \citenamefont
  {Moraga}(1995)}]{Han1995sigmoid}%
  \BibitemOpen
  \bibfield  {author} {\bibinfo {author} {\bibfnamefont {J.}~\bibnamefont
  {Han}}\ and\ \bibinfo {author} {\bibfnamefont {C.}~\bibnamefont {Moraga}},\
  }\bibfield  {title} {\bibinfo {title} {The influence of the sigmoid function
  parameters on the speed of backpropagation learning},\ }in\ \href@noop {}
  {\emph {\bibinfo {booktitle} {From Natural to Artificial Neural
  Computation}}},\ \bibinfo {editor} {edited by\ \bibinfo {editor}
  {\bibfnamefont {J.}~\bibnamefont {Mira}}\ and\ \bibinfo {editor}
  {\bibfnamefont {F.}~\bibnamefont {Sandoval}}}\ (\bibinfo  {publisher}
  {Springer Berlin Heidelberg},\ \bibinfo {address} {Berlin, Heidelberg},\
  \bibinfo {year} {1995})\ pp.\ \bibinfo {pages} {195--201}\BibitemShut
  {NoStop}%
\bibitem [{\citenamefont {Amico}\ \emph {et~al.}(2008)\citenamefont {Amico},
  \citenamefont {Fazio}, \citenamefont {Osterloh},\ and\ \citenamefont
  {Vedral}}]{Amico2008}%
  \BibitemOpen
  \bibfield  {author} {\bibinfo {author} {\bibfnamefont {L.}~\bibnamefont
  {Amico}}, \bibinfo {author} {\bibfnamefont {R.}~\bibnamefont {Fazio}},
  \bibinfo {author} {\bibfnamefont {A.}~\bibnamefont {Osterloh}},\ and\
  \bibinfo {author} {\bibfnamefont {V.}~\bibnamefont {Vedral}},\ }\bibfield
  {title} {\bibinfo {title} {Entanglement in many-body systems},\ }\href
  {https://doi.org/10.1103/RevModPhys.80.517} {\bibfield  {journal} {\bibinfo
  {journal} {Rev. Mod. Phys.}\ }\textbf {\bibinfo {volume} {80}},\ \bibinfo
  {pages} {517} (\bibinfo {year} {2008})}\BibitemShut {NoStop}%
\bibitem [{\citenamefont {Eisert}\ \emph {et~al.}(2010)\citenamefont {Eisert},
  \citenamefont {Cramer},\ and\ \citenamefont {Plenio}}]{Eisert2010}%
  \BibitemOpen
  \bibfield  {author} {\bibinfo {author} {\bibfnamefont {J.}~\bibnamefont
  {Eisert}}, \bibinfo {author} {\bibfnamefont {M.}~\bibnamefont {Cramer}},\
  and\ \bibinfo {author} {\bibfnamefont {M.~B.}\ \bibnamefont {Plenio}},\
  }\bibfield  {title} {\bibinfo {title} {Colloquium: Area laws for the
  entanglement entropy},\ }\href {https://doi.org/10.1103/RevModPhys.82.277}
  {\bibfield  {journal} {\bibinfo  {journal} {Rev. Mod. Phys.}\ }\textbf
  {\bibinfo {volume} {82}},\ \bibinfo {pages} {277} (\bibinfo {year}
  {2010})}\BibitemShut {NoStop}%
\bibitem [{\citenamefont {Zhang}\ \emph
  {et~al.}(2020{\natexlab{b}})\citenamefont {Zhang}, \citenamefont {Ginsparg},\
  and\ \citenamefont {Kim}}]{Zhang2020}%
  \BibitemOpen
  \bibfield  {author} {\bibinfo {author} {\bibfnamefont {Y.}~\bibnamefont
  {Zhang}}, \bibinfo {author} {\bibfnamefont {P.}~\bibnamefont {Ginsparg}},\
  and\ \bibinfo {author} {\bibfnamefont {E.-A.}\ \bibnamefont {Kim}},\
  }\bibfield  {title} {\bibinfo {title} {Interpreting machine learning of
  topological quantum phase transitions},\ }\href
  {https://doi.org/10.1103/PhysRevResearch.2.023283} {\bibfield  {journal}
  {\bibinfo  {journal} {Phys. Rev. Research}\ }\textbf {\bibinfo {volume}
  {2}},\ \bibinfo {pages} {023283} (\bibinfo {year}
  {2020}{\natexlab{b}})}\BibitemShut {NoStop}%
\bibitem [{\citenamefont {Ribeiro}\ \emph {et~al.}(2016)\citenamefont
  {Ribeiro}, \citenamefont {Singh},\ and\ \citenamefont
  {Guestrin}}]{Ribeiro2016}%
  \BibitemOpen
  \bibfield  {author} {\bibinfo {author} {\bibfnamefont {M.~T.}\ \bibnamefont
  {Ribeiro}}, \bibinfo {author} {\bibfnamefont {S.}~\bibnamefont {Singh}},\
  and\ \bibinfo {author} {\bibfnamefont {C.}~\bibnamefont {Guestrin}},\
  }\bibfield  {title} {\bibinfo {title} {"why should i trust you?": Explaining
  the predictions of any classifier},\ }in\ \href
  {https://doi.org/10.1145/2939672.2939778} {\emph {\bibinfo {booktitle}
  {Proceedings of the 22nd ACM SIGKDD International Conference on Knowledge
  Discovery and Data Mining}}},\ \bibinfo {series and number} {KDD '16}\
  (\bibinfo  {publisher} {Association for Computing Machinery},\ \bibinfo
  {address} {New York, NY, USA},\ \bibinfo {year} {2016})\ p.\ \bibinfo {pages}
  {1135–1144}\BibitemShut {NoStop}%
\bibitem [{\citenamefont {Barredo~Arrieta}\ \emph {et~al.}(2020)\citenamefont
  {Barredo~Arrieta}, \citenamefont {D\'{\i}az-Rodr\'{\i}guez}, \citenamefont
  {Del~Ser}, \citenamefont {Bennetot}, \citenamefont {Tabik}, \citenamefont
  {Barbado}, \citenamefont {Garcia}, \citenamefont {Gil-Lopez}, \citenamefont
  {Molina}, \citenamefont {Benjamins}, \citenamefont {Chatila},\ and\
  \citenamefont {Herrera}}]{Arrieta2020}%
  \BibitemOpen
  \bibfield  {author} {\bibinfo {author} {\bibfnamefont {A.}~\bibnamefont
  {Barredo~Arrieta}}, \bibinfo {author} {\bibfnamefont {N.}~\bibnamefont
  {D\'{\i}az-Rodr\'{\i}guez}}, \bibinfo {author} {\bibfnamefont
  {J.}~\bibnamefont {Del~Ser}}, \bibinfo {author} {\bibfnamefont
  {A.}~\bibnamefont {Bennetot}}, \bibinfo {author} {\bibfnamefont
  {S.}~\bibnamefont {Tabik}}, \bibinfo {author} {\bibfnamefont
  {A.}~\bibnamefont {Barbado}}, \bibinfo {author} {\bibfnamefont
  {S.}~\bibnamefont {Garcia}}, \bibinfo {author} {\bibfnamefont
  {S.}~\bibnamefont {Gil-Lopez}}, \bibinfo {author} {\bibfnamefont
  {D.}~\bibnamefont {Molina}}, \bibinfo {author} {\bibfnamefont
  {R.}~\bibnamefont {Benjamins}}, \bibinfo {author} {\bibfnamefont
  {R.}~\bibnamefont {Chatila}},\ and\ \bibinfo {author} {\bibfnamefont
  {F.}~\bibnamefont {Herrera}},\ }\bibfield  {title} {\bibinfo {title}
  {Explainable artificial intelligence (xai): Concepts, taxonomies,
  opportunities and challenges toward responsible ai},\ }\href
  {https://doi.org/10.1016/j.inffus.2019.12.012} {\bibfield  {journal}
  {\bibinfo  {journal} {Inf. Fusion}\ }\textbf {\bibinfo {volume} {58}},\
  \bibinfo {pages} {82–115} (\bibinfo {year} {2020})}\BibitemShut {NoStop}%
\bibitem [{Note1()}]{Note1}%
  \BibitemOpen
  \bibinfo {note} {We may generalize Eq. \ref {eq:ham} to non-Hermitian FNNs
  with $t_{rr'}\neq t_{r'r}^*$ \cite {nhermi2020, nhermi2021} and closer
  resemblance to classical ANNs with unidirectional information
  flow.}\BibitemShut {Stop}%
\bibitem [{\citenamefont {Hopfield}(1982)}]{Hopfield1982}%
  \BibitemOpen
  \bibfield  {author} {\bibinfo {author} {\bibfnamefont {J.~J.}\ \bibnamefont
  {Hopfield}},\ }\bibfield  {title} {\bibinfo {title} {Neural networks and
  physical systems with emergent collective computational abilities.},\ }\href
  {https://doi.org/10.1073/pnas.79.8.2554} {\bibfield  {journal} {\bibinfo
  {journal} {Proceedings of the National Academy of Sciences}\ }\textbf
  {\bibinfo {volume} {79}},\ \bibinfo {pages} {2554} (\bibinfo {year}
  {1982})}\BibitemShut {NoStop}%
\bibitem [{\citenamefont {Croy}\ \emph {et~al.}(2006)\citenamefont {Croy},
  \citenamefont {R{\"o}mer},\ and\ \citenamefont {Schreiber}}]{RGF2006}%
  \BibitemOpen
  \bibfield  {author} {\bibinfo {author} {\bibfnamefont {A.}~\bibnamefont
  {Croy}}, \bibinfo {author} {\bibfnamefont {R.~A.}\ \bibnamefont
  {R{\"o}mer}},\ and\ \bibinfo {author} {\bibfnamefont {M.}~\bibnamefont
  {Schreiber}},\ }\bibfield  {title} {\bibinfo {title} {Localization of
  electronic states in amorphous materials: Recursive green's function method
  and the metal-insulator transition at $e\neq0$},\ }in\ \href@noop {} {\emph
  {\bibinfo {booktitle} {Parallel Algorithms and Cluster Computing}}},\
  \bibinfo {editor} {edited by\ \bibinfo {editor} {\bibfnamefont {K.~H.}\
  \bibnamefont {Hoffmann}}\ and\ \bibinfo {editor} {\bibfnamefont
  {A.}~\bibnamefont {Meyer}}}\ (\bibinfo  {publisher} {Springer Berlin
  Heidelberg},\ \bibinfo {address} {Berlin, Heidelberg},\ \bibinfo {year}
  {2006})\ pp.\ \bibinfo {pages} {203--226}\BibitemShut {NoStop}%
\bibitem [{\citenamefont {Krogh}\ and\ \citenamefont {Hertz}(1991)}]{wd1991}%
  \BibitemOpen
  \bibfield  {author} {\bibinfo {author} {\bibfnamefont {A.}~\bibnamefont
  {Krogh}}\ and\ \bibinfo {author} {\bibfnamefont {J.~A.}\ \bibnamefont
  {Hertz}},\ }\bibfield  {title} {\bibinfo {title} {A simple weight decay can
  improve generalization},\ }in\ \href@noop {} {\emph {\bibinfo {booktitle}
  {Proceedings of the 4th International Conference on Neural Information
  Processing Systems}}},\ \bibinfo {series and number} {NIPS'91}\ (\bibinfo
  {publisher} {Morgan Kaufmann Publishers Inc.},\ \bibinfo {address} {San
  Francisco, CA, USA},\ \bibinfo {year} {1991})\ p.\ \bibinfo {pages}
  {950–957}\BibitemShut {NoStop}%
\bibitem [{Note2()}]{Note2}%
  \BibitemOpen
  \bibinfo {note} {The state-of-the-art 99.87\% accuracy is achieved by
  ensemble learning with homogeneous vector capsules and convolutional neural
  networks, please refer to \protect \url
  {https://paperswithcode.com/sota/image-classification-on-mnist} for details.
  Likewise, FNNs have much room for faster convergence and better accuracy,
  achievable with proper hyper-parameter settings and architecture expansions;
  see Appendix \ref {app:architecture}}\BibitemShut {NoStop}%
\bibitem [{Note3()}]{Note3}%
  \BibitemOpen
  \bibinfo {note} {See benchmark ANN accuracy at \protect \url
  {http://yann.lecun.com/exdb/mnist/}.}\BibitemShut {Stop}%
\bibitem [{\citenamefont {Kitaev}(2006)}]{Kitaev2006}%
  \BibitemOpen
  \bibfield  {author} {\bibinfo {author} {\bibfnamefont {A.}~\bibnamefont
  {Kitaev}},\ }\bibfield  {title} {\bibinfo {title} {Anyons in an exactly
  solved model and beyond},\ }\href
  {https://doi.org/http://dx.doi.org/10.1016/j.aop.2005.10.005} {\bibfield
  {journal} {\bibinfo  {journal} {Annals of Physics}\ }\textbf {\bibinfo
  {volume} {321}},\ \bibinfo {pages} {2 } (\bibinfo {year} {2006})}\BibitemShut
  {NoStop}%
\bibitem [{\citenamefont {Bradley}(1997)}]{AUC1997}%
  \BibitemOpen
  \bibfield  {author} {\bibinfo {author} {\bibfnamefont {A.~P.}\ \bibnamefont
  {Bradley}},\ }\bibfield  {title} {\bibinfo {title} {The use of the area under
  the roc curve in the evaluation of machine learning algorithms},\ }\href
  {https://doi.org/10.1016/S0031-3203(96)00142-2} {\bibfield  {journal}
  {\bibinfo  {journal} {Pattern Recogn.}\ }\textbf {\bibinfo {volume} {30}},\
  \bibinfo {pages} {1145–1159} (\bibinfo {year} {1997})}\BibitemShut
  {NoStop}%
\bibitem [{\citenamefont {Fawcett}(2006)}]{ROC2006}%
  \BibitemOpen
  \bibfield  {author} {\bibinfo {author} {\bibfnamefont {T.}~\bibnamefont
  {Fawcett}},\ }\bibfield  {title} {\bibinfo {title} {An introduction to roc
  analysis},\ }\href {https://doi.org/10.1016/j.patrec.2005.10.010} {\bibfield
  {journal} {\bibinfo  {journal} {Pattern Recogn. Lett.}\ }\textbf {\bibinfo
  {volume} {27}},\ \bibinfo {pages} {861–874} (\bibinfo {year}
  {2006})}\BibitemShut {NoStop}%
\bibitem [{\citenamefont {Hettler}\ \emph {et~al.}(1998)\citenamefont
  {Hettler}, \citenamefont {Tahvildar-Zadeh}, \citenamefont {Jarrell},
  \citenamefont {Pruschke},\ and\ \citenamefont {Krishnamurthy}}]{Hettler1998}%
  \BibitemOpen
  \bibfield  {author} {\bibinfo {author} {\bibfnamefont {M.~H.}\ \bibnamefont
  {Hettler}}, \bibinfo {author} {\bibfnamefont {A.~N.}\ \bibnamefont
  {Tahvildar-Zadeh}}, \bibinfo {author} {\bibfnamefont {M.}~\bibnamefont
  {Jarrell}}, \bibinfo {author} {\bibfnamefont {T.}~\bibnamefont {Pruschke}},\
  and\ \bibinfo {author} {\bibfnamefont {H.~R.}\ \bibnamefont
  {Krishnamurthy}},\ }\bibfield  {title} {\bibinfo {title} {Nonlocal dynamical
  correlations of strongly interacting electron systems},\ }\href
  {https://doi.org/10.1103/PhysRevB.58.R7475} {\bibfield  {journal} {\bibinfo
  {journal} {Phys. Rev. B}\ }\textbf {\bibinfo {volume} {58}},\ \bibinfo
  {pages} {R7475} (\bibinfo {year} {1998})}\BibitemShut {NoStop}%
\bibitem [{\citenamefont {Freericks}\ and\ \citenamefont
  {Zlati\ifmmode~\acute{c}\else \'{c}\fi{}}(2003)}]{FK2003}%
  \BibitemOpen
  \bibfield  {author} {\bibinfo {author} {\bibfnamefont {J.~K.}\ \bibnamefont
  {Freericks}}\ and\ \bibinfo {author} {\bibfnamefont {V.}~\bibnamefont
  {Zlati\ifmmode~\acute{c}\else \'{c}\fi{}}},\ }\bibfield  {title} {\bibinfo
  {title} {Exact dynamical mean-field theory of the falicov-kimball model},\
  }\href {https://doi.org/10.1103/RevModPhys.75.1333} {\bibfield  {journal}
  {\bibinfo  {journal} {Rev. Mod. Phys.}\ }\textbf {\bibinfo {volume} {75}},\
  \bibinfo {pages} {1333} (\bibinfo {year} {2003})}\BibitemShut {NoStop}%
\bibitem [{\citenamefont {Brandt}\ and\ \citenamefont
  {Mielsch}(1989)}]{FKBrandt1989}%
  \BibitemOpen
  \bibfield  {author} {\bibinfo {author} {\bibfnamefont {U.}~\bibnamefont
  {Brandt}}\ and\ \bibinfo {author} {\bibfnamefont {C.}~\bibnamefont
  {Mielsch}},\ }\bibfield  {title} {\bibinfo {title} {Thermodynamics and
  correlation functions of the falicov-kimball model in large dimensions},\
  }\href {https://doi.org/10.1007/BF01321824} {\bibfield  {journal} {\bibinfo
  {journal} {Zeitschrift f{\"u}r Physik B Condensed Matter}\ }\textbf {\bibinfo
  {volume} {75}},\ \bibinfo {pages} {365} (\bibinfo {year} {1989})}\BibitemShut
  {NoStop}%
\bibitem [{\citenamefont {Brandt}\ and\ \citenamefont
  {Mielsch}(1990)}]{FKBrandt1990}%
  \BibitemOpen
  \bibfield  {author} {\bibinfo {author} {\bibfnamefont {U.}~\bibnamefont
  {Brandt}}\ and\ \bibinfo {author} {\bibfnamefont {C.}~\bibnamefont
  {Mielsch}},\ }\bibfield  {title} {\bibinfo {title} {Thermodynamics of the
  falicov-kimball model in large dimensions ii},\ }\href
  {https://doi.org/10.1007/BF01406598} {\bibfield  {journal} {\bibinfo
  {journal} {Zeitschrift f{\"u}r Physik B Condensed Matter}\ }\textbf {\bibinfo
  {volume} {79}},\ \bibinfo {pages} {295} (\bibinfo {year} {1990})}\BibitemShut
  {NoStop}%
\bibitem [{\citenamefont {Maska}\ and\ \citenamefont
  {Czajka}(2006)}]{FKModelQMC2006}%
  \BibitemOpen
  \bibfield  {author} {\bibinfo {author} {\bibfnamefont {M.~M.}\ \bibnamefont
  {Maska}}\ and\ \bibinfo {author} {\bibfnamefont {K.}~\bibnamefont {Czajka}},\
  }\bibfield  {title} {\bibinfo {title} {Thermodynamics of the two-dimensional
  falicov-kimball model: A classical monte carlo study},\ }\href
  {https://doi.org/10.1103/PhysRevB.74.035109} {\bibfield  {journal} {\bibinfo
  {journal} {Phys. Rev. B}\ }\textbf {\bibinfo {volume} {74}},\ \bibinfo
  {pages} {035109} (\bibinfo {year} {2006})}\BibitemShut {NoStop}%
\bibitem [{\citenamefont {Trivedi}\ and\ \citenamefont
  {Randeria}(1995)}]{Nandini1995}%
  \BibitemOpen
  \bibfield  {author} {\bibinfo {author} {\bibfnamefont {N.}~\bibnamefont
  {Trivedi}}\ and\ \bibinfo {author} {\bibfnamefont {M.}~\bibnamefont
  {Randeria}},\ }\bibfield  {title} {\bibinfo {title} {Deviations from
  fermi-liquid behavior above ${T}_{c}$ in 2d short coherence length
  superconductors},\ }\href {https://doi.org/10.1103/PhysRevLett.75.312}
  {\bibfield  {journal} {\bibinfo  {journal} {Phys. Rev. Lett.}\ }\textbf
  {\bibinfo {volume} {75}},\ \bibinfo {pages} {312} (\bibinfo {year}
  {1995})}\BibitemShut {NoStop}%
\bibitem [{Note4()}]{Note4}%
  \BibitemOpen
  \bibinfo {note} {The number of nearest neighbors in the original 2D plane
  remains limited even after coupling to FNNs; therefore, we may generalize
  DMFT to local clusters, i.e., dynamical cluster approximation \cite
  {Hettler1998}, for better-controlled analysis.}\BibitemShut {Stop}%
\bibitem [{\citenamefont {Fukushima}(1980)}]{Fukushima1980}%
  \BibitemOpen
  \bibfield  {author} {\bibinfo {author} {\bibfnamefont {K.}~\bibnamefont
  {Fukushima}},\ }\bibfield  {title} {\bibinfo {title} {Neocognitron: A
  self-organizing neural network model for a mechanism of pattern recognition
  unaffected by shift in position},\ }\href
  {https://doi.org/10.1007/BF00344251} {\bibfield  {journal} {\bibinfo
  {journal} {Biological Cybernetics}\ }\textbf {\bibinfo {volume} {36}},\
  \bibinfo {pages} {193} (\bibinfo {year} {1980})}\BibitemShut {NoStop}%
\bibitem [{\citenamefont {McClean}\ \emph {et~al.}(2018)\citenamefont
  {McClean}, \citenamefont {Boixo}, \citenamefont {Smelyanskiy}, \citenamefont
  {Babbush},\ and\ \citenamefont {Neven}}]{McClean2018}%
  \BibitemOpen
  \bibfield  {author} {\bibinfo {author} {\bibfnamefont {J.~R.}\ \bibnamefont
  {McClean}}, \bibinfo {author} {\bibfnamefont {S.}~\bibnamefont {Boixo}},
  \bibinfo {author} {\bibfnamefont {V.~N.}\ \bibnamefont {Smelyanskiy}},
  \bibinfo {author} {\bibfnamefont {R.}~\bibnamefont {Babbush}},\ and\ \bibinfo
  {author} {\bibfnamefont {H.}~\bibnamefont {Neven}},\ }\bibfield  {title}
  {\bibinfo {title} {Barren plateaus in quantum neural network training
  landscapes},\ }\href {https://doi.org/10.1038/s41467-018-07090-4} {\bibfield
  {journal} {\bibinfo  {journal} {Nature Communications}\ }\textbf {\bibinfo
  {volume} {9}},\ \bibinfo {pages} {4812} (\bibinfo {year} {2018})}\BibitemShut
  {NoStop}%
\bibitem [{\citenamefont {Cerezo}\ \emph
  {et~al.}(2021{\natexlab{a}})\citenamefont {Cerezo}, \citenamefont {Sone},
  \citenamefont {Volkoff}, \citenamefont {Cincio},\ and\ \citenamefont
  {Coles}}]{Cerezo2021BP}%
  \BibitemOpen
  \bibfield  {author} {\bibinfo {author} {\bibfnamefont {M.}~\bibnamefont
  {Cerezo}}, \bibinfo {author} {\bibfnamefont {A.}~\bibnamefont {Sone}},
  \bibinfo {author} {\bibfnamefont {T.}~\bibnamefont {Volkoff}}, \bibinfo
  {author} {\bibfnamefont {L.}~\bibnamefont {Cincio}},\ and\ \bibinfo {author}
  {\bibfnamefont {P.~J.}\ \bibnamefont {Coles}},\ }\bibfield  {title} {\bibinfo
  {title} {Cost function dependent barren plateaus in shallow parametrized
  quantum circuits},\ }\href {https://doi.org/10.1038/s41467-021-21728-w}
  {\bibfield  {journal} {\bibinfo  {journal} {Nature Communications}\ }\textbf
  {\bibinfo {volume} {12}},\ \bibinfo {pages} {1791} (\bibinfo {year}
  {2021}{\natexlab{a}})}\BibitemShut {NoStop}%
\bibitem [{\citenamefont {Wang}\ \emph {et~al.}(2021)\citenamefont {Wang},
  \citenamefont {Fontana}, \citenamefont {Cerezo}, \citenamefont {Sharma},
  \citenamefont {Sone}, \citenamefont {Cincio},\ and\ \citenamefont
  {Coles}}]{Wang2021}%
  \BibitemOpen
  \bibfield  {author} {\bibinfo {author} {\bibfnamefont {S.}~\bibnamefont
  {Wang}}, \bibinfo {author} {\bibfnamefont {E.}~\bibnamefont {Fontana}},
  \bibinfo {author} {\bibfnamefont {M.}~\bibnamefont {Cerezo}}, \bibinfo
  {author} {\bibfnamefont {K.}~\bibnamefont {Sharma}}, \bibinfo {author}
  {\bibfnamefont {A.}~\bibnamefont {Sone}}, \bibinfo {author} {\bibfnamefont
  {L.}~\bibnamefont {Cincio}},\ and\ \bibinfo {author} {\bibfnamefont {P.~J.}\
  \bibnamefont {Coles}},\ }\bibfield  {title} {\bibinfo {title} {Noise-induced
  barren plateaus in variational quantum algorithms},\ }\href
  {https://doi.org/10.1038/s41467-021-27045-6} {\bibfield  {journal} {\bibinfo
  {journal} {Nature Communications}\ }\textbf {\bibinfo {volume} {12}},\
  \bibinfo {pages} {6961} (\bibinfo {year} {2021})}\BibitemShut {NoStop}%
\bibitem [{\citenamefont {Adami}\ and\ \citenamefont {Cerf}(1997)}]{Adami1997}%
  \BibitemOpen
  \bibfield  {author} {\bibinfo {author} {\bibfnamefont {C.}~\bibnamefont
  {Adami}}\ and\ \bibinfo {author} {\bibfnamefont {N.~J.}\ \bibnamefont
  {Cerf}},\ }\bibfield  {title} {\bibinfo {title} {von neumann capacity of
  noisy quantum channels},\ }\href {https://doi.org/10.1103/PhysRevA.56.3470}
  {\bibfield  {journal} {\bibinfo  {journal} {Phys. Rev. A}\ }\textbf {\bibinfo
  {volume} {56}},\ \bibinfo {pages} {3470} (\bibinfo {year}
  {1997})}\BibitemShut {NoStop}%
\bibitem [{\citenamefont {Groisman}\ \emph {et~al.}(2005)\citenamefont
  {Groisman}, \citenamefont {Popescu},\ and\ \citenamefont
  {Winter}}]{Groisman2005}%
  \BibitemOpen
  \bibfield  {author} {\bibinfo {author} {\bibfnamefont {B.}~\bibnamefont
  {Groisman}}, \bibinfo {author} {\bibfnamefont {S.}~\bibnamefont {Popescu}},\
  and\ \bibinfo {author} {\bibfnamefont {A.}~\bibnamefont {Winter}},\
  }\bibfield  {title} {\bibinfo {title} {Quantum, classical, and total amount
  of correlations in a quantum state},\ }\href
  {https://doi.org/10.1103/PhysRevA.72.032317} {\bibfield  {journal} {\bibinfo
  {journal} {Phys. Rev. A}\ }\textbf {\bibinfo {volume} {72}},\ \bibinfo
  {pages} {032317} (\bibinfo {year} {2005})}\BibitemShut {NoStop}%
\bibitem [{\citenamefont {Rombach}\ \emph {et~al.}(2022)\citenamefont
  {Rombach}, \citenamefont {Blattmann}, \citenamefont {Lorenz}, \citenamefont
  {Esser},\ and\ \citenamefont {Ommer}}]{StableDiffusion_2022}%
  \BibitemOpen
  \bibfield  {author} {\bibinfo {author} {\bibfnamefont {R.}~\bibnamefont
  {Rombach}}, \bibinfo {author} {\bibfnamefont {A.}~\bibnamefont {Blattmann}},
  \bibinfo {author} {\bibfnamefont {D.}~\bibnamefont {Lorenz}}, \bibinfo
  {author} {\bibfnamefont {P.}~\bibnamefont {Esser}},\ and\ \bibinfo {author}
  {\bibfnamefont {B.}~\bibnamefont {Ommer}},\ }\bibfield  {title} {\bibinfo
  {title} {High-resolution image synthesis with latent diffusion models},\ }in\
  \href@noop {} {\emph {\bibinfo {booktitle} {Proceedings of the IEEE/CVF
  Conference on Computer Vision and Pattern Recognition (CVPR)}}}\ (\bibinfo
  {year} {2022})\ pp.\ \bibinfo {pages} {10684--10695}\BibitemShut {NoStop}%
\bibitem [{\citenamefont {Harrow}\ and\ \citenamefont
  {Montanaro}(2017)}]{Harrow2017}%
  \BibitemOpen
  \bibfield  {author} {\bibinfo {author} {\bibfnamefont {A.~W.}\ \bibnamefont
  {Harrow}}\ and\ \bibinfo {author} {\bibfnamefont {A.}~\bibnamefont
  {Montanaro}},\ }\bibfield  {title} {\bibinfo {title} {Quantum computational
  supremacy},\ }\href {https://doi.org/10.1038/nature23458} {\bibfield
  {journal} {\bibinfo  {journal} {Nature}\ }\textbf {\bibinfo {volume} {549}},\
  \bibinfo {pages} {203} (\bibinfo {year} {2017})}\BibitemShut {NoStop}%
\bibitem [{\citenamefont {Shor}(1997)}]{Shor1997}%
  \BibitemOpen
  \bibfield  {author} {\bibinfo {author} {\bibfnamefont {P.~W.}\ \bibnamefont
  {Shor}},\ }\bibfield  {title} {\bibinfo {title} {Polynomial-time algorithms
  for prime factorization and discrete logarithms on a quantum computer},\
  }\href {https://doi.org/10.1137/S0097539795293172} {\bibfield  {journal}
  {\bibinfo  {journal} {SIAM J. Comput.}\ }\textbf {\bibinfo {volume} {26}},\
  \bibinfo {pages} {1484–1509} (\bibinfo {year} {1997})}\BibitemShut
  {NoStop}%
\bibitem [{\citenamefont {Terhal}\ and\ \citenamefont
  {DiVincenzo}(2004)}]{Terhal2004}%
  \BibitemOpen
  \bibfield  {author} {\bibinfo {author} {\bibfnamefont {B.~M.}\ \bibnamefont
  {Terhal}}\ and\ \bibinfo {author} {\bibfnamefont {D.~P.}\ \bibnamefont
  {DiVincenzo}},\ }\bibfield  {title} {\bibinfo {title} {Adptive quantum
  computation, constant depth quantum circuits and arthur-merlin games},\
  }\href@noop {} {\bibfield  {journal} {\bibinfo  {journal} {Quantum Info.
  Comput.}\ }\textbf {\bibinfo {volume} {4}},\ \bibinfo {pages} {134–145}
  (\bibinfo {year} {2004})}\BibitemShut {NoStop}%
\bibitem [{\citenamefont {Aaronson}\ and\ \citenamefont
  {Arkhipov}(2013)}]{Aaronson2013}%
  \BibitemOpen
  \bibfield  {author} {\bibinfo {author} {\bibfnamefont {S.}~\bibnamefont
  {Aaronson}}\ and\ \bibinfo {author} {\bibfnamefont {A.}~\bibnamefont
  {Arkhipov}},\ }\bibfield  {title} {\bibinfo {title} {The computational
  complexity of linear optics},\ }\href
  {https://doi.org/10.4086/toc.2013.v009a004} {\bibfield  {journal} {\bibinfo
  {journal} {Theory of Computing}\ }\textbf {\bibinfo {volume} {9}},\ \bibinfo
  {pages} {143} (\bibinfo {year} {2013})}\BibitemShut {NoStop}%
\bibitem [{\citenamefont {Aaronson}\ and\ \citenamefont
  {Chen}(2017)}]{Aaronson2017}%
  \BibitemOpen
  \bibfield  {author} {\bibinfo {author} {\bibfnamefont {S.}~\bibnamefont
  {Aaronson}}\ and\ \bibinfo {author} {\bibfnamefont {L.}~\bibnamefont
  {Chen}},\ }\bibfield  {title} {\bibinfo {title} {Complexity-theoretic
  foundations of quantum supremacy experiments},\ }in\ \href@noop {} {\emph
  {\bibinfo {booktitle} {Proceedings of the 32nd Computational Complexity
  Conference}}},\ \bibinfo {series and number} {CCC '17}\ (\bibinfo
  {publisher} {Schloss Dagstuhl--Leibniz-Zentrum fuer Informatik},\ \bibinfo
  {address} {Dagstuhl, DEU},\ \bibinfo {year} {2017})\BibitemShut {NoStop}%
\bibitem [{\citenamefont {Bravyi}\ \emph {et~al.}(2018)\citenamefont {Bravyi},
  \citenamefont {Gosset},\ and\ \citenamefont {Konig}}]{Bravyi2018}%
  \BibitemOpen
  \bibfield  {author} {\bibinfo {author} {\bibfnamefont {S.}~\bibnamefont
  {Bravyi}}, \bibinfo {author} {\bibfnamefont {D.}~\bibnamefont {Gosset}},\
  and\ \bibinfo {author} {\bibfnamefont {R.}~\bibnamefont {Konig}},\ }\bibfield
   {title} {\bibinfo {title} {Quantum advantage with shallow circuits},\ }\href
  {https://doi.org/10.1126/science.aar3106} {\bibfield  {journal} {\bibinfo
  {journal} {Science}\ }\textbf {\bibinfo {volume} {362}},\ \bibinfo {pages}
  {308} (\bibinfo {year} {2018})}\BibitemShut {NoStop}%
\bibitem [{\citenamefont {Liu}\ \emph {et~al.}(2021)\citenamefont {Liu},
  \citenamefont {Arunachalam},\ and\ \citenamefont {Temme}}]{Liu2021}%
  \BibitemOpen
  \bibfield  {author} {\bibinfo {author} {\bibfnamefont {Y.}~\bibnamefont
  {Liu}}, \bibinfo {author} {\bibfnamefont {S.}~\bibnamefont {Arunachalam}},\
  and\ \bibinfo {author} {\bibfnamefont {K.}~\bibnamefont {Temme}},\ }\bibfield
   {title} {\bibinfo {title} {A rigorous and robust quantum speed-up in
  supervised machine learning},\ }\href
  {https://doi.org/10.1038/s41567-021-01287-z} {\bibfield  {journal} {\bibinfo
  {journal} {Nature Physics}\ }\textbf {\bibinfo {volume} {17}},\ \bibinfo
  {pages} {1013} (\bibinfo {year} {2021})}\BibitemShut {NoStop}%
\bibitem [{\citenamefont {Huang}\ \emph {et~al.}(2021)\citenamefont {Huang},
  \citenamefont {Kueng},\ and\ \citenamefont {Preskill}}]{Huang2021QA}%
  \BibitemOpen
  \bibfield  {author} {\bibinfo {author} {\bibfnamefont {H.-Y.}\ \bibnamefont
  {Huang}}, \bibinfo {author} {\bibfnamefont {R.}~\bibnamefont {Kueng}},\ and\
  \bibinfo {author} {\bibfnamefont {J.}~\bibnamefont {Preskill}},\ }\bibfield
  {title} {\bibinfo {title} {Information-theoretic bounds on quantum advantage
  in machine learning},\ }\href
  {https://doi.org/10.1103/PhysRevLett.126.190505} {\bibfield  {journal}
  {\bibinfo  {journal} {Phys. Rev. Lett.}\ }\textbf {\bibinfo {volume} {126}},\
  \bibinfo {pages} {190505} (\bibinfo {year} {2021})}\BibitemShut {NoStop}%
\bibitem [{\citenamefont {Huang}\ \emph
  {et~al.}(2022{\natexlab{b}})\citenamefont {Huang}, \citenamefont {Broughton},
  \citenamefont {Cotler}, \citenamefont {Chen}, \citenamefont {Li},
  \citenamefont {Mohseni}, \citenamefont {Neven}, \citenamefont {Babbush},
  \citenamefont {Kueng}, \citenamefont {Preskill},\ and\ \citenamefont
  {McClean}}]{Huang2022QA}%
  \BibitemOpen
  \bibfield  {author} {\bibinfo {author} {\bibfnamefont {H.-Y.}\ \bibnamefont
  {Huang}}, \bibinfo {author} {\bibfnamefont {M.}~\bibnamefont {Broughton}},
  \bibinfo {author} {\bibfnamefont {J.}~\bibnamefont {Cotler}}, \bibinfo
  {author} {\bibfnamefont {S.}~\bibnamefont {Chen}}, \bibinfo {author}
  {\bibfnamefont {J.}~\bibnamefont {Li}}, \bibinfo {author} {\bibfnamefont
  {M.}~\bibnamefont {Mohseni}}, \bibinfo {author} {\bibfnamefont
  {H.}~\bibnamefont {Neven}}, \bibinfo {author} {\bibfnamefont
  {R.}~\bibnamefont {Babbush}}, \bibinfo {author} {\bibfnamefont
  {R.}~\bibnamefont {Kueng}}, \bibinfo {author} {\bibfnamefont
  {J.}~\bibnamefont {Preskill}},\ and\ \bibinfo {author} {\bibfnamefont
  {J.~R.}\ \bibnamefont {McClean}},\ }\bibfield  {title} {\bibinfo {title}
  {Quantum advantage in learning from experiments},\ }\href
  {https://doi.org/10.1126/science.abn7293} {\bibfield  {journal} {\bibinfo
  {journal} {Science}\ }\textbf {\bibinfo {volume} {376}},\ \bibinfo {pages}
  {1182} (\bibinfo {year} {2022}{\natexlab{b}})}\BibitemShut {NoStop}%
\bibitem [{\citenamefont {Arute}\ \emph {et~al.}(2019)\citenamefont {Arute},
  \citenamefont {Arya}, \citenamefont {Babbush}, \citenamefont {Bacon},
  \citenamefont {Bardin}, \citenamefont {Barends}, \citenamefont {Biswas},
  \citenamefont {Boixo}, \citenamefont {Brandao}, \citenamefont {Buell},
  \citenamefont {Burkett}, \citenamefont {Chen}, \citenamefont {Chen},
  \citenamefont {Chiaro}, \citenamefont {Collins}, \citenamefont {Courtney},
  \citenamefont {Dunsworth}, \citenamefont {Farhi}, \citenamefont {Foxen},
  \citenamefont {Fowler}, \citenamefont {Gidney}, \citenamefont {Giustina},
  \citenamefont {Graff}, \citenamefont {Guerin}, \citenamefont {Habegger},
  \citenamefont {Harrigan}, \citenamefont {Hartmann}, \citenamefont {Ho},
  \citenamefont {Hoffmann}, \citenamefont {Huang}, \citenamefont {Humble},
  \citenamefont {Isakov}, \citenamefont {Jeffrey}, \citenamefont {Jiang},
  \citenamefont {Kafri}, \citenamefont {Kechedzhi}, \citenamefont {Kelly},
  \citenamefont {Klimov}, \citenamefont {Knysh}, \citenamefont {Korotkov},
  \citenamefont {Kostritsa}, \citenamefont {Landhuis}, \citenamefont
  {Lindmark}, \citenamefont {Lucero}, \citenamefont {Lyakh}, \citenamefont
  {Mandr{\`a}}, \citenamefont {McClean}, \citenamefont {McEwen}, \citenamefont
  {Megrant}, \citenamefont {Mi}, \citenamefont {Michielsen}, \citenamefont
  {Mohseni}, \citenamefont {Mutus}, \citenamefont {Naaman}, \citenamefont
  {Neeley}, \citenamefont {Neill}, \citenamefont {Niu}, \citenamefont {Ostby},
  \citenamefont {Petukhov}, \citenamefont {Platt}, \citenamefont {Quintana},
  \citenamefont {Rieffel}, \citenamefont {Roushan}, \citenamefont {Rubin},
  \citenamefont {Sank}, \citenamefont {Satzinger}, \citenamefont {Smelyanskiy},
  \citenamefont {Sung}, \citenamefont {Trevithick}, \citenamefont
  {Vainsencher}, \citenamefont {Villalonga}, \citenamefont {White},
  \citenamefont {Yao}, \citenamefont {Yeh}, \citenamefont {Zalcman},
  \citenamefont {Neven},\ and\ \citenamefont {Martinis}}]{Arute2019}%
  \BibitemOpen
  \bibfield  {author} {\bibinfo {author} {\bibfnamefont {F.}~\bibnamefont
  {Arute}}, \bibinfo {author} {\bibfnamefont {K.}~\bibnamefont {Arya}},
  \bibinfo {author} {\bibfnamefont {R.}~\bibnamefont {Babbush}}, \bibinfo
  {author} {\bibfnamefont {D.}~\bibnamefont {Bacon}}, \bibinfo {author}
  {\bibfnamefont {J.~C.}\ \bibnamefont {Bardin}}, \bibinfo {author}
  {\bibfnamefont {R.}~\bibnamefont {Barends}}, \bibinfo {author} {\bibfnamefont
  {R.}~\bibnamefont {Biswas}}, \bibinfo {author} {\bibfnamefont
  {S.}~\bibnamefont {Boixo}}, \bibinfo {author} {\bibfnamefont {F.~G. S.~L.}\
  \bibnamefont {Brandao}}, \bibinfo {author} {\bibfnamefont {D.~A.}\
  \bibnamefont {Buell}}, \bibinfo {author} {\bibfnamefont {B.}~\bibnamefont
  {Burkett}}, \bibinfo {author} {\bibfnamefont {Y.}~\bibnamefont {Chen}},
  \bibinfo {author} {\bibfnamefont {Z.}~\bibnamefont {Chen}}, \bibinfo {author}
  {\bibfnamefont {B.}~\bibnamefont {Chiaro}}, \bibinfo {author} {\bibfnamefont
  {R.}~\bibnamefont {Collins}}, \bibinfo {author} {\bibfnamefont
  {W.}~\bibnamefont {Courtney}}, \bibinfo {author} {\bibfnamefont
  {A.}~\bibnamefont {Dunsworth}}, \bibinfo {author} {\bibfnamefont
  {E.}~\bibnamefont {Farhi}}, \bibinfo {author} {\bibfnamefont
  {B.}~\bibnamefont {Foxen}}, \bibinfo {author} {\bibfnamefont
  {A.}~\bibnamefont {Fowler}}, \bibinfo {author} {\bibfnamefont
  {C.}~\bibnamefont {Gidney}}, \bibinfo {author} {\bibfnamefont
  {M.}~\bibnamefont {Giustina}}, \bibinfo {author} {\bibfnamefont
  {R.}~\bibnamefont {Graff}}, \bibinfo {author} {\bibfnamefont
  {K.}~\bibnamefont {Guerin}}, \bibinfo {author} {\bibfnamefont
  {S.}~\bibnamefont {Habegger}}, \bibinfo {author} {\bibfnamefont {M.~P.}\
  \bibnamefont {Harrigan}}, \bibinfo {author} {\bibfnamefont {M.~J.}\
  \bibnamefont {Hartmann}}, \bibinfo {author} {\bibfnamefont {A.}~\bibnamefont
  {Ho}}, \bibinfo {author} {\bibfnamefont {M.}~\bibnamefont {Hoffmann}},
  \bibinfo {author} {\bibfnamefont {T.}~\bibnamefont {Huang}}, \bibinfo
  {author} {\bibfnamefont {T.~S.}\ \bibnamefont {Humble}}, \bibinfo {author}
  {\bibfnamefont {S.~V.}\ \bibnamefont {Isakov}}, \bibinfo {author}
  {\bibfnamefont {E.}~\bibnamefont {Jeffrey}}, \bibinfo {author} {\bibfnamefont
  {Z.}~\bibnamefont {Jiang}}, \bibinfo {author} {\bibfnamefont
  {D.}~\bibnamefont {Kafri}}, \bibinfo {author} {\bibfnamefont
  {K.}~\bibnamefont {Kechedzhi}}, \bibinfo {author} {\bibfnamefont
  {J.}~\bibnamefont {Kelly}}, \bibinfo {author} {\bibfnamefont {P.~V.}\
  \bibnamefont {Klimov}}, \bibinfo {author} {\bibfnamefont {S.}~\bibnamefont
  {Knysh}}, \bibinfo {author} {\bibfnamefont {A.}~\bibnamefont {Korotkov}},
  \bibinfo {author} {\bibfnamefont {F.}~\bibnamefont {Kostritsa}}, \bibinfo
  {author} {\bibfnamefont {D.}~\bibnamefont {Landhuis}}, \bibinfo {author}
  {\bibfnamefont {M.}~\bibnamefont {Lindmark}}, \bibinfo {author}
  {\bibfnamefont {E.}~\bibnamefont {Lucero}}, \bibinfo {author} {\bibfnamefont
  {D.}~\bibnamefont {Lyakh}}, \bibinfo {author} {\bibfnamefont
  {S.}~\bibnamefont {Mandr{\`a}}}, \bibinfo {author} {\bibfnamefont {J.~R.}\
  \bibnamefont {McClean}}, \bibinfo {author} {\bibfnamefont {M.}~\bibnamefont
  {McEwen}}, \bibinfo {author} {\bibfnamefont {A.}~\bibnamefont {Megrant}},
  \bibinfo {author} {\bibfnamefont {X.}~\bibnamefont {Mi}}, \bibinfo {author}
  {\bibfnamefont {K.}~\bibnamefont {Michielsen}}, \bibinfo {author}
  {\bibfnamefont {M.}~\bibnamefont {Mohseni}}, \bibinfo {author} {\bibfnamefont
  {J.}~\bibnamefont {Mutus}}, \bibinfo {author} {\bibfnamefont
  {O.}~\bibnamefont {Naaman}}, \bibinfo {author} {\bibfnamefont
  {M.}~\bibnamefont {Neeley}}, \bibinfo {author} {\bibfnamefont
  {C.}~\bibnamefont {Neill}}, \bibinfo {author} {\bibfnamefont {M.~Y.}\
  \bibnamefont {Niu}}, \bibinfo {author} {\bibfnamefont {E.}~\bibnamefont
  {Ostby}}, \bibinfo {author} {\bibfnamefont {A.}~\bibnamefont {Petukhov}},
  \bibinfo {author} {\bibfnamefont {J.~C.}\ \bibnamefont {Platt}}, \bibinfo
  {author} {\bibfnamefont {C.}~\bibnamefont {Quintana}}, \bibinfo {author}
  {\bibfnamefont {E.~G.}\ \bibnamefont {Rieffel}}, \bibinfo {author}
  {\bibfnamefont {P.}~\bibnamefont {Roushan}}, \bibinfo {author} {\bibfnamefont
  {N.~C.}\ \bibnamefont {Rubin}}, \bibinfo {author} {\bibfnamefont
  {D.}~\bibnamefont {Sank}}, \bibinfo {author} {\bibfnamefont {K.~J.}\
  \bibnamefont {Satzinger}}, \bibinfo {author} {\bibfnamefont {V.}~\bibnamefont
  {Smelyanskiy}}, \bibinfo {author} {\bibfnamefont {K.~J.}\ \bibnamefont
  {Sung}}, \bibinfo {author} {\bibfnamefont {M.~D.}\ \bibnamefont
  {Trevithick}}, \bibinfo {author} {\bibfnamefont {A.}~\bibnamefont
  {Vainsencher}}, \bibinfo {author} {\bibfnamefont {B.}~\bibnamefont
  {Villalonga}}, \bibinfo {author} {\bibfnamefont {T.}~\bibnamefont {White}},
  \bibinfo {author} {\bibfnamefont {Z.~J.}\ \bibnamefont {Yao}}, \bibinfo
  {author} {\bibfnamefont {P.}~\bibnamefont {Yeh}}, \bibinfo {author}
  {\bibfnamefont {A.}~\bibnamefont {Zalcman}}, \bibinfo {author} {\bibfnamefont
  {H.}~\bibnamefont {Neven}},\ and\ \bibinfo {author} {\bibfnamefont {J.~M.}\
  \bibnamefont {Martinis}},\ }\bibfield  {title} {\bibinfo {title} {Quantum
  supremacy using a programmable superconducting processor},\ }\href
  {https://doi.org/10.1038/s41586-019-1666-5} {\bibfield  {journal} {\bibinfo
  {journal} {Nature}\ }\textbf {\bibinfo {volume} {574}},\ \bibinfo {pages}
  {505} (\bibinfo {year} {2019})}\BibitemShut {NoStop}%
\bibitem [{\citenamefont {Zhong}\ \emph {et~al.}(2020)\citenamefont {Zhong},
  \citenamefont {Wang}, \citenamefont {Deng}, \citenamefont {Chen},
  \citenamefont {Peng}, \citenamefont {Luo}, \citenamefont {Qin}, \citenamefont
  {Wu}, \citenamefont {Ding}, \citenamefont {Hu}, \citenamefont {Hu},
  \citenamefont {Yang}, \citenamefont {Zhang}, \citenamefont {Li},
  \citenamefont {Li}, \citenamefont {Jiang}, \citenamefont {Gan}, \citenamefont
  {Yang}, \citenamefont {You}, \citenamefont {Wang}, \citenamefont {Li},
  \citenamefont {Liu}, \citenamefont {Lu},\ and\ \citenamefont
  {Pan}}]{Zhong2020}%
  \BibitemOpen
  \bibfield  {author} {\bibinfo {author} {\bibfnamefont {H.-S.}\ \bibnamefont
  {Zhong}}, \bibinfo {author} {\bibfnamefont {H.}~\bibnamefont {Wang}},
  \bibinfo {author} {\bibfnamefont {Y.-H.}\ \bibnamefont {Deng}}, \bibinfo
  {author} {\bibfnamefont {M.-C.}\ \bibnamefont {Chen}}, \bibinfo {author}
  {\bibfnamefont {L.-C.}\ \bibnamefont {Peng}}, \bibinfo {author}
  {\bibfnamefont {Y.-H.}\ \bibnamefont {Luo}}, \bibinfo {author} {\bibfnamefont
  {J.}~\bibnamefont {Qin}}, \bibinfo {author} {\bibfnamefont {D.}~\bibnamefont
  {Wu}}, \bibinfo {author} {\bibfnamefont {X.}~\bibnamefont {Ding}}, \bibinfo
  {author} {\bibfnamefont {Y.}~\bibnamefont {Hu}}, \bibinfo {author}
  {\bibfnamefont {P.}~\bibnamefont {Hu}}, \bibinfo {author} {\bibfnamefont
  {X.-Y.}\ \bibnamefont {Yang}}, \bibinfo {author} {\bibfnamefont {W.-J.}\
  \bibnamefont {Zhang}}, \bibinfo {author} {\bibfnamefont {H.}~\bibnamefont
  {Li}}, \bibinfo {author} {\bibfnamefont {Y.}~\bibnamefont {Li}}, \bibinfo
  {author} {\bibfnamefont {X.}~\bibnamefont {Jiang}}, \bibinfo {author}
  {\bibfnamefont {L.}~\bibnamefont {Gan}}, \bibinfo {author} {\bibfnamefont
  {G.}~\bibnamefont {Yang}}, \bibinfo {author} {\bibfnamefont {L.}~\bibnamefont
  {You}}, \bibinfo {author} {\bibfnamefont {Z.}~\bibnamefont {Wang}}, \bibinfo
  {author} {\bibfnamefont {L.}~\bibnamefont {Li}}, \bibinfo {author}
  {\bibfnamefont {N.-L.}\ \bibnamefont {Liu}}, \bibinfo {author} {\bibfnamefont
  {C.-Y.}\ \bibnamefont {Lu}},\ and\ \bibinfo {author} {\bibfnamefont {J.-W.}\
  \bibnamefont {Pan}},\ }\bibfield  {title} {\bibinfo {title} {Quantum
  computational advantage using photons},\ }\href
  {https://doi.org/10.1126/science.abe8770} {\bibfield  {journal} {\bibinfo
  {journal} {Science}\ }\textbf {\bibinfo {volume} {370}},\ \bibinfo {pages}
  {1460} (\bibinfo {year} {2020})}\BibitemShut {NoStop}%
\bibitem [{\citenamefont {Wu}\ \emph {et~al.}(2021)\citenamefont {Wu},
  \citenamefont {Bao}, \citenamefont {Cao}, \citenamefont {Chen}, \citenamefont
  {Chen}, \citenamefont {Chen}, \citenamefont {Chung}, \citenamefont {Deng},
  \citenamefont {Du}, \citenamefont {Fan}, \citenamefont {Gong}, \citenamefont
  {Guo}, \citenamefont {Guo}, \citenamefont {Guo}, \citenamefont {Han},
  \citenamefont {Hong}, \citenamefont {Huang}, \citenamefont {Huo},
  \citenamefont {Li}, \citenamefont {Li}, \citenamefont {Li}, \citenamefont
  {Li}, \citenamefont {Liang}, \citenamefont {Lin}, \citenamefont {Lin},
  \citenamefont {Qian}, \citenamefont {Qiao}, \citenamefont {Rong},
  \citenamefont {Su}, \citenamefont {Sun}, \citenamefont {Wang}, \citenamefont
  {Wang}, \citenamefont {Wu}, \citenamefont {Xu}, \citenamefont {Yan},
  \citenamefont {Yang}, \citenamefont {Yang}, \citenamefont {Ye}, \citenamefont
  {Yin}, \citenamefont {Ying}, \citenamefont {Yu}, \citenamefont {Zha},
  \citenamefont {Zhang}, \citenamefont {Zhang}, \citenamefont {Zhang},
  \citenamefont {Zhang}, \citenamefont {Zhao}, \citenamefont {Zhao},
  \citenamefont {Zhou}, \citenamefont {Zhu}, \citenamefont {Lu}, \citenamefont
  {Peng}, \citenamefont {Zhu},\ and\ \citenamefont {Pan}}]{Wu2021}%
  \BibitemOpen
  \bibfield  {author} {\bibinfo {author} {\bibfnamefont {Y.}~\bibnamefont
  {Wu}}, \bibinfo {author} {\bibfnamefont {W.-S.}\ \bibnamefont {Bao}},
  \bibinfo {author} {\bibfnamefont {S.}~\bibnamefont {Cao}}, \bibinfo {author}
  {\bibfnamefont {F.}~\bibnamefont {Chen}}, \bibinfo {author} {\bibfnamefont
  {M.-C.}\ \bibnamefont {Chen}}, \bibinfo {author} {\bibfnamefont
  {X.}~\bibnamefont {Chen}}, \bibinfo {author} {\bibfnamefont {T.-H.}\
  \bibnamefont {Chung}}, \bibinfo {author} {\bibfnamefont {H.}~\bibnamefont
  {Deng}}, \bibinfo {author} {\bibfnamefont {Y.}~\bibnamefont {Du}}, \bibinfo
  {author} {\bibfnamefont {D.}~\bibnamefont {Fan}}, \bibinfo {author}
  {\bibfnamefont {M.}~\bibnamefont {Gong}}, \bibinfo {author} {\bibfnamefont
  {C.}~\bibnamefont {Guo}}, \bibinfo {author} {\bibfnamefont {C.}~\bibnamefont
  {Guo}}, \bibinfo {author} {\bibfnamefont {S.}~\bibnamefont {Guo}}, \bibinfo
  {author} {\bibfnamefont {L.}~\bibnamefont {Han}}, \bibinfo {author}
  {\bibfnamefont {L.}~\bibnamefont {Hong}}, \bibinfo {author} {\bibfnamefont
  {H.-L.}\ \bibnamefont {Huang}}, \bibinfo {author} {\bibfnamefont {Y.-H.}\
  \bibnamefont {Huo}}, \bibinfo {author} {\bibfnamefont {L.}~\bibnamefont
  {Li}}, \bibinfo {author} {\bibfnamefont {N.}~\bibnamefont {Li}}, \bibinfo
  {author} {\bibfnamefont {S.}~\bibnamefont {Li}}, \bibinfo {author}
  {\bibfnamefont {Y.}~\bibnamefont {Li}}, \bibinfo {author} {\bibfnamefont
  {F.}~\bibnamefont {Liang}}, \bibinfo {author} {\bibfnamefont
  {C.}~\bibnamefont {Lin}}, \bibinfo {author} {\bibfnamefont {J.}~\bibnamefont
  {Lin}}, \bibinfo {author} {\bibfnamefont {H.}~\bibnamefont {Qian}}, \bibinfo
  {author} {\bibfnamefont {D.}~\bibnamefont {Qiao}}, \bibinfo {author}
  {\bibfnamefont {H.}~\bibnamefont {Rong}}, \bibinfo {author} {\bibfnamefont
  {H.}~\bibnamefont {Su}}, \bibinfo {author} {\bibfnamefont {L.}~\bibnamefont
  {Sun}}, \bibinfo {author} {\bibfnamefont {L.}~\bibnamefont {Wang}}, \bibinfo
  {author} {\bibfnamefont {S.}~\bibnamefont {Wang}}, \bibinfo {author}
  {\bibfnamefont {D.}~\bibnamefont {Wu}}, \bibinfo {author} {\bibfnamefont
  {Y.}~\bibnamefont {Xu}}, \bibinfo {author} {\bibfnamefont {K.}~\bibnamefont
  {Yan}}, \bibinfo {author} {\bibfnamefont {W.}~\bibnamefont {Yang}}, \bibinfo
  {author} {\bibfnamefont {Y.}~\bibnamefont {Yang}}, \bibinfo {author}
  {\bibfnamefont {Y.}~\bibnamefont {Ye}}, \bibinfo {author} {\bibfnamefont
  {J.}~\bibnamefont {Yin}}, \bibinfo {author} {\bibfnamefont {C.}~\bibnamefont
  {Ying}}, \bibinfo {author} {\bibfnamefont {J.}~\bibnamefont {Yu}}, \bibinfo
  {author} {\bibfnamefont {C.}~\bibnamefont {Zha}}, \bibinfo {author}
  {\bibfnamefont {C.}~\bibnamefont {Zhang}}, \bibinfo {author} {\bibfnamefont
  {H.}~\bibnamefont {Zhang}}, \bibinfo {author} {\bibfnamefont
  {K.}~\bibnamefont {Zhang}}, \bibinfo {author} {\bibfnamefont
  {Y.}~\bibnamefont {Zhang}}, \bibinfo {author} {\bibfnamefont
  {H.}~\bibnamefont {Zhao}}, \bibinfo {author} {\bibfnamefont {Y.}~\bibnamefont
  {Zhao}}, \bibinfo {author} {\bibfnamefont {L.}~\bibnamefont {Zhou}}, \bibinfo
  {author} {\bibfnamefont {Q.}~\bibnamefont {Zhu}}, \bibinfo {author}
  {\bibfnamefont {C.-Y.}\ \bibnamefont {Lu}}, \bibinfo {author} {\bibfnamefont
  {C.-Z.}\ \bibnamefont {Peng}}, \bibinfo {author} {\bibfnamefont
  {X.}~\bibnamefont {Zhu}},\ and\ \bibinfo {author} {\bibfnamefont {J.-W.}\
  \bibnamefont {Pan}},\ }\bibfield  {title} {\bibinfo {title} {Strong quantum
  computational advantage using a superconducting quantum processor},\ }\href
  {https://doi.org/10.1103/PhysRevLett.127.180501} {\bibfield  {journal}
  {\bibinfo  {journal} {Phys. Rev. Lett.}\ }\textbf {\bibinfo {volume} {127}},\
  \bibinfo {pages} {180501} (\bibinfo {year} {2021})}\BibitemShut {NoStop}%
\bibitem [{\citenamefont {Tersoff}\ and\ \citenamefont
  {Hamann}(1985)}]{Tersoff1985}%
  \BibitemOpen
  \bibfield  {author} {\bibinfo {author} {\bibfnamefont {J.}~\bibnamefont
  {Tersoff}}\ and\ \bibinfo {author} {\bibfnamefont {D.~R.}\ \bibnamefont
  {Hamann}},\ }\bibfield  {title} {\bibinfo {title} {Theory of the scanning
  tunneling microscope},\ }\href {https://doi.org/10.1103/PhysRevB.31.805}
  {\bibfield  {journal} {\bibinfo  {journal} {Phys. Rev. B}\ }\textbf {\bibinfo
  {volume} {31}},\ \bibinfo {pages} {805} (\bibinfo {year} {1985})}\BibitemShut
  {NoStop}%
\bibitem [{\citenamefont {Crommie}\ \emph {et~al.}(1993)\citenamefont
  {Crommie}, \citenamefont {Lutz},\ and\ \citenamefont {Eigler}}]{Crommie1993}%
  \BibitemOpen
  \bibfield  {author} {\bibinfo {author} {\bibfnamefont {M.~F.}\ \bibnamefont
  {Crommie}}, \bibinfo {author} {\bibfnamefont {C.~P.}\ \bibnamefont {Lutz}},\
  and\ \bibinfo {author} {\bibfnamefont {D.~M.}\ \bibnamefont {Eigler}},\
  }\bibfield  {title} {\bibinfo {title} {Confinement of electrons to quantum
  corrals on a metal surface},\ }\href
  {https://doi.org/10.1126/science.262.5131.218} {\bibfield  {journal}
  {\bibinfo  {journal} {Science}\ }\textbf {\bibinfo {volume} {262}},\ \bibinfo
  {pages} {218} (\bibinfo {year} {1993})}\BibitemShut {NoStop}%
\bibitem [{\citenamefont {van Wees}\ \emph {et~al.}(1988)\citenamefont {van
  Wees}, \citenamefont {van Houten}, \citenamefont {Beenakker}, \citenamefont
  {Williamson}, \citenamefont {Kouwenhoven}, \citenamefont {van~der Marel},\
  and\ \citenamefont {Foxon}}]{Wees1988}%
  \BibitemOpen
  \bibfield  {author} {\bibinfo {author} {\bibfnamefont {B.~J.}\ \bibnamefont
  {van Wees}}, \bibinfo {author} {\bibfnamefont {H.}~\bibnamefont {van
  Houten}}, \bibinfo {author} {\bibfnamefont {C.~W.~J.}\ \bibnamefont
  {Beenakker}}, \bibinfo {author} {\bibfnamefont {J.~G.}\ \bibnamefont
  {Williamson}}, \bibinfo {author} {\bibfnamefont {L.~P.}\ \bibnamefont
  {Kouwenhoven}}, \bibinfo {author} {\bibfnamefont {D.}~\bibnamefont {van~der
  Marel}},\ and\ \bibinfo {author} {\bibfnamefont {C.~T.}\ \bibnamefont
  {Foxon}},\ }\bibfield  {title} {\bibinfo {title} {Quantized conductance of
  point contacts in a two-dimensional electron gas},\ }\href
  {https://doi.org/10.1103/PhysRevLett.60.848} {\bibfield  {journal} {\bibinfo
  {journal} {Phys. Rev. Lett.}\ }\textbf {\bibinfo {volume} {60}},\ \bibinfo
  {pages} {848} (\bibinfo {year} {1988})}\BibitemShut {NoStop}%
\bibitem [{\citenamefont {Biamonte}\ \emph {et~al.}(2017)\citenamefont
  {Biamonte}, \citenamefont {Wittek}, \citenamefont {Pancotti}, \citenamefont
  {Rebentrost}, \citenamefont {Wiebe},\ and\ \citenamefont
  {Lloyd}}]{biamonte2017}%
  \BibitemOpen
  \bibfield  {author} {\bibinfo {author} {\bibfnamefont {J.}~\bibnamefont
  {Biamonte}}, \bibinfo {author} {\bibfnamefont {P.}~\bibnamefont {Wittek}},
  \bibinfo {author} {\bibfnamefont {N.}~\bibnamefont {Pancotti}}, \bibinfo
  {author} {\bibfnamefont {P.}~\bibnamefont {Rebentrost}}, \bibinfo {author}
  {\bibfnamefont {N.}~\bibnamefont {Wiebe}},\ and\ \bibinfo {author}
  {\bibfnamefont {S.}~\bibnamefont {Lloyd}},\ }\bibfield  {title} {\bibinfo
  {title} {Quantum machine learning},\ }\href@noop {} {\bibfield  {journal}
  {\bibinfo  {journal} {Nature}\ }\textbf {\bibinfo {volume} {549}},\ \bibinfo
  {pages} {195} (\bibinfo {year} {2017})}\BibitemShut {NoStop}%
\bibitem [{\citenamefont {Cerezo}\ \emph
  {et~al.}(2021{\natexlab{b}})\citenamefont {Cerezo}, \citenamefont
  {Arrasmith}, \citenamefont {Babbush}, \citenamefont {Benjamin}, \citenamefont
  {Endo}, \citenamefont {Fujii}, \citenamefont {McClean}, \citenamefont
  {Mitarai}, \citenamefont {Yuan}, \citenamefont {Cincio},\ and\ \citenamefont
  {Coles}}]{Cerezo2021}%
  \BibitemOpen
  \bibfield  {author} {\bibinfo {author} {\bibfnamefont {M.}~\bibnamefont
  {Cerezo}}, \bibinfo {author} {\bibfnamefont {A.}~\bibnamefont {Arrasmith}},
  \bibinfo {author} {\bibfnamefont {R.}~\bibnamefont {Babbush}}, \bibinfo
  {author} {\bibfnamefont {S.~C.}\ \bibnamefont {Benjamin}}, \bibinfo {author}
  {\bibfnamefont {S.}~\bibnamefont {Endo}}, \bibinfo {author} {\bibfnamefont
  {K.}~\bibnamefont {Fujii}}, \bibinfo {author} {\bibfnamefont {J.~R.}\
  \bibnamefont {McClean}}, \bibinfo {author} {\bibfnamefont {K.}~\bibnamefont
  {Mitarai}}, \bibinfo {author} {\bibfnamefont {X.}~\bibnamefont {Yuan}},
  \bibinfo {author} {\bibfnamefont {L.}~\bibnamefont {Cincio}},\ and\ \bibinfo
  {author} {\bibfnamefont {P.~J.}\ \bibnamefont {Coles}},\ }\bibfield  {title}
  {\bibinfo {title} {Variational quantum algorithms},\ }\href
  {https://doi.org/10.1038/s42254-021-00348-9} {\bibfield  {journal} {\bibinfo
  {journal} {Nature Reviews Physics}\ }\textbf {\bibinfo {volume} {3}},\
  \bibinfo {pages} {625} (\bibinfo {year} {2021}{\natexlab{b}})}\BibitemShut
  {NoStop}%
\bibitem [{\citenamefont {Mitarai}\ \emph {et~al.}(2018)\citenamefont
  {Mitarai}, \citenamefont {Negoro}, \citenamefont {Kitagawa},\ and\
  \citenamefont {Fujii}}]{ParaShift2018}%
  \BibitemOpen
  \bibfield  {author} {\bibinfo {author} {\bibfnamefont {K.}~\bibnamefont
  {Mitarai}}, \bibinfo {author} {\bibfnamefont {M.}~\bibnamefont {Negoro}},
  \bibinfo {author} {\bibfnamefont {M.}~\bibnamefont {Kitagawa}},\ and\
  \bibinfo {author} {\bibfnamefont {K.}~\bibnamefont {Fujii}},\ }\bibfield
  {title} {\bibinfo {title} {Quantum circuit learning},\ }\href
  {https://doi.org/10.1103/PhysRevA.98.032309} {\bibfield  {journal} {\bibinfo
  {journal} {Phys. Rev. A}\ }\textbf {\bibinfo {volume} {98}},\ \bibinfo
  {pages} {032309} (\bibinfo {year} {2018})}\BibitemShut {NoStop}%
\bibitem [{\citenamefont {Schuld}\ \emph {et~al.}(2019)\citenamefont {Schuld},
  \citenamefont {Bergholm}, \citenamefont {Gogolin}, \citenamefont {Izaac},\
  and\ \citenamefont {Killoran}}]{ParaShift2019}%
  \BibitemOpen
  \bibfield  {author} {\bibinfo {author} {\bibfnamefont {M.}~\bibnamefont
  {Schuld}}, \bibinfo {author} {\bibfnamefont {V.}~\bibnamefont {Bergholm}},
  \bibinfo {author} {\bibfnamefont {C.}~\bibnamefont {Gogolin}}, \bibinfo
  {author} {\bibfnamefont {J.}~\bibnamefont {Izaac}},\ and\ \bibinfo {author}
  {\bibfnamefont {N.}~\bibnamefont {Killoran}},\ }\bibfield  {title} {\bibinfo
  {title} {Evaluating analytic gradients on quantum hardware},\ }\href
  {https://doi.org/10.1103/PhysRevA.99.032331} {\bibfield  {journal} {\bibinfo
  {journal} {Phys. Rev. A}\ }\textbf {\bibinfo {volume} {99}},\ \bibinfo
  {pages} {032331} (\bibinfo {year} {2019})}\BibitemShut {NoStop}%
\bibitem [{\citenamefont {Cong}\ \emph {et~al.}(2019)\citenamefont {Cong},
  \citenamefont {Choi},\ and\ \citenamefont {Lukin}}]{Cong2019}%
  \BibitemOpen
  \bibfield  {author} {\bibinfo {author} {\bibfnamefont {I.}~\bibnamefont
  {Cong}}, \bibinfo {author} {\bibfnamefont {S.}~\bibnamefont {Choi}},\ and\
  \bibinfo {author} {\bibfnamefont {M.~D.}\ \bibnamefont {Lukin}},\ }\bibfield
  {title} {\bibinfo {title} {Quantum convolutional neural networks},\ }\href
  {https://doi.org/10.1038/s41567-019-0648-8} {\bibfield  {journal} {\bibinfo
  {journal} {Nature Physics}\ }\textbf {\bibinfo {volume} {15}},\ \bibinfo
  {pages} {1273} (\bibinfo {year} {2019})}\BibitemShut {NoStop}%
\bibitem [{\citenamefont {Bittel}\ and\ \citenamefont
  {Kliesch}(2021)}]{Bittel2021}%
  \BibitemOpen
  \bibfield  {author} {\bibinfo {author} {\bibfnamefont {L.}~\bibnamefont
  {Bittel}}\ and\ \bibinfo {author} {\bibfnamefont {M.}~\bibnamefont
  {Kliesch}},\ }\bibfield  {title} {\bibinfo {title} {Training variational
  quantum algorithms is np-hard},\ }\href
  {https://doi.org/10.1103/PhysRevLett.127.120502} {\bibfield  {journal}
  {\bibinfo  {journal} {Phys. Rev. Lett.}\ }\textbf {\bibinfo {volume} {127}},\
  \bibinfo {pages} {120502} (\bibinfo {year} {2021})}\BibitemShut {NoStop}%
\bibitem [{\citenamefont {Abbas}\ \emph {et~al.}(2021)\citenamefont {Abbas},
  \citenamefont {Sutter}, \citenamefont {Zoufal}, \citenamefont {Lucchi},
  \citenamefont {Figalli},\ and\ \citenamefont {Woerner}}]{Abbas2021}%
  \BibitemOpen
  \bibfield  {author} {\bibinfo {author} {\bibfnamefont {A.}~\bibnamefont
  {Abbas}}, \bibinfo {author} {\bibfnamefont {D.}~\bibnamefont {Sutter}},
  \bibinfo {author} {\bibfnamefont {C.}~\bibnamefont {Zoufal}}, \bibinfo
  {author} {\bibfnamefont {A.}~\bibnamefont {Lucchi}}, \bibinfo {author}
  {\bibfnamefont {A.}~\bibnamefont {Figalli}},\ and\ \bibinfo {author}
  {\bibfnamefont {S.}~\bibnamefont {Woerner}},\ }\bibfield  {title} {\bibinfo
  {title} {The power of quantum neural networks},\ }\href
  {https://doi.org/10.1038/s43588-021-00084-1} {\bibfield  {journal} {\bibinfo
  {journal} {Nature Computational Science}\ }\textbf {\bibinfo {volume} {1}},\
  \bibinfo {pages} {403} (\bibinfo {year} {2021})}\BibitemShut {NoStop}%
\bibitem [{\citenamefont {Cerezo}\ \emph {et~al.}(2022)\citenamefont {Cerezo},
  \citenamefont {Verdon}, \citenamefont {Huang}, \citenamefont {Cincio},\ and\
  \citenamefont {Coles}}]{Cerezo2022}%
  \BibitemOpen
  \bibfield  {author} {\bibinfo {author} {\bibfnamefont {M.}~\bibnamefont
  {Cerezo}}, \bibinfo {author} {\bibfnamefont {G.}~\bibnamefont {Verdon}},
  \bibinfo {author} {\bibfnamefont {H.-Y.}\ \bibnamefont {Huang}}, \bibinfo
  {author} {\bibfnamefont {L.}~\bibnamefont {Cincio}},\ and\ \bibinfo {author}
  {\bibfnamefont {P.~J.}\ \bibnamefont {Coles}},\ }\bibfield  {title} {\bibinfo
  {title} {Challenges and opportunities in quantum machine learning},\ }\href
  {https://doi.org/10.1038/s43588-022-00311-3} {\bibfield  {journal} {\bibinfo
  {journal} {Nature Computational Science}\ }\textbf {\bibinfo {volume} {2}},\
  \bibinfo {pages} {567} (\bibinfo {year} {2022})}\BibitemShut {NoStop}%
\bibitem [{\citenamefont {Wang}\ \emph {et~al.}(2023)\citenamefont {Wang},
  \citenamefont {Zheng}, \citenamefont {Wu},\ and\ \citenamefont
  {Zhang}}]{Wang2023}%
  \BibitemOpen
  \bibfield  {author} {\bibinfo {author} {\bibfnamefont {Z.}~\bibnamefont
  {Wang}}, \bibinfo {author} {\bibfnamefont {P.-L.}\ \bibnamefont {Zheng}},
  \bibinfo {author} {\bibfnamefont {B.}~\bibnamefont {Wu}},\ and\ \bibinfo
  {author} {\bibfnamefont {Y.}~\bibnamefont {Zhang}},\ }\bibfield  {title}
  {\bibinfo {title} {Quantum dropout: On and over the hardness of quantum
  approximate optimization algorithm},\ }\href
  {https://doi.org/10.1103/PhysRevResearch.5.023171} {\bibfield  {journal}
  {\bibinfo  {journal} {Phys. Rev. Res.}\ }\textbf {\bibinfo {volume} {5}},\
  \bibinfo {pages} {023171} (\bibinfo {year} {2023})}\BibitemShut {NoStop}%
\bibitem [{\citenamefont {Buluta}\ and\ \citenamefont
  {Nori}(2009)}]{Buluta2009}%
  \BibitemOpen
  \bibfield  {author} {\bibinfo {author} {\bibfnamefont {I.}~\bibnamefont
  {Buluta}}\ and\ \bibinfo {author} {\bibfnamefont {F.}~\bibnamefont {Nori}},\
  }\bibfield  {title} {\bibinfo {title} {Quantum simulators},\ }\href
  {https://doi.org/10.1126/science.1177838} {\bibfield  {journal} {\bibinfo
  {journal} {Science}\ }\textbf {\bibinfo {volume} {326}},\ \bibinfo {pages}
  {108} (\bibinfo {year} {2009})}\BibitemShut {NoStop}%
\bibitem [{\citenamefont {Georgescu}\ \emph {et~al.}(2014)\citenamefont
  {Georgescu}, \citenamefont {Ashhab},\ and\ \citenamefont
  {Nori}}]{Georgescu2014}%
  \BibitemOpen
  \bibfield  {author} {\bibinfo {author} {\bibfnamefont {I.~M.}\ \bibnamefont
  {Georgescu}}, \bibinfo {author} {\bibfnamefont {S.}~\bibnamefont {Ashhab}},\
  and\ \bibinfo {author} {\bibfnamefont {F.}~\bibnamefont {Nori}},\ }\bibfield
  {title} {\bibinfo {title} {Quantum simulation},\ }\href
  {https://doi.org/10.1103/RevModPhys.86.153} {\bibfield  {journal} {\bibinfo
  {journal} {Rev. Mod. Phys.}\ }\textbf {\bibinfo {volume} {86}},\ \bibinfo
  {pages} {153} (\bibinfo {year} {2014})}\BibitemShut {NoStop}%
\bibitem [{\citenamefont {Barthelemy}\ and\ \citenamefont
  {Vandersypen}(2013)}]{Barthelemy2013}%
  \BibitemOpen
  \bibfield  {author} {\bibinfo {author} {\bibfnamefont {P.}~\bibnamefont
  {Barthelemy}}\ and\ \bibinfo {author} {\bibfnamefont {L.~M.~K.}\ \bibnamefont
  {Vandersypen}},\ }\bibfield  {title} {\bibinfo {title} {Quantum dot systems:
  a versatile platform for quantum simulations},\ }\href
  {https://doi.org/10.1002/andp.201300124} {\bibfield  {journal} {\bibinfo
  {journal} {Annalen der Physik}\ }\textbf {\bibinfo {volume} {525}},\ \bibinfo
  {pages} {808} (\bibinfo {year} {2013})}\BibitemShut {NoStop}%
\bibitem [{\citenamefont {Browaeys}\ and\ \citenamefont
  {Lahaye}(2020)}]{Browaeys2020}%
  \BibitemOpen
  \bibfield  {author} {\bibinfo {author} {\bibfnamefont {A.}~\bibnamefont
  {Browaeys}}\ and\ \bibinfo {author} {\bibfnamefont {T.}~\bibnamefont
  {Lahaye}},\ }\bibfield  {title} {\bibinfo {title} {Many-body physics with
  individually controlled rydberg atoms},\ }\href
  {https://doi.org/10.1038/s41567-019-0733-z} {\bibfield  {journal} {\bibinfo
  {journal} {Nature Physics}\ }\textbf {\bibinfo {volume} {16}},\ \bibinfo
  {pages} {132} (\bibinfo {year} {2020})}\BibitemShut {NoStop}%
\bibitem [{\citenamefont {Scholl}\ \emph {et~al.}(2021)\citenamefont {Scholl},
  \citenamefont {Schuler}, \citenamefont {Williams}, \citenamefont
  {Eberharter}, \citenamefont {Barredo}, \citenamefont {Schymik}, \citenamefont
  {Lienhard}, \citenamefont {Henry}, \citenamefont {Lang}, \citenamefont
  {Lahaye}, \citenamefont {L{\"a}uchli},\ and\ \citenamefont
  {Browaeys}}]{Scholl2021}%
  \BibitemOpen
  \bibfield  {author} {\bibinfo {author} {\bibfnamefont {P.}~\bibnamefont
  {Scholl}}, \bibinfo {author} {\bibfnamefont {M.}~\bibnamefont {Schuler}},
  \bibinfo {author} {\bibfnamefont {H.~J.}\ \bibnamefont {Williams}}, \bibinfo
  {author} {\bibfnamefont {A.~A.}\ \bibnamefont {Eberharter}}, \bibinfo
  {author} {\bibfnamefont {D.}~\bibnamefont {Barredo}}, \bibinfo {author}
  {\bibfnamefont {K.-N.}\ \bibnamefont {Schymik}}, \bibinfo {author}
  {\bibfnamefont {V.}~\bibnamefont {Lienhard}}, \bibinfo {author}
  {\bibfnamefont {L.-P.}\ \bibnamefont {Henry}}, \bibinfo {author}
  {\bibfnamefont {T.~C.}\ \bibnamefont {Lang}}, \bibinfo {author}
  {\bibfnamefont {T.}~\bibnamefont {Lahaye}}, \bibinfo {author} {\bibfnamefont
  {A.~M.}\ \bibnamefont {L{\"a}uchli}},\ and\ \bibinfo {author} {\bibfnamefont
  {A.}~\bibnamefont {Browaeys}},\ }\bibfield  {title} {\bibinfo {title}
  {Quantum simulation of 2d antiferromagnets with hundreds of rydberg atoms},\
  }\href {https://doi.org/10.1038/s41586-021-03585-1} {\bibfield  {journal}
  {\bibinfo  {journal} {Nature}\ }\textbf {\bibinfo {volume} {595}},\ \bibinfo
  {pages} {233} (\bibinfo {year} {2021})}\BibitemShut {NoStop}%
\bibitem [{\citenamefont {Ebadi}\ \emph {et~al.}(2022)\citenamefont {Ebadi},
  \citenamefont {Keesling}, \citenamefont {Cain}, \citenamefont {Wang},
  \citenamefont {Levine}, \citenamefont {Bluvstein}, \citenamefont {Semeghini},
  \citenamefont {Omran}, \citenamefont {Liu}, \citenamefont {Samajdar},
  \citenamefont {Luo}, \citenamefont {Nash}, \citenamefont {Gao}, \citenamefont
  {Barak}, \citenamefont {Farhi}, \citenamefont {Sachdev}, \citenamefont
  {Gemelke}, \citenamefont {Zhou}, \citenamefont {Choi}, \citenamefont
  {Pichler}, \citenamefont {Wang}, \citenamefont {Greiner}, \citenamefont
  {Vuleti{\'{c}}},\ and\ \citenamefont {Lukin}}]{Ebadi2022}%
  \BibitemOpen
  \bibfield  {author} {\bibinfo {author} {\bibfnamefont {S.}~\bibnamefont
  {Ebadi}}, \bibinfo {author} {\bibfnamefont {A.}~\bibnamefont {Keesling}},
  \bibinfo {author} {\bibfnamefont {M.}~\bibnamefont {Cain}}, \bibinfo {author}
  {\bibfnamefont {T.~T.}\ \bibnamefont {Wang}}, \bibinfo {author}
  {\bibfnamefont {H.}~\bibnamefont {Levine}}, \bibinfo {author} {\bibfnamefont
  {D.}~\bibnamefont {Bluvstein}}, \bibinfo {author} {\bibfnamefont
  {G.}~\bibnamefont {Semeghini}}, \bibinfo {author} {\bibfnamefont
  {A.}~\bibnamefont {Omran}}, \bibinfo {author} {\bibfnamefont {J.-G.}\
  \bibnamefont {Liu}}, \bibinfo {author} {\bibfnamefont {R.}~\bibnamefont
  {Samajdar}}, \bibinfo {author} {\bibfnamefont {X.-Z.}\ \bibnamefont {Luo}},
  \bibinfo {author} {\bibfnamefont {B.}~\bibnamefont {Nash}}, \bibinfo {author}
  {\bibfnamefont {X.}~\bibnamefont {Gao}}, \bibinfo {author} {\bibfnamefont
  {B.}~\bibnamefont {Barak}}, \bibinfo {author} {\bibfnamefont
  {E.}~\bibnamefont {Farhi}}, \bibinfo {author} {\bibfnamefont
  {S.}~\bibnamefont {Sachdev}}, \bibinfo {author} {\bibfnamefont
  {N.}~\bibnamefont {Gemelke}}, \bibinfo {author} {\bibfnamefont
  {L.}~\bibnamefont {Zhou}}, \bibinfo {author} {\bibfnamefont {S.}~\bibnamefont
  {Choi}}, \bibinfo {author} {\bibfnamefont {H.}~\bibnamefont {Pichler}},
  \bibinfo {author} {\bibfnamefont {S.-T.}\ \bibnamefont {Wang}}, \bibinfo
  {author} {\bibfnamefont {M.}~\bibnamefont {Greiner}}, \bibinfo {author}
  {\bibfnamefont {V.}~\bibnamefont {Vuleti{\'{c}}}},\ and\ \bibinfo {author}
  {\bibfnamefont {M.~D.}\ \bibnamefont {Lukin}},\ }\bibfield  {title} {\bibinfo
  {title} {Quantum optimization of maximum independent set using rydberg atom
  arrays},\ }\href {https://doi.org/10.1126/science.abo6587} {\bibfield
  {journal} {\bibinfo  {journal} {Science}\ }\textbf {\bibinfo {volume}
  {376}},\ \bibinfo {pages} {1209} (\bibinfo {year} {2022})}\BibitemShut
  {NoStop}%
\bibitem [{\citenamefont {Bluvstein}\ \emph {et~al.}(2021)\citenamefont
  {Bluvstein}, \citenamefont {Omran}, \citenamefont {Levine}, \citenamefont
  {Keesling}, \citenamefont {Semeghini}, \citenamefont {Ebadi}, \citenamefont
  {Wang}, \citenamefont {Michailidis}, \citenamefont {Maskara}, \citenamefont
  {Ho}, \citenamefont {Choi}, \citenamefont {Serbyn}, \citenamefont {Greiner},
  \citenamefont {Vuleti{\'{c}}},\ and\ \citenamefont {Lukin}}]{Bluvstein2021}%
  \BibitemOpen
  \bibfield  {author} {\bibinfo {author} {\bibfnamefont {D.}~\bibnamefont
  {Bluvstein}}, \bibinfo {author} {\bibfnamefont {A.}~\bibnamefont {Omran}},
  \bibinfo {author} {\bibfnamefont {H.}~\bibnamefont {Levine}}, \bibinfo
  {author} {\bibfnamefont {A.}~\bibnamefont {Keesling}}, \bibinfo {author}
  {\bibfnamefont {G.}~\bibnamefont {Semeghini}}, \bibinfo {author}
  {\bibfnamefont {S.}~\bibnamefont {Ebadi}}, \bibinfo {author} {\bibfnamefont
  {T.~T.}\ \bibnamefont {Wang}}, \bibinfo {author} {\bibfnamefont {A.~A.}\
  \bibnamefont {Michailidis}}, \bibinfo {author} {\bibfnamefont
  {N.}~\bibnamefont {Maskara}}, \bibinfo {author} {\bibfnamefont {W.~W.}\
  \bibnamefont {Ho}}, \bibinfo {author} {\bibfnamefont {S.}~\bibnamefont
  {Choi}}, \bibinfo {author} {\bibfnamefont {M.}~\bibnamefont {Serbyn}},
  \bibinfo {author} {\bibfnamefont {M.}~\bibnamefont {Greiner}}, \bibinfo
  {author} {\bibfnamefont {V.}~\bibnamefont {Vuleti{\'{c}}}},\ and\ \bibinfo
  {author} {\bibfnamefont {M.~D.}\ \bibnamefont {Lukin}},\ }\bibfield  {title}
  {\bibinfo {title} {Controlling quantum many-body dynamics in driven rydberg
  atom arrays},\ }\href {https://doi.org/10.1126/science.abg2530} {\bibfield
  {journal} {\bibinfo  {journal} {Science}\ }\textbf {\bibinfo {volume}
  {371}},\ \bibinfo {pages} {1355} (\bibinfo {year} {2021})}\BibitemShut
  {NoStop}%
\bibitem [{\citenamefont {Hanson}\ \emph {et~al.}(2007)\citenamefont {Hanson},
  \citenamefont {Kouwenhoven}, \citenamefont {Petta}, \citenamefont {Tarucha},\
  and\ \citenamefont {Vandersypen}}]{Hanson2007}%
  \BibitemOpen
  \bibfield  {author} {\bibinfo {author} {\bibfnamefont {R.}~\bibnamefont
  {Hanson}}, \bibinfo {author} {\bibfnamefont {L.~P.}\ \bibnamefont
  {Kouwenhoven}}, \bibinfo {author} {\bibfnamefont {J.~R.}\ \bibnamefont
  {Petta}}, \bibinfo {author} {\bibfnamefont {S.}~\bibnamefont {Tarucha}},\
  and\ \bibinfo {author} {\bibfnamefont {L.~M.~K.}\ \bibnamefont
  {Vandersypen}},\ }\bibfield  {title} {\bibinfo {title} {Spins in few-electron
  quantum dots},\ }\href {https://doi.org/10.1103/RevModPhys.79.1217}
  {\bibfield  {journal} {\bibinfo  {journal} {Rev. Mod. Phys.}\ }\textbf
  {\bibinfo {volume} {79}},\ \bibinfo {pages} {1217} (\bibinfo {year}
  {2007})}\BibitemShut {NoStop}%
\bibitem [{\citenamefont {Byrnes}\ \emph {et~al.}(2008)\citenamefont {Byrnes},
  \citenamefont {Kim}, \citenamefont {Kusudo},\ and\ \citenamefont
  {Yamamoto}}]{Byrnes2008}%
  \BibitemOpen
  \bibfield  {author} {\bibinfo {author} {\bibfnamefont {T.}~\bibnamefont
  {Byrnes}}, \bibinfo {author} {\bibfnamefont {N.~Y.}\ \bibnamefont {Kim}},
  \bibinfo {author} {\bibfnamefont {K.}~\bibnamefont {Kusudo}},\ and\ \bibinfo
  {author} {\bibfnamefont {Y.}~\bibnamefont {Yamamoto}},\ }\bibfield  {title}
  {\bibinfo {title} {Quantum simulation of fermi-hubbard models in
  semiconductor quantum-dot arrays},\ }\href
  {https://doi.org/10.1103/PhysRevB.78.075320} {\bibfield  {journal} {\bibinfo
  {journal} {Phys. Rev. B}\ }\textbf {\bibinfo {volume} {78}},\ \bibinfo
  {pages} {075320} (\bibinfo {year} {2008})}\BibitemShut {NoStop}%
\bibitem [{\citenamefont {Braakman}\ \emph {et~al.}(2013)\citenamefont
  {Braakman}, \citenamefont {Barthelemy}, \citenamefont {Reichl}, \citenamefont
  {Wegscheider},\ and\ \citenamefont {Vandersypen}}]{Braakman2013}%
  \BibitemOpen
  \bibfield  {author} {\bibinfo {author} {\bibfnamefont {F.~R.}\ \bibnamefont
  {Braakman}}, \bibinfo {author} {\bibfnamefont {P.}~\bibnamefont
  {Barthelemy}}, \bibinfo {author} {\bibfnamefont {C.}~\bibnamefont {Reichl}},
  \bibinfo {author} {\bibfnamefont {W.}~\bibnamefont {Wegscheider}},\ and\
  \bibinfo {author} {\bibfnamefont {L.~M.~K.}\ \bibnamefont {Vandersypen}},\
  }\bibfield  {title} {\bibinfo {title} {Long-distance coherent coupling in a
  quantum dot array},\ }\href {https://doi.org/10.1038/nnano.2013.67}
  {\bibfield  {journal} {\bibinfo  {journal} {Nature Nanotechnology}\ }\textbf
  {\bibinfo {volume} {8}},\ \bibinfo {pages} {432} (\bibinfo {year}
  {2013})}\BibitemShut {NoStop}%
\bibitem [{\citenamefont {Hensgens}\ \emph {et~al.}(2017)\citenamefont
  {Hensgens}, \citenamefont {Fujita}, \citenamefont {Janssen}, \citenamefont
  {Li}, \citenamefont {Van~Diepen}, \citenamefont {Reichl}, \citenamefont
  {Wegscheider}, \citenamefont {Das~Sarma},\ and\ \citenamefont
  {Vandersypen}}]{Hensgens2017}%
  \BibitemOpen
  \bibfield  {author} {\bibinfo {author} {\bibfnamefont {T.}~\bibnamefont
  {Hensgens}}, \bibinfo {author} {\bibfnamefont {T.}~\bibnamefont {Fujita}},
  \bibinfo {author} {\bibfnamefont {L.}~\bibnamefont {Janssen}}, \bibinfo
  {author} {\bibfnamefont {X.}~\bibnamefont {Li}}, \bibinfo {author}
  {\bibfnamefont {C.~J.}\ \bibnamefont {Van~Diepen}}, \bibinfo {author}
  {\bibfnamefont {C.}~\bibnamefont {Reichl}}, \bibinfo {author} {\bibfnamefont
  {W.}~\bibnamefont {Wegscheider}}, \bibinfo {author} {\bibfnamefont
  {S.}~\bibnamefont {Das~Sarma}},\ and\ \bibinfo {author} {\bibfnamefont
  {L.~M.~K.}\ \bibnamefont {Vandersypen}},\ }\bibfield  {title} {\bibinfo
  {title} {Quantum simulation of a fermi--hubbard model using a semiconductor
  quantum dot array},\ }\href {https://doi.org/10.1038/nature23022} {\bibfield
  {journal} {\bibinfo  {journal} {Nature}\ }\textbf {\bibinfo {volume} {548}},\
  \bibinfo {pages} {70} (\bibinfo {year} {2017})}\BibitemShut {NoStop}%
\bibitem [{\citenamefont {van Diepen}\ \emph {et~al.}(2021)\citenamefont {van
  Diepen}, \citenamefont {Hsiao}, \citenamefont {Mukhopadhyay}, \citenamefont
  {Reichl}, \citenamefont {Wegscheider},\ and\ \citenamefont
  {Vandersypen}}]{Diepen2021}%
  \BibitemOpen
  \bibfield  {author} {\bibinfo {author} {\bibfnamefont {C.~J.}\ \bibnamefont
  {van Diepen}}, \bibinfo {author} {\bibfnamefont {T.-K.}\ \bibnamefont
  {Hsiao}}, \bibinfo {author} {\bibfnamefont {U.}~\bibnamefont {Mukhopadhyay}},
  \bibinfo {author} {\bibfnamefont {C.}~\bibnamefont {Reichl}}, \bibinfo
  {author} {\bibfnamefont {W.}~\bibnamefont {Wegscheider}},\ and\ \bibinfo
  {author} {\bibfnamefont {L.~M.~K.}\ \bibnamefont {Vandersypen}},\ }\bibfield
  {title} {\bibinfo {title} {Quantum simulation of antiferromagnetic heisenberg
  chain with gate-defined quantum dots},\ }\href
  {https://doi.org/10.1103/PhysRevX.11.041025} {\bibfield  {journal} {\bibinfo
  {journal} {Phys. Rev. X}\ }\textbf {\bibinfo {volume} {11}},\ \bibinfo
  {pages} {041025} (\bibinfo {year} {2021})}\BibitemShut {NoStop}%
\bibitem [{\citenamefont {Saffman}\ \emph {et~al.}(2010)\citenamefont
  {Saffman}, \citenamefont {Walker},\ and\ \citenamefont
  {M\o{}lmer}}]{Saffman2010}%
  \BibitemOpen
  \bibfield  {author} {\bibinfo {author} {\bibfnamefont {M.}~\bibnamefont
  {Saffman}}, \bibinfo {author} {\bibfnamefont {T.~G.}\ \bibnamefont
  {Walker}},\ and\ \bibinfo {author} {\bibfnamefont {K.}~\bibnamefont
  {M\o{}lmer}},\ }\bibfield  {title} {\bibinfo {title} {Quantum information
  with rydberg atoms},\ }\href {https://doi.org/10.1103/RevModPhys.82.2313}
  {\bibfield  {journal} {\bibinfo  {journal} {Rev. Mod. Phys.}\ }\textbf
  {\bibinfo {volume} {82}},\ \bibinfo {pages} {2313} (\bibinfo {year}
  {2010})}\BibitemShut {NoStop}%
\bibitem [{\citenamefont {Weimer}\ \emph {et~al.}(2010)\citenamefont {Weimer},
  \citenamefont {M{\"u}ller}, \citenamefont {Lesanovsky}, \citenamefont
  {Zoller},\ and\ \citenamefont {B{\"u}chler}}]{Weimer2010}%
  \BibitemOpen
  \bibfield  {author} {\bibinfo {author} {\bibfnamefont {H.}~\bibnamefont
  {Weimer}}, \bibinfo {author} {\bibfnamefont {M.}~\bibnamefont {M{\"u}ller}},
  \bibinfo {author} {\bibfnamefont {I.}~\bibnamefont {Lesanovsky}}, \bibinfo
  {author} {\bibfnamefont {P.}~\bibnamefont {Zoller}},\ and\ \bibinfo {author}
  {\bibfnamefont {H.~P.}\ \bibnamefont {B{\"u}chler}},\ }\bibfield  {title}
  {\bibinfo {title} {A rydberg quantum simulator},\ }\href
  {https://doi.org/10.1038/nphys1614} {\bibfield  {journal} {\bibinfo
  {journal} {Nature Physics}\ }\textbf {\bibinfo {volume} {6}},\ \bibinfo
  {pages} {382} (\bibinfo {year} {2010})}\BibitemShut {NoStop}%
\bibitem [{\citenamefont {Omran}\ \emph {et~al.}(2019)\citenamefont {Omran},
  \citenamefont {Levine}, \citenamefont {Keesling}, \citenamefont {Semeghini},
  \citenamefont {Wang}, \citenamefont {Ebadi}, \citenamefont {Bernien},
  \citenamefont {Zibrov}, \citenamefont {Pichler}, \citenamefont {Choi},
  \citenamefont {Cui}, \citenamefont {Rossignolo}, \citenamefont {Rembold},
  \citenamefont {Montangero}, \citenamefont {Calarco}, \citenamefont {Endres},
  \citenamefont {Greiner}, \citenamefont {Vuleti{\'{c}}},\ and\ \citenamefont
  {Lukin}}]{Omran2019}%
  \BibitemOpen
  \bibfield  {author} {\bibinfo {author} {\bibfnamefont {A.}~\bibnamefont
  {Omran}}, \bibinfo {author} {\bibfnamefont {H.}~\bibnamefont {Levine}},
  \bibinfo {author} {\bibfnamefont {A.}~\bibnamefont {Keesling}}, \bibinfo
  {author} {\bibfnamefont {G.}~\bibnamefont {Semeghini}}, \bibinfo {author}
  {\bibfnamefont {T.~T.}\ \bibnamefont {Wang}}, \bibinfo {author}
  {\bibfnamefont {S.}~\bibnamefont {Ebadi}}, \bibinfo {author} {\bibfnamefont
  {H.}~\bibnamefont {Bernien}}, \bibinfo {author} {\bibfnamefont {A.~S.}\
  \bibnamefont {Zibrov}}, \bibinfo {author} {\bibfnamefont {H.}~\bibnamefont
  {Pichler}}, \bibinfo {author} {\bibfnamefont {S.}~\bibnamefont {Choi}},
  \bibinfo {author} {\bibfnamefont {J.}~\bibnamefont {Cui}}, \bibinfo {author}
  {\bibfnamefont {M.}~\bibnamefont {Rossignolo}}, \bibinfo {author}
  {\bibfnamefont {P.}~\bibnamefont {Rembold}}, \bibinfo {author} {\bibfnamefont
  {S.}~\bibnamefont {Montangero}}, \bibinfo {author} {\bibfnamefont
  {T.}~\bibnamefont {Calarco}}, \bibinfo {author} {\bibfnamefont
  {M.}~\bibnamefont {Endres}}, \bibinfo {author} {\bibfnamefont
  {M.}~\bibnamefont {Greiner}}, \bibinfo {author} {\bibfnamefont
  {V.}~\bibnamefont {Vuleti{\'{c}}}},\ and\ \bibinfo {author} {\bibfnamefont
  {M.~D.}\ \bibnamefont {Lukin}},\ }\bibfield  {title} {\bibinfo {title}
  {Generation and manipulation of schr{\"o}dinger cat states in rydberg atom
  arrays},\ }\href {https://doi.org/10.1126/science.aax9743} {\bibfield
  {journal} {\bibinfo  {journal} {Science}\ }\textbf {\bibinfo {volume}
  {365}},\ \bibinfo {pages} {570} (\bibinfo {year} {2019})}\BibitemShut
  {NoStop}%
\bibitem [{\citenamefont {Moreno-Cardoner}\ \emph {et~al.}(2021)\citenamefont
  {Moreno-Cardoner}, \citenamefont {Goncalves},\ and\ \citenamefont
  {Chang}}]{Moreno2021}%
  \BibitemOpen
  \bibfield  {author} {\bibinfo {author} {\bibfnamefont {M.}~\bibnamefont
  {Moreno-Cardoner}}, \bibinfo {author} {\bibfnamefont {D.}~\bibnamefont
  {Goncalves}},\ and\ \bibinfo {author} {\bibfnamefont {D.~E.}\ \bibnamefont
  {Chang}},\ }\bibfield  {title} {\bibinfo {title} {Quantum nonlinear optics
  based on two-dimensional rydberg atom arrays},\ }\href
  {https://doi.org/10.1103/PhysRevLett.127.263602} {\bibfield  {journal}
  {\bibinfo  {journal} {Phys. Rev. Lett.}\ }\textbf {\bibinfo {volume} {127}},\
  \bibinfo {pages} {263602} (\bibinfo {year} {2021})}\BibitemShut {NoStop}%
\bibitem [{\citenamefont {Gonz{\'a}lez-Cuadra}\ \emph
  {et~al.}(2023)\citenamefont {Gonz{\'a}lez-Cuadra}, \citenamefont {Bluvstein},
  \citenamefont {Kalinowski}, \citenamefont {Kaubruegger}, \citenamefont
  {Maskara}, \citenamefont {Naldesi}, \citenamefont {Zache}, \citenamefont
  {Kaufman}, \citenamefont {Lukin}, \citenamefont {Pichler}, \citenamefont
  {Vermersch}, \citenamefont {Ye},\ and\ \citenamefont {Zoller}}]{Cuadra2023}%
  \BibitemOpen
  \bibfield  {author} {\bibinfo {author} {\bibfnamefont {D.}~\bibnamefont
  {Gonz{\'a}lez-Cuadra}}, \bibinfo {author} {\bibfnamefont {D.}~\bibnamefont
  {Bluvstein}}, \bibinfo {author} {\bibfnamefont {M.}~\bibnamefont
  {Kalinowski}}, \bibinfo {author} {\bibfnamefont {R.}~\bibnamefont
  {Kaubruegger}}, \bibinfo {author} {\bibfnamefont {N.}~\bibnamefont
  {Maskara}}, \bibinfo {author} {\bibfnamefont {P.}~\bibnamefont {Naldesi}},
  \bibinfo {author} {\bibfnamefont {T.~V.}\ \bibnamefont {Zache}}, \bibinfo
  {author} {\bibfnamefont {A.~M.}\ \bibnamefont {Kaufman}}, \bibinfo {author}
  {\bibfnamefont {M.~D.}\ \bibnamefont {Lukin}}, \bibinfo {author}
  {\bibfnamefont {H.}~\bibnamefont {Pichler}}, \bibinfo {author} {\bibfnamefont
  {B.}~\bibnamefont {Vermersch}}, \bibinfo {author} {\bibfnamefont
  {J.}~\bibnamefont {Ye}},\ and\ \bibinfo {author} {\bibfnamefont
  {P.}~\bibnamefont {Zoller}},\ }\bibfield  {title} {\bibinfo {title}
  {Fermionic quantum processing with programmable neutral atom arrays},\ }\href
  {https://doi.org/10.1073/pnas.2304294120} {\bibfield  {journal} {\bibinfo
  {journal} {Proceedings of the National Academy of Sciences}\ }\textbf
  {\bibinfo {volume} {120}},\ \bibinfo {pages} {e2304294120} (\bibinfo {year}
  {2023})}\BibitemShut {NoStop}%
\bibitem [{Note5()}]{Note5}%
  \BibitemOpen
  \bibinfo {note} {\protect \url
  {https://github.com/PeilinZHENG/FQNN}}\BibitemShut {NoStop}%
\bibitem [{\citenamefont {Srivastava}\ \emph {et~al.}(2014)\citenamefont
  {Srivastava}, \citenamefont {Hinton}, \citenamefont {Krizhevsky},
  \citenamefont {Sutskever},\ and\ \citenamefont
  {Salakhutdinov}}]{dropout2014}%
  \BibitemOpen
  \bibfield  {author} {\bibinfo {author} {\bibfnamefont {N.}~\bibnamefont
  {Srivastava}}, \bibinfo {author} {\bibfnamefont {G.}~\bibnamefont {Hinton}},
  \bibinfo {author} {\bibfnamefont {A.}~\bibnamefont {Krizhevsky}}, \bibinfo
  {author} {\bibfnamefont {I.}~\bibnamefont {Sutskever}},\ and\ \bibinfo
  {author} {\bibfnamefont {R.}~\bibnamefont {Salakhutdinov}},\ }\bibfield
  {title} {\bibinfo {title} {Dropout: A simple way to prevent neural networks
  from overfitting},\ }\href@noop {} {\bibfield  {journal} {\bibinfo  {journal}
  {J. Mach. Learn. Res.}\ }\textbf {\bibinfo {volume} {15}},\ \bibinfo {pages}
  {1929–1958} (\bibinfo {year} {2014})}\BibitemShut {NoStop}%
\bibitem [{\citenamefont {Smith}\ and\ \citenamefont
  {Erdman}(1974)}]{ComMat1974}%
  \BibitemOpen
  \bibfield  {author} {\bibinfo {author} {\bibfnamefont {W.}~\bibnamefont
  {Smith}}\ and\ \bibinfo {author} {\bibfnamefont {S.}~\bibnamefont {Erdman}},\
  }\bibfield  {title} {\bibinfo {title} {A note on the inversion of complex
  matrices},\ }\href {https://doi.org/10.1109/TAC.1974.1100466} {\bibfield
  {journal} {\bibinfo  {journal} {IEEE Transactions on Automatic Control}\
  }\textbf {\bibinfo {volume} {19}},\ \bibinfo {pages} {64} (\bibinfo {year}
  {1974})}\BibitemShut {NoStop}%
\bibitem [{\citenamefont {Petersen}\ and\ \citenamefont
  {Pedersen}(2012)}]{MatCook2012}%
  \BibitemOpen
  \bibfield  {author} {\bibinfo {author} {\bibfnamefont {K.~B.}\ \bibnamefont
  {Petersen}}\ and\ \bibinfo {author} {\bibfnamefont {M.~S.}\ \bibnamefont
  {Pedersen}},\ }\href {http://www2.compute.dtu.dk/pubdb/pubs/3274-full.html}
  {\bibinfo {title} {The matrix cookbook}} (\bibinfo {year} {2012}),\ \bibinfo
  {note} {version 20121115}\BibitemShut {NoStop}%
\bibitem [{\citenamefont {Huang}\ and\ \citenamefont {Wang}(2017)}]{Huang2017}%
  \BibitemOpen
  \bibfield  {author} {\bibinfo {author} {\bibfnamefont {L.}~\bibnamefont
  {Huang}}\ and\ \bibinfo {author} {\bibfnamefont {L.}~\bibnamefont {Wang}},\
  }\bibfield  {title} {\bibinfo {title} {Accelerated monte carlo simulations
  with restricted boltzmann machines},\ }\href
  {https://doi.org/10.1103/PhysRevB.95.035105} {\bibfield  {journal} {\bibinfo
  {journal} {Phys. Rev. B}\ }\textbf {\bibinfo {volume} {95}},\ \bibinfo
  {pages} {035105} (\bibinfo {year} {2017})}\BibitemShut {NoStop}%
\bibitem [{\citenamefont {Peschel}\ and\ \citenamefont
  {Eisler}(2009)}]{Peschel2009}%
  \BibitemOpen
  \bibfield  {author} {\bibinfo {author} {\bibfnamefont {I.}~\bibnamefont
  {Peschel}}\ and\ \bibinfo {author} {\bibfnamefont {V.}~\bibnamefont
  {Eisler}},\ }\bibfield  {title} {\bibinfo {title} {Reduced density matrices
  and entanglement entropy in free lattice models},\ }\href
  {https://doi.org/10.1088/1751-8113/42/50/504003} {\bibfield  {journal}
  {\bibinfo  {journal} {Journal of Physics A: Mathematical and Theoretical}\
  }\textbf {\bibinfo {volume} {42}},\ \bibinfo {pages} {504003} (\bibinfo
  {year} {2009})}\BibitemShut {NoStop}%
\bibitem [{\citenamefont {Peschel}(2003)}]{Peschel2003}%
  \BibitemOpen
  \bibfield  {author} {\bibinfo {author} {\bibfnamefont {I.}~\bibnamefont
  {Peschel}},\ }\bibfield  {title} {\bibinfo {title} {Calculation of reduced
  density matrices from correlation functions},\ }\href
  {https://doi.org/10.1088/0305-4470/36/14/101} {\bibfield  {journal} {\bibinfo
   {journal} {Journal of Physics A: Mathematical and General}\ }\textbf
  {\bibinfo {volume} {36}},\ \bibinfo {pages} {L205} (\bibinfo {year}
  {2003})}\BibitemShut {NoStop}%
\bibitem [{\citenamefont {Turner}\ \emph {et~al.}(2010)\citenamefont {Turner},
  \citenamefont {Zhang},\ and\ \citenamefont {Vishwanath}}]{Turner2010}%
  \BibitemOpen
  \bibfield  {author} {\bibinfo {author} {\bibfnamefont {A.~M.}\ \bibnamefont
  {Turner}}, \bibinfo {author} {\bibfnamefont {Y.}~\bibnamefont {Zhang}},\ and\
  \bibinfo {author} {\bibfnamefont {A.}~\bibnamefont {Vishwanath}},\ }\bibfield
   {title} {\bibinfo {title} {Entanglement and inversion symmetry in
  topological insulators},\ }\href {https://doi.org/10.1103/PhysRevB.82.241102}
  {\bibfield  {journal} {\bibinfo  {journal} {Phys. Rev. B}\ }\textbf {\bibinfo
  {volume} {82}},\ \bibinfo {pages} {241102} (\bibinfo {year}
  {2010})}\BibitemShut {NoStop}%
\bibitem [{\citenamefont {Frankle}\ and\ \citenamefont
  {Carbin}(2019)}]{Frankle2018}%
  \BibitemOpen
  \bibfield  {author} {\bibinfo {author} {\bibfnamefont {J.}~\bibnamefont
  {Frankle}}\ and\ \bibinfo {author} {\bibfnamefont {M.}~\bibnamefont
  {Carbin}},\ }\bibfield  {title} {\bibinfo {title} {The lottery ticket
  hypothesis: Finding sparse, trainable neural networks},\ }in\ \href
  {https://openreview.net/forum?id=rJl-b3RcF7} {\emph {\bibinfo {booktitle}
  {International Conference on Learning Representations}}}\ (\bibinfo {year}
  {2019})\BibitemShut {NoStop}%
\bibitem [{\citenamefont {Blalock}\ \emph {et~al.}(2020)\citenamefont
  {Blalock}, \citenamefont {Gonzalez~Ortiz}, \citenamefont {Frankle},\ and\
  \citenamefont {Guttag}}]{Blalock2020}%
  \BibitemOpen
  \bibfield  {author} {\bibinfo {author} {\bibfnamefont {D.}~\bibnamefont
  {Blalock}}, \bibinfo {author} {\bibfnamefont {J.~J.}\ \bibnamefont
  {Gonzalez~Ortiz}}, \bibinfo {author} {\bibfnamefont {J.}~\bibnamefont
  {Frankle}},\ and\ \bibinfo {author} {\bibfnamefont {J.}~\bibnamefont
  {Guttag}},\ }\bibfield  {title} {\bibinfo {title} {What is the state of
  neural network pruning?},\ }\href@noop {} {\bibfield  {journal} {\bibinfo
  {journal} {Proceedings of machine learning and systems}\ }\textbf {\bibinfo
  {volume} {2}},\ \bibinfo {pages} {129} (\bibinfo {year} {2020})}\BibitemShut
  {NoStop}%
\bibitem [{\citenamefont {Ashida}\ \emph {et~al.}(2020)\citenamefont {Ashida},
  \citenamefont {Gong},\ and\ \citenamefont {Ueda}}]{nhermi2020}%
  \BibitemOpen
  \bibfield  {author} {\bibinfo {author} {\bibfnamefont {Y.}~\bibnamefont
  {Ashida}}, \bibinfo {author} {\bibfnamefont {Z.}~\bibnamefont {Gong}},\ and\
  \bibinfo {author} {\bibfnamefont {M.}~\bibnamefont {Ueda}},\ }\bibfield
  {title} {\bibinfo {title} {Non-hermitian physics},\ }\href
  {https://doi.org/10.1080/00018732.2021.1876991} {\bibfield  {journal}
  {\bibinfo  {journal} {Advances in Physics}\ }\textbf {\bibinfo {volume}
  {69}},\ \bibinfo {pages} {249} (\bibinfo {year} {2020})}\BibitemShut
  {NoStop}%
\bibitem [{\citenamefont {Bergholtz}\ \emph {et~al.}(2021)\citenamefont
  {Bergholtz}, \citenamefont {Budich},\ and\ \citenamefont
  {Kunst}}]{nhermi2021}%
  \BibitemOpen
  \bibfield  {author} {\bibinfo {author} {\bibfnamefont {E.~J.}\ \bibnamefont
  {Bergholtz}}, \bibinfo {author} {\bibfnamefont {J.~C.}\ \bibnamefont
  {Budich}},\ and\ \bibinfo {author} {\bibfnamefont {F.~K.}\ \bibnamefont
  {Kunst}},\ }\bibfield  {title} {\bibinfo {title} {Exceptional topology of
  non-hermitian systems},\ }\href
  {https://doi.org/10.1103/RevModPhys.93.015005} {\bibfield  {journal}
  {\bibinfo  {journal} {Rev. Mod. Phys.}\ }\textbf {\bibinfo {volume} {93}},\
  \bibinfo {pages} {015005} (\bibinfo {year} {2021})}\BibitemShut {NoStop}%
\end{thebibliography}%

\end{document}